\documentclass[fleqn,usenatbib]{mnras}
\usepackage{newtxtext,newtxmath}
 
\usepackage[T1]{fontenc}

\DeclareRobustCommand{\VAN}[3]{#2}
\let\VANthebibliography\thebibliography
\def\thebibliography{\DeclareRobustCommand{\VAN}[3]{##3}\VANthebibliography} 
\usepackage{graphicx}	
\usepackage{amsmath}	

\usepackage{threeparttable}

\newcommand{\CIVdblt}{\textrm{C}\,\textsc{\large{iv}}~\lambda\lambda 1548, 1550}

\newcommand{\CII}{\textrm{C}\,\textsc{\large{ii}}}

\newcommand{\CIII}{\textrm{C}\,\textsc{\large{iii}}}
\newcommand{\CIV}{\textrm{C}\,\textsc{\large{iv}}}
\newcommand{\CIVl}{\textrm{C}\,\textsc{\Large{iv}}}
\newcommand{\CIVm}{\textrm{C}\,\textsc{\mdseries{iv}}}
\def\H{\textrm{H}}
\newcommand{\HI}{\textrm{H}\,\textsc{\large{i}}}
\newcommand{\HIm}{\textrm{H}\,\textsc{\textmd{i}}}
\newcommand{\HIl}{\textrm{H}\,\textsc{\Large{i}}}

\newcommand{\Lya}{\textrm{Ly}\alpha}
\newcommand{\Lyb}{\textrm{Ly}\beta}

\newcommand{\MgII}{\textrm{Mg}\,\textsc{\large{ii}}}

\newcommand{\NV}{\textrm{N}\,\textsc{\large{v}}}

\newcommand{\NII}{\textrm{N}\,\textsc{\large{ii}}}

\newcommand{\OII}{\textrm{O}\,\textsc{\large{ii}}}

\newcommand{\OVI}{\textrm{O}\,\textsc{\large{vi}}}

\newcommand{\SiII}{\textrm{Si}\,\textsc{\large{ii}}}
\newcommand{\SiIII}{\textrm{Si}\,\textsc{\large{iii}}}

\newcommand{\SiIV}{\textrm{Si}\,\textsc{\large{iv}}}

\newcommand{\NeVIII}{\textrm{Ne}\,\textsc{\large{viii}}}
\newcommand{\kms}{\textrm{km\,s}^{-1}}
\newcommand{\cmsq}{\textrm{cm}^{-2}}
\newcommand{\cc}{\textrm{cm}^{-3}}

\graphicspath{{plots/}}

\title[C\,{\textsc{\Large{iv}}} absorbers in the low-$z$ Universe]{The COS-legacy survey of C\,\textsc{\Huge iv} absorbers: properties and origins of the intervening systems}

\author[A. Manuwal et al.]{Aditya Manuwal,$^{1,2}$\thanks{E-mail: aditya.manuwal@icrar.org}
Anand Narayanan,$^{2}$
~Purvi Udhwani,$^{2}$
~Raghunathan Srianand,$^{3}$
~Blair D. Savage,$^{4}$
\newauthor
Jane C. Charlton$^{5}$
~and Toru Misawa$^{6}$
\\
$^{1}$International Centre for Radio Astronomy Research, University of Western Australia, 35 Stirling Highway, Crawley, WA 6009, Australia\\
$^{2}$Department of Earth and Space Sciences, Indian Institute of Space Science \& Technology, Thiruvananthapuram 695547, Kerala, India\\
$^{3}$Inter-University Centre for Astronomy and Astrophysics, Post Bag 4, Pune 411007, India\\
$^{4}$Department of Astronomy, The University of Wisconsin-Madison, 5534 Sterling Hall, 475 N. Charter Street, Madison, WI 53706-1582, USA\\
$^{5}$The Pennsylvania State University, 413 Davey Lab, University Park, State College, PA 16802, USA\\
$^{6}$School of General Education, Shinshu University, 3-1-1 Asahi, Matsumoto, Nagano 390-8621, Japan
}

\date{Accepted 2021 May 26. Received 2021 May 26; in original form 2020 September 8}

\pubyear{2021}

\begin{document}
\label{firstpage}
\pagerange{\pageref{firstpage}--\pageref{lastpage}}
\maketitle

\begin{abstract}
We present here results from a survey of intervening {\CIVl} absorbers at $z < 0.16$ conducted using 223 sightlines from the Hubble Spectroscopic Legacy Archive. Most systems ($83\%$) out of the total sample of 69 have simple kinematics with 1 or 2 {\CIVl} components. In the 22 {\CIVl} systems with well constrained {\HIl} column densities, the temperatures from the $b$-values imply predominantly photoionized plasma ($T\leq 10^5$~K) and non-thermal dynamics. These systems also have solar or higher metallicities. We obtain a {\CIVl} line density of $d\mathcal{N}/dX = 5.1\pm 1.0$ for $\log [N(\CIVl)~(\cmsq)]\geq12.9$, and $\Omega_{\CIVm}=(8.01\pm 1.62) \times 10^{-8}$ for $12.9 \leq \log [N(\CIVl)~(\cmsq)] \leq 15.0$. The {\CIVl} bearing diffuse gas in the $z < 0.16$ Universe has a metallicity of
$(2.07~{\pm}~0.43)~\times~10^{-3}$~Z$_{\odot}$, an order of magnitude more than the metal abundances in the IGM at high redshifts ($z \gtrsim 5$), and consistent with the slow build-up of metals in the diffuse circum/intergalactic space with cosmic time. For $z<0.015$ (complete above $L>0.01L^\star$), the Sloan Digital Sky Survey provides a tentative evidence of declining covering fraction for strong {\CIVl} ($N>10^{13.5}~\cmsq$) with $\rho$ (impact parameter) and $\rho/R_\mathrm{vir}$. However, the increase at high separations suggests that strong systems are not necessarily coincident with such galaxies. We also find that strong {\CIVl} absorption at $z<0.051$ is not coincident with galaxy over-dense regions complete for $L>0.13L^\star$.
\end{abstract}




\section{INTRODUCTION}

In the past few decades, results from simulations of galaxy formation in the frame-work of large cosmological volumes \citep{Cen1999,Dave1999,Dave2001} and observations \citep{Fukugita1998,Fukugita2004,Shull2012,Stocke2013,Werk2014,Prochaska2017} have established that most of the baryons reside outside the stellar disks of galaxies in the circumgalactic (CGM) or intergalactic medium (IGM). The gaseous clouds containing these baryons exist in a multi-phase structure and have substantial influence on the evolution of galaxies. The IGM contributes baryons to the galaxies through gravitational infall of gas clouds. Galaxies in turn enrich the CGM and the IGM through metal rich outflows initiated by supernova/starburst driven galactic scale winds, or tidal interactions \citep{Heckman1990,Heckman2001,Martin2002,Scannapieco2002,Hopkins2006,Keeney2006,Songaila2006,Rupke2011,Tripp2011,Tumlinson2011,Stocke2013,Muzahid2015,Keeney2017,Pratt2018}.

Even at low redshifts ($z \lesssim 0.5$), the mass fraction of baryons in the cool ($T \sim 10^{4}$~K), warm ($T \sim 10^{5} - 10^{6}$~K) and in the hot ($T > 10^6$~K) phase outside of the star forming regions of galaxies far outweighs the mass fraction inside galaxies \citep{Salucci1999,Dave2001,Fukugita2004}. An observational census of the baryons in the low redshift Universe is incomplete without properly accounting for the warm-hot phase. Since the gas at $T<10^{6}$~K can be detected using absorption lines in spectra of background sources at high redshift (like GRBs or QSOs), a plethora of absorption line studies have been focused on detecting these baryonic reservoirs in the circumgalactic and intergalactic space around galaxies, characterizing their physical properties and chemical abundances, and understanding their origins \citep[e.g][and references therein]{Fang2010,Danforth2011,Narayanan2011,Burchett2018}. 

The cool phase also possesses a significant fraction of baryonic matter in the local Universe ($\sim 30\%$) \citep{Penton2004,Lehner2007} and has therefore been explored extensively through {\MgII} \citep[e.g.][]{Narayanan2005,Chen2010,Churchill2013,Nielsen2015,Bowen2016}, {\CIV} \citep[e.g.][]{Schaye2003,Songaila2006,Schaye2007,Cooksey2010,Burchett2013}, and also {\CII} and {\SiII} which are used as proxies for {\MgII}. {\CIV} traces a higher ionization, more diffuse phase than {\MgII} with densities that are lower by 1 - 2 orders of magnitude \citep{Ding2003,Charlton2003}. Multiple, but closely separated, lines of sight observations show this phase to be more smoothly distributed over sub-kiloparsec to several kiloparsec scales; potentially tracing the diffuse photoionized halos of galaxies \citep{Lopez1999,Rauch2001}. Based on these, {\MgII} seems to be tracing compact pockets of high density gas, often having high metallicity (near-solar or solar), and with sizes of a few tens of parsec, whereas {\CIV} covers a larger dynamic range.

It has been shown that cool gas clouds can serve as tools to comprehend chemical enrichment, physical conditions and the effect of environment on the CGM \citep{Stocke2013,Yoon2013,Burchett2016,Keeney2017,Lehner2018}. In galaxy dense environments like rich groups and clusters, most of the gas is shock heated to temperatures of $T > 10^{7}$~K \citep{Mushotzky1978,Loken2002,Roncarelli2006,DeGrandi2004} and cool gas clouds can arise in intergalactic/intracluster space through thermal instability \citep{Maller2004,Thompson2016}, or stripping mechanisms like galaxy mergers \citep{Hani2017,Burchett2018}, tidal interactions \citep{Morris1993} and ram pressure \citep{Abadi1999}. This implies that the chemical and physical properties of cool gas clouds, along with their large scale environment, can be used to gain insights about their origins, and the physical mechanisms involved in interactions of the galaxies with their surrounding medium which play key roles in galaxy evolution. It is better to carry out such studies at low redshifts since that allows exploring relationships between absorbers and nearby galaxies, especially sub-$L^\star$ galaxies which contribute significantly to feedback processes.

{\CIV} or triply ionized carbon is an excellent tracer of enriched, diffuse cool gas \citep{Buson1990,Jenkins2003,Schaye2003,Simcoe2004,Songaila2006,Fox2007,Schaye2007,Tejos2009,Ranquist2012,Burchett2013,Burchett2016,Burchett2018}. Most of the earlier studies were focused on absorbers at high redshifts ($z\gtrsim1.5$). \citet{Rauch1998} used simulations of the $z\sim3$ universe and predicted that {\CIV} is a good tracer of filaments and small protogalactic clouds. Surveys of {\CIV} absorbers at high redshifts claim the gas to be mostly photoionized with median [C/H]~$\sim-3$ \citep{Schaye2003,Simcoe2004}, and dominated by thermal energy \citep{Ranquist2012}. \citet{Songaila2006} finds most low column density absorbers to reside in the IGM and suggests that half of the high column density systems are possibly produced by outflows. The metal rich systems at these redshifts are compact and short lived, which also suggests transport of metals into the IGM \citep{Schaye2007}. 

The advent of the Cosmic Origins Spectrograph (COS) onboard the Hubble Space Telescope (\textit{HST}) has offered a unique opportunity for conducting such surveys at low redshift owing to its unprecedented sensitivity and performance. The studies that followed show that {\CIV} absorbers at low redshifts can represent both collisional and photoionized gas \citep{Danforth2008}, and the ones within galactic halos tend to be metal rich \citep{Stocke2013}. As expected, they are also associated with star forming galaxies and strong outflows are required to explain the observations \citep{Bordoloi2014}. Such clouds are also seen to be accreted by galaxies \citep{Burchett2013}. The largest blind survey of {\CIV} absorbers yet at $z < 0.16$ was done using the \textit{HST}/COS spectra towards $89$ QSO sightlines \citep{Burchett2015,Burchett2016} but their results based on ion-ion relations are not statistically significant since the survey was limited to a small sample size and a study with a larger sample is required; as also emphasized by \citet{Burchett2015}.

In this study, we use the data from the Hubble Spectroscopic Legacy Archive\footnote{\url{https://archive.stsci.edu/hst/spectral\_legacy/}} \citep{Peeples2017} to identify {\CIV} absorbers at $z < 0.16$ in the FUV spectra of 223 background quasars. Through this work we aim to leverage better statistics provided by the archive and thus conduct the largest survey of such absorption systems at low redshift. We infer the physical conditions and chemical enrichment of the clouds by drawing interpretations from ionization modelling and considering their galaxy environments and use these as contexts to hypothesize on the origins of {\CIV} absorbers. This paper is divided into 7 sections. Sec.~\ref{select} delineates the methodology behind selecting the systems in our sample. The completeness of this survey is limited by the achieved spectral signal-to-noise ratio, this has been addressed in Sec.~\ref{complete}. The techniques involved in measuring absorption line properties and results from those measurements are described in Sec.~\ref{measure}. The absorber statistics and related parameters are derived in Sec.~\ref{stats}. In Sec.~\ref{models}, we describe the ionization modelling of absorbers and the cloud properties derived from it. Sec.~\ref{galaxies} explores the distribution and properties of galaxies around the absorbers. We summarize the results in Sec.~\ref{summary}. Additionally, we hypothesize about the plausible origins of some specific clouds in a document provided as supplementary material, and we recommend the reader to go through it as well. Throughout the paper, we adopt the cosmology with $H_0 = 69.6~\kms$~Mpc$^{-1}$, $\Omega_\textrm{m} = 0.286$ and $\Omega_{\Lambda} = 0.714$ from \citet{Bennett2014}. All the logarithmic values mentioned are in base-10. The python routines used for data analysis in this study can be found at our online repository\footnote{\url{https://github.com/adimanuwal/ROQALS}}.

\section{SYSTEM SELECTION}\label{select}

For generating a statistically significant sample of {\CIV} absorbers, we have used \textit{HST}/COS FUV archival spectra of 223 quasars. The spectroscopic abilities of COS and its in-flight performances are given in \citet{Green2012} and \citet{Osterman2011}. The FUV wavelength coverage of the data lies in the range of $1136 - 1795$~\AA~at medium spectral resolutions (FWHM) of $\sim (17-20)$~$\kms$. The data acquisition involved both G130M and G160M gratings for 216 sightlines, and only G160M grating for the other 7 sightlines. This limits the {\CIV} survey redshift to $z_\mathrm{max} \sim 0.157$ requiring coverage of both members of the {\CIV} doublet. Continuum models were generated for each spectrum by fitting low order polynomials locally to $\sim 10 - 20$~{\AA} wavelength blocks, excluding regions with absorption features, and covering broad emission lines.

The search for {\CIV} lines was carried out with the following conditions: (1) the spectrum should cover both $\lambda1548$ and $\lambda1550$ transitions of the doublet with visually similar profiles, (2) the $\CIV~\lambda1548$ line should have a formal detection significance of $\ge 3\sigma$, (3) the equivalent width ratio for the unsaturated $\CIVdblt$ lines that are not blended with some other absorption should be approximately consistent with their expected value of $\approx 2:1$ within the measurement uncertainty, and (4) the absorber should be at $|\Delta v| > 5000$~$\kms$ from the emission redshift of the background quasar, for it to be not associated with the quasar \citep{Foltz1986,Misawa2007,Ganguly2013,Muzahid2013}. The profiles were checked for contaminations by comparing the apparent column density profiles of the doublet transitions, and using the line identifications\footnote{\url{https://archive.stsci.edu/prepds/igm/}} provided by \citet{Danforth2016}. An absorber is included in our sample only if all of the above criteria are satisfied. A search was carried out through each sightline by blindly assuming every absorption feature redward of $1548$~{\AA} to be a $\CIV~\lambda1548$ line. The corresponding $1550$~{\AA} line was then searched for at its redshifted wavelength. For every positive matching of the doublets, the apparent column density profiles were also inspected for agreement. As additional validation, the presence of other metal lines and {\HI} at this redshift were also considered, though this was not a necessary condition for claiming a {\CIV} detection. The pixel in the {\CIV}~$\lambda1548$ feature with the maximum optical depth was used to define the redshift of the system. For saturated {\CIV} components, we adopt the redshift corresponding to the saturated pixel with the lowest velocity. 

\section{SURVEY COMPLETENESS}\label{complete}

The 223 archival QSO spectra were obtained as part of different observing programs and hence have varying levels of spectral sensitivity towards the detection of lines. The sample constitutes spectroscopic data that was available through the Hubble Spectroscopic Legacy Archive as of June 2017. From this larger sample, we excluded 10 sightlines that did not offer any path length for {\CIV} systems based on our search criteria explained later. To assess the completeness of our survey and to better understand the statistical implications of our results, cumulative redshift path lengths for {\CIV} over the column density range of $N(\CIV) = 10^{12.8} - 10^{14.9}$~$\cmsq$ are determined (which encompasses the \textit{total} {\CIV} column density in our systems\footnote{The $z_\mathrm{abs}=0.08940$ system towards QSO B1435-0645 has a saturated {\CIV} component that is excluded from analysis in this paper}). The path length along a given sightline is estimated by summing $\Delta z$ over only those pixels where a {\CIV} detection is possible, avoiding wavelengths where search is impossible due to the presence of Galactic or spectral features intrinsic to the QSO, gaps in the data, and a few narrow regions with sharp variations in the quasar continuum due to noise. Some of the sightlines were observed to probe IGM/CGM around known galaxies. Since this is a blind survey, we also exclude the spectral regions within $|\Delta v|<500~\kms$ of the targeted galaxies in such sightlines when computing pathlengths or absorber statistics. We use the formalism given in \citet{Burchett2015} for calculating the rest-frame equivalent width threshold required for a $3\sigma$ detection of the $\CIV~\lambda1548$ line with a column density of $N(\CIV)$. The rest equivalent width limit $W_\mathrm{lim}$ for a $N_\mathrm{\sigma}$ detection of the feature centered at a pixel located at the wavelength corresponding to redshift $z$ is
\begin{equation}\label{thres}
W_\mathrm{lim}=\frac{N_\mathrm{\sigma}~\sigma_{w}}{1+z},
\end{equation}
\noindent where $N_\mathrm{\sigma} = 3$ is the desired detection significance, and $\sigma_{w}$ is the uncertainty in equivalent width calculated over a $102~\kms$ wide region, which is the mean width of \textit{total} $\CIV$ absorption in our systems. It is determined as
\begin{equation}
\sigma_{w}=\sqrt{\sum_{i}\left(\frac{\sigma_{I_i}\Delta\lambda_i}{I_i}\right)^2},
\end{equation}
where $I_i$ is the continuum flux at the $i$th pixel, $\sigma_{I_i}$ is the uncertainty in $I_i$, and  $\Delta\lambda_i$ is wavelength interval at the pixel. The $W_\mathrm{lim}$ value is converted to corresponding column density threshold $N_\mathrm{lim}$ assuming the line to be on the Curve of Growth (CoG) for $b(\CIV)= 14~\kms$; the mean $b$-value for our sample (the CoG is generated using the \textsc{\Large v}\textsc{oigt}\textsc{\Large f}\textsc{it} python module\footnote{\url{https://github.com/jkrogager/VoigtFit}}. Along a given sightline, we only examine pixels at $z > 0.00167$ considering that the line should arise from an extragalactic cloud displaced by $|\Delta v| > 500$~$\kms$ from $z = 0$ \citep{Mcclure2009}. This corresponds to an observed wavelength of $\sim 1550.78$~{\AA} for $\lambda 1548$ which is well within the FUV coverage of COS. We should mention here that all the 223 QSOs have $z_\mathrm{em} > 0.01852$, or $\Delta v>500~\kms$ with respect to $z=0.00167$. The maximum redshift for consideration depends on two factors: (1) non-association with the QSO ($|\Delta v| > 5000$~$\kms$ from the quasar redshift) and, (2) coverage of the $\lambda 1550$ line of the doublet.

\begin{figure}
    \begin{center}
        \includegraphics[width=230pt,height=170pt,trim={0.3cm 0.4cm 0.15cm 0.3cm},clip=true]{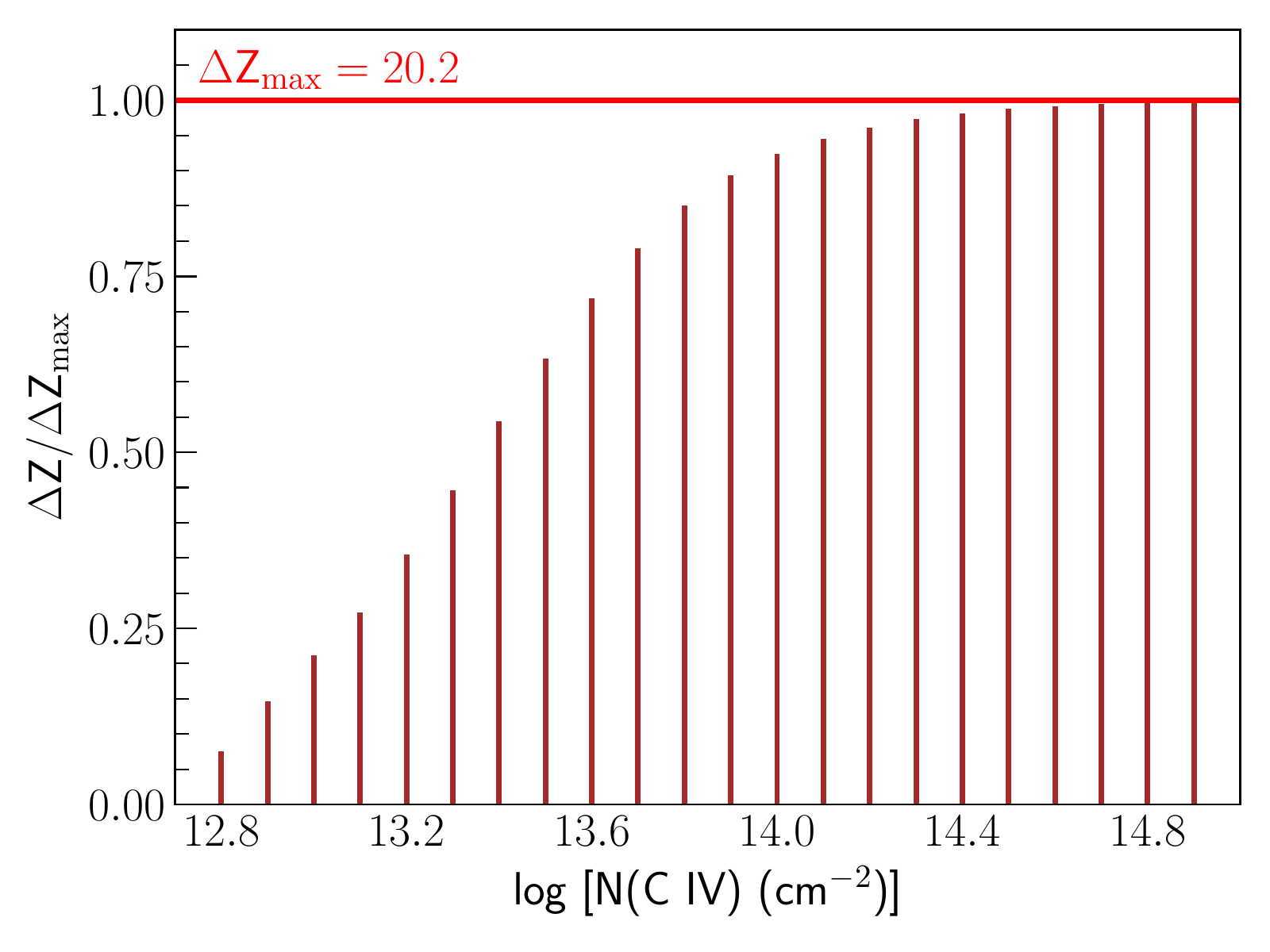}
    \caption{The cumulative redshift path length for different {\CIV} column densities of $\lambda 1548$ normalized by $\Delta Z_\mathrm{max} = 20.2$, which corresponds to $\Delta Z$ for $N(\CIV) = 10^{14.9}~\cmsq$.}
    \label{1}
    \end{center}
\end{figure}

\begin{table}
\caption{Pathlengths for different {\CIV} column densities}
\begin{center}
\begin{threeparttable}
\begin{tabular}{lcccc}
\hline
$\log[N(\CIV)~(\cmsq)]$ & $W_\mathrm{r}(\lambda1548)$ (m\AA) & $\Delta Z$ & $\Delta X$ & $\mathcal{N}$ \\
(1) & (2) & (3) & (4) &  (5) \\
\hline
$12.8$ & $23.7$ & $1.6$ & $1.8$ & $0$\\
$12.9$ & $29.3$ & $3.0$ & $3.3$ & $2$\\
$13.0$ & $36.0$ & $4.4$ & $4.8$ & $0$\\
$13.1$ & $44.1$ & $5.6$ & $6.2$ & $3$\\
$13.2$ & $53.5$ & $7.3$ & $8.1$ & $1$\\
$13.3$ & $64.5$ & $9.2$ & $10.1$ & $1$\\
$13.4$ & $77.0$ & $11.2$ & $12.4$ & $3$\\
$13.5$ & $91.0$ & $12.9$ & $14.3$ & $3$\\
$13.6$ & $105.7$ & $14.7$ & $16.2$ & $3$\\
$13.7$ & $121.4$ & $16.1$ & $17.8$ & $6$\\
$13.8$ & $137.2$ & $17.3$ & $19.2$ & $4$\\
$13.9$ & $152.8$ & $18.1$ & $20.1$ & $4$\\
$14.0$ & $167.5$ & $18.7$ & $20.8$ & $2$\\
$14.1$ & $181.1$ & $19.1$ & $21.3$ & $4$\\
$14.2$ & $193.7$ & $19.4$ & $21.6$ & $8$\\
$14.3$ & $205.1$ & $19.7$ & $21.9$ & $5$\\
$14.4$ & $215.7$ & $19.8$ & $22.1$ & $2$\\
$14.5$ & $225.7$ & $19.9$ & $22.2$ & $1$\\
$14.6$ & $235.1$ & $20.0$ & $22.3$ & $0$\\
$14.7$ & $244.2$ & $20.1$ & $22.3$ & $0$\\
$14.8$ & $252.8$ & $20.1$ & $22.4$ & $0$\\
$14.9$ & $261.2$ & $20.2$ & $22.5$ & $0$\\
\hline
\end{tabular}
\label{pathtab}
\begin{tablenotes}
\item[] (1) {\CIV} column density; (2) Equivalent width correspond to the column density according to $\lambda1548$ CoG for $b(\CIV)=14~\kms$; (3) Redshift pathlength for $\geq 3\sigma$ detection based on the formalism of Eq.~\ref{thres} and Eq.~\ref{cumpath}; (4) Co-moving pathlength; (5) Number of absorbers within $\pm 0.05$~dex of the logarithmic column density (excluding the systems around targeted galaxies).
\end{tablenotes}
\end{threeparttable}
\end{center}
\end{table}

A line with a given $N(\CIV)$ will be a 3$\sigma$ detection only at those pixels where $N_\mathrm{lim}<N(\CIV)$. For a $k$-th sightline, the total pathlength $\Delta z_k$ for $N(\CIV)$ is the sum of all such pixels. The cumulative redshift path length $\Delta Z$ for this $N(\CIV)$ is determined by coadding such redshift path lengths for the entire sample of 223 QSOs in the following manner,
\begin{equation}\label{cumpath}
\Delta Z=\sum_{k=1}^{M}{\Delta z}_k,
\end{equation}
\noindent where $M=223$ is the number of sightlines. The resultant pathlengths are tabulated  in Table.~\ref{pathtab}. The cumulative redshift path lengths for a $3\sigma~\CIV$ detection suggest that the survey reaches $\Delta Z_\mathrm{max} = 20.2$ for $N(\CIV) = 10^{14.9}$~$\cmsq$. This is almost a two fold increase compared to earlier such surveys by \citet{Burchett2015} and \citet{Danforth2016} , which had $\Delta Z = 12.9$ and $\Delta Z = 8.85$, respectively. 

We discovered 69 {\CIV} absorbers spanning the redshift interval $0.002 < z < 0.154$ that satisfied the selection criteria mentioned in Sec.~\ref{select}. This sample is larger by 27 compared to \citet{Burchett2015}, and by 1 compared to \citet{Danforth2016}, which searched for systems in the same redshift range. The basic properties of all absorbers are summarized in Table~\ref{tab1} and Table~\ref{tab2}. Our sample has an overlap of 3 systems with \citet{Cooksey2010}, 18 systems with that of \citet{Burchett2015}, and 26 with \citet{Danforth2016}. The reader may refer to Table~\ref{tab1} for individual cases.

\section{ABSORBER LINE MEASUREMENTS}\label{measure} 

The line measurements are carried out using two methods: (1) the integrated apparent optical depth (AOD) method given by \citet{Savage1991} and (2) Voigt  profile fitting using the \textsc{\large{vpfit}} routine \citep[version 12.1;][]{Carswell2014}, where the profiles are convolved with the relevant model COS instrumental spread functions\footnote{\url{https://www.stsci.edu/hst/instrumentation/cos/performance/spectral-resolution}}. We have accounted for COS lifetime positions while selecting the line spread functions. In addition to constraining the column density, the best-fit models also provide the centroids of the absorption features and the line widths in terms of the Doppler parameter $b$ in $\kms$. Wavelength calibration of COS is known to possess residual offsets as large as $|\Delta v|= 40$~$\kms$ (see appendix of \citealt{Wakker2015}). Velocity offsets of $\leq 40$~$\kms$ between lines in an individual absorber are therefore treated as within this systematic uncertainty. 

\begin{figure}
    \begin{center}
       \includegraphics[width=230pt,height=170pt,trim={0.3cm 0.4cm 0.15cm 0.3cm},clip=true]{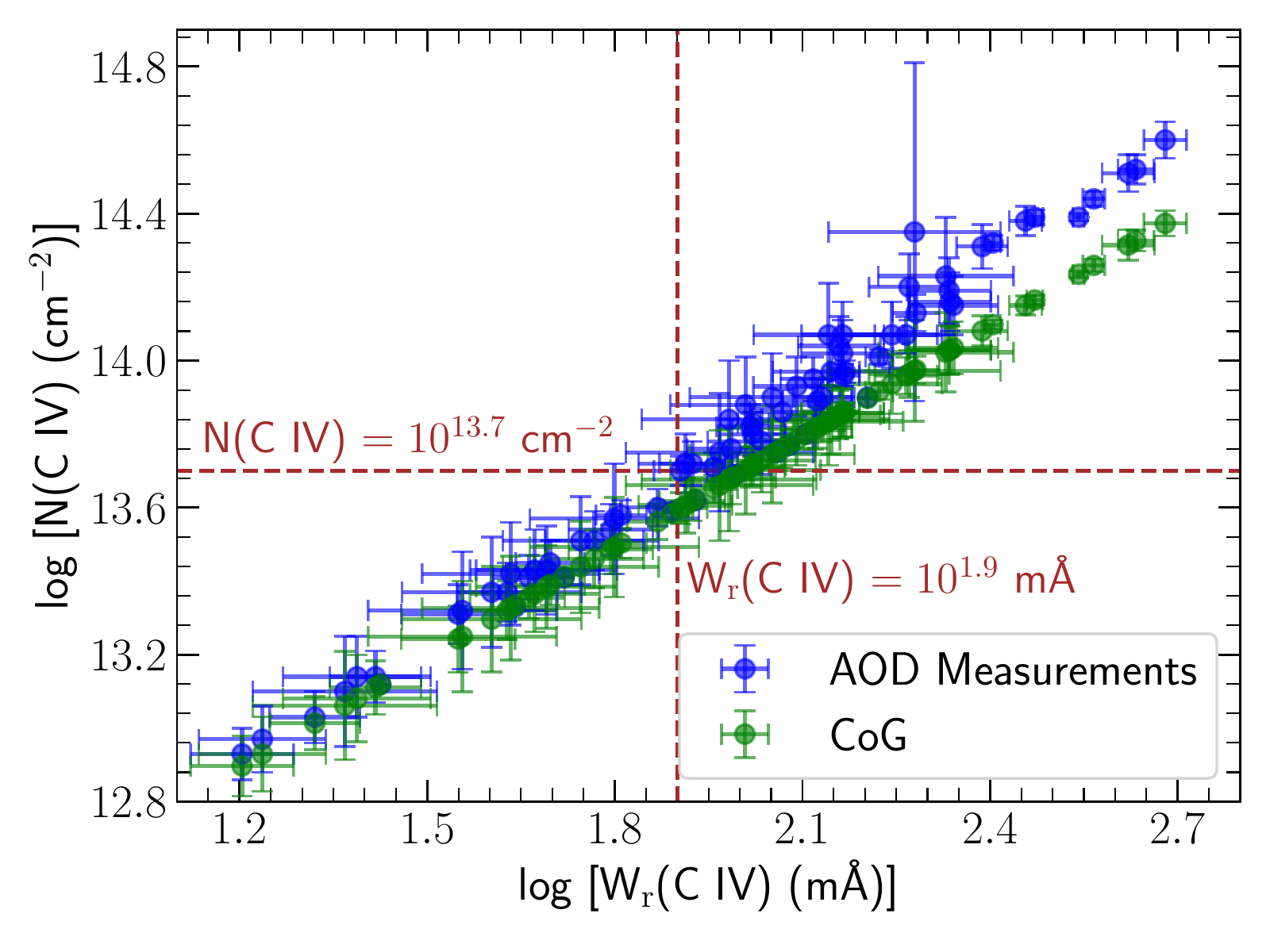}
    \caption{Comparison of $\CIV~\lambda1550$ column densities measured using the AOD method with those predicted by the linear part of CoG for the measured rest-frame equivalent widths. The comparison shows that line saturation begins for $N(\CIV) \gtrsim 10^{13.7}~\cmsq$.}
    \label{2}
    \end{center}
\end{figure}

Profiles are simultaneously fitted to all available absorption lines from the same ion. For two ions that are known to trace gas of similar ionization (e.g., {\CII} and {\SiII}, or {\CIV} and {\SiIV}), the model fits to the lines were guided fits (in terms of the number of components, and their velocity centroids), if adequate constraints are otherwise not available. Only those lines that are detected at 3$\sigma$, with no obvious indication of contamination\footnote{Based on the comparison of apparent column density profiles of multiple transitions from the same ion, and features identified by \citet{Danforth2016}}, are included in the fitting. \textsc{\large{vpfit}} compensates for mild levels of line saturation, especially while fitting multiple lines from the same ion. $\CIV~\lambda1550$ is a weaker line than $\CIV~\lambda1548$ and therefore an accurate estimation of {\CIV} column density is governed by the saturation level of $\lambda1550$. It is possible to have a constrained measurement for {\CIV} in a system with a saturated $\lambda1548$ if the $\lambda1550$ is moderate to mildly saturated, because both lines of the doublet are fit together with Voigt profile models, which provides a unique solution of $b$ and $N$. We assess the saturation level of our profiles through a combination of visual inspection and comparison with the linear part of CoG (Fig.~\ref{2}). We consider a line to be saturated if the measured column density and that based on the linear CoG are not consistent within their uncertainties. The lowest column density where this happens is $N(\CIV) = 10^{13.7}$~$\cmsq$. This line has $W_\mathrm{r}=79.43$~m{\AA} which implies a $b(\CIV)\approx15$, close to the median $b(\CIV)$ in our sample.

In addition to the Doppler $b$ parameter, the overall kinematics of the {\CIV} profiles are estimated through $\Delta v_{90}$ \citep{Prochaska1997}. The $\Delta v_{90}$ width is calculated as the velocity range of the absorption profile containing $90\%$ of the integrated apparent optical depth. In other words, this value corresponds to the velocity difference between the two pixels across the full range of absorption whose cumulative optical depths are $5\%$ and $95\%$ of the total.

\subsection{Column density of saturated H\,{\textsc{\large i}}}\label{nhidet}

The COS spectra mainly cover only $\Lya$ amongst the {\HI} lines for the redshift range of {\CIV} absorbers in this survey. In most cases, the $\Lya$ components are significantly saturated. In the 69 absorbers featured in this paper, $\Lyb$ is covered only in 11, which are near to the redshift upper limit of the sample of $z \approx 0.16$. In cases where the $\Lyb$ is unsaturated, accurate estimations of $N(\HI)$ are possible. In an additional 11 absorbers, components of $\Lya$ itself are unsaturated or mildly saturated. Voigt profile fits predict a narrow range of $N(\HI)$ for these as well. Thus, altogether there are 22 absorbers for which well-constrained {\HI} measurements are possible. We refer to these as the ``sample with secure {\HI}''. For the rest, the profile models do not yield a unique solution for the $N$, $b$ and $v$ of the absorbing component(s) due to line saturation. But, however large, there can only be a finite number of combinations of $N$ and $b$ that can reproduce a given saturated profile, especially at the profile edges. Moreover, if the {\CIV} (or any other metal line) is devoid of saturation and spans the same velocity as the saturated {\HI}, it is likely that the {\CIV} and the {\HI} represent the same cloud. In such cases, we also adopt for {\HI} the same component structure as {\CIV}, with the caveat that {\HI} can be more widely distributed and kinematically more complex than the metal lines. We implement these ideas to constrain the $b(\HI)$ and the $N(\HI)$ for such saturated $\Lya$ components as explained below. 

We begin by adopting the well constrained Doppler $b$ parameter for {\CIV} based on a simultaneous free profile fit to the doublet lines. The lower and upper limits on $b(\HI)$ are then found by assuming the two extremes in line broadening; the entirely non-thermal [$b(\HI) = b(\CIV)$] scenario, and the fully thermal [$b(\HI) = 3.452~\times~b(\CIV)$] scenario. This is done for aligned {\HI} and {\CIV} components, in cases where the {\HI} component structure is evident. For strongly saturated {\HI} lines, the component structure is assumed to be the same as {\CIV}. The limiting $b$-values for {\HI} thus obtained are used to generate model Voigt profiles using \textsc{\large{vpfit}} for the saturated {\HI}. A plausible range for the {\HI} column density is finally inferred by accepting those $N$ and $b$ combinations that yield fits consistent with the observed {\HI}. The column density and $b$-parameter thus inferred carry a wide uncertainty range for $N$ and $b$ compared to the sample with secure {\HI}. It should be noted that the $N(\HI)$ range thus derived represents the broadest possible range for {\HI} in the cloud that has the {\CIV}. The actual range might be narrower, as some {\HI} might not be associated with {\CIV} but rather correspond to a phase with different ionization conditions compared to the phase where the {\CIV} originates.

\begin{table*}
\caption{Basic properties of the {\CIV} absorbers}
\begin{center}
\begin{threeparttable}
\begin{tabular}{lcccccccc}
\hline
QSO & $z_\mathrm{em}$     & $z_\mathrm{abs}$     & Heavier Metals & Low Ions & High Ions & No. of {\HI} & No. of {\CIV} & $\log[\Sigma N(\CIV)~(\cmsq)]$ \\
(1) &     (2)      &    (3)        &     (4)      &      (5)     &      (6)      &      (7)  & (8) & (9)    \\
\hline
1H 1613-097 & $0.33500$ & $0.06274$ & N & N & N & $2$ & $2$ & $13.48$ \\
2MASS J10155924-2748289 & $0.24125$ & $0.00819$ & Y & Y & N & $3$ & $3$ & $14.18$ \\
3C 263  & $0.39758$ & $0.06355^\mathrm{F}$  & Y & Y & N & $4$ & $3$ & $14.06$ \\
3C 263  & $0.39758$ & $0.11391^\mathrm{\star F}$  & Y & N & Y & $3$ & $1$ & $13.35$ \\
FBQS 0751+2919  & $0.91572$ & $0.06020^\mathrm{D}$ & Y & N & N & $5$ & $3$ & $13.66$ \\
HE 0056-3622  & $0.16298$ & $0.04344^\mathrm{F}$  & N & N & N & $2$ & $1$ & $13.59$ \\
HE 0439-5254  & $1.05300$ & $0.00546^\mathrm{H}$  & N & N & N & $4$ & $3$ & $13.97$ \\
IRAS F00040+4325 & $0.16363$ & $0.12247^\star$  & Y & N & Y & $2$ & $2$ & $13.73$ \\
IRAS F04250-5718 & $0.10457$  & $0.00377$  & Y & Y & N & $4$ & $2$ & $12.88$ \\
J141949.39+060654.0 & $1.64890$ & $0.04872$  & Y & Y & Y & $3$ & $2$ & $14.45$ \\
LBQS 0107-0235 & $0.95703$ & $0.11548^\star$  & Y & N & N & $2$ & $2$ & $14.22$ \\
LBQS 1435-0134  & $1.30791$ & $0.13842^\mathrm{\star D}$ & Y & N & Y & $1$ & $1$ & $13.50$ \\
PG 0003+158  & $0.45047$ & $0.09106^\mathrm{\star F}$ & N & N & N & $3$ & $2$ & $13.84$ \\
PG 0832+251  & $0.32865$ & $0.01749^\mathrm{HF}$ & Y & Y & N & $6$ & $6$ & $14.74$ \\                  
PG 0953+414  & $0.22312$ & $0.06808^\mathrm{\star EF}$ & N & N & N & $2$ & $1$ & $13.72$ \\
PG 1116+215  & $0.17466$ & $0.13850^\mathrm{\star F}$ & Y & Y & Y & $1^\mathrm{A}$ & $1$ & $13.12$ \\
PG 1148+549 & $0.97598$  & $0.00346^\mathrm{D}$ & N & N & N & $2$ & $1$ & $13.59$ \\
PG 1202+280 & $0.16530$  & $0.05144^\mathrm{H}$ & Y & Y & N & $3$ & $1$ & $13.66$ \\
PG 1216+069  & $0.33243$ & $0.12360^\mathrm{\star F}$ & Y & N & Y & $4$ & $4$ & $14.29$ \\
PG 1222+216  & $0.43460$ & $0.05456^\mathrm{F}$ & Y & N & N & $3$ & $2$ & $13.65$ \\
PG 1222+216  & $0.43460$ & $0.09870^\mathrm{F}$ & N & N & N & $1$ & $1$  & $13.99$ \\
PG 1352+183  & $0.15200$ & $0.09407$ & N & N & N & $3$ & $3$  & $13.99$ \\
PG 1407+265 & $0.94000$  & $0.07225^\mathrm{\star D}$ & N & N & N & $2$ & $2$ & $13.67$ \\
PG 1424+240 & $0.60100^\mathrm{B}$ & $0.12119^\mathrm{\star F}$ & Y & Y & Y & $2$ & $2$ & $13.77$ \\
PG 1424+240 & $0.60100^\mathrm{B}$ & $0.14702^\mathrm{\star F}$ & Y & Y & Y & $4$ & $2$ & $14.15$ \\
PG 1522+101 & $1.32801$  & $0.07519^\mathrm{\star D}$ & Y & N & Y & $1$ & $1$ & $13.38$ \\
PHL 1811  & $0.19168$ & $0.08093^\mathrm{EF}$  & Y & Y & N & $2$ & $2$ & $13.89$ \\
PHL 1811  & $0.19168$ & $0.13542^\mathrm{\star F}$ & N & N & N & $1$ & $1$ & $13.14$ \\
PHL 2525  & $0.19975$ & $0.06668^\mathrm{F}$ & Y & N & N & $2$ & $1$ & $13.72$ \\ 
PKS 1302-102  & $0.27840$ & $0.00439^\mathrm{EF}$ & Y & N & N & $1$ & $1$ & $12.94$ \\ 
PMN J1103-2329 & $0.18617$ & $0.00400^\mathrm{HF}$ & Y & N & N & $2$ & $1$ & $14.06$ \\
PMN J1103-2329 & $0.18617$ & $0.08354^\mathrm{F}$ & Y & Y & N & $3$ & $2$ & $14.18$ \\
Q 2251+155  & $0.85900$ & $0.05054^\star$ & N & N & N & $1$ & $1$ & $13.40$ \\
Q 2251+155  & $0.85900$ & $0.15373$ & Y & Y & N & $4$ & $2$ & $14.17$ \\ 
QSO B1435-0645  & $0.12900$ & $0.08940$ & Y & N & Y & $4$ & $3$ & $14.17^\mathrm{G}$ \\
RBS 1666  & $0.07963$ & $0.04235$  & Y & Y & N & $2$ & $2$ & $13.94$ \\
RXJ 1230.8+0115  & $0.11642$ & $0.00574$  & Y & Y & N & $2$ & $2$ & $13.25$ \\
RXJ 1230.8+0115  & $0.11642$ & $0.07810$  & Y & N & Y & $2$ & $2$ & $14.31$ \\
RXJ 1230.8+0115  & $0.11642$ & $0.09513^{\star +}$ & N & N & N & $2$ & $1$ & $13.05$ \\
RXJ 2154.1-4414 & $0.34366$ & $0.06229^\mathrm{F}$ & Y & Y & Y & $2$ & $2$ & $14.43$ \\
SBS 1108+560  & $0.76662$ & $0.00218^\mathrm{HF}$  & Y & Y & N & $0$ & $2$ & $14.51$ \\
SBS 1122+594 & $0.85142$ & $0.00402^\mathrm{HF}$ & Y & Y & N & $1$ & $1$ & $14.49$ \\
SBS 1122+594  & $0.85142$ & $0.06016^\mathrm{F}$  & Y & N & N & $2$ & $1$ & $13.50$ \\
SDSS J004222.29-103743.8 & $0.42312$ & $0.09509^\mathrm{H}$ & Y & Y & Y & $5$ & $3$ & $14.76$ \\
SDSS J021218.32-073719.8  & $0.17194$ & $0.01594$ & Y & Y & N & $2$ & $2$ & $14.25$ \\
SDSS J082633.51+074248.3  & $0.30956$ & $0.05126^\mathrm{H}$ & Y & Y & N & $2$ & $1$ & $14.11$ \\
SDSS J084349.49+411741.6 & $0.99079$ & $0.03016^\mathrm{\star +H}$ & Y & N & N & $1$ & $1$ & $13.83$ \\
SDSS J091029.75+101413.6  & $0.46141$ & $0.14237$ & Y & Y & N & $3$ & $2$ & $14.19$ \\
SDSS J092554.43+453544.4  & $0.32948$ & $0.01422^\mathrm{HF}$ & N & N & N & $3$ & $2$ & $13.50$ \\
SDSS J094952.91+390203.9  & $0.36347$ & $0.01815^\mathrm{HF}$ & Y & N & N & $3$ & $3$ & $14.62$ \\
SDSS J095915.65+050355.1  & $0.16230$ & $0.05901^\mathrm{H}$ & Y & Y & N & $3$ & $3$ & $14.69$ \\
SDSS J104741.75+151332.2  & $0.38508$ & $0.00231^\star$ & Y & Y & N & $2$ & $2$ & $13.89$ \\
SDSS J105945.23+144142.9  & $0.63171$ & $0.00243^\mathrm{\star D}$ & Y & Y & Y & $4$ & $1$ & $14.05$ \\
SDSS J105958.82+251708.8  & $0.66191$ & $0.11884^\mathrm{D}$  & Y & N & N & $2$ & $1$ & $14.32$ \\
SDSS J110406.94+314111.4  & $0.43336$ & $0.06209^\mathrm{D}$ & N & N & N & $2$ & $1$ & $14.23$ \\ 
SDSS J111754.31+263416.6  & $0.42047$ & $0.04751^\mathrm{D}$  & Y & Y & N & $2$ & $2$ & $14.31$ \\
SDSS J112244.89+575543.0  & $0.90685$ & $0.05324^\mathrm{D}$ & Y & N & N & $2$ & $1$ & $13.75$ \\
SDSS J121037.56+315706.0  & $0.39018$ & $0.05983^\mathrm{D}$  & Y & Y & Y & $1$ & $1$ & $14.20$ \\
SDSS J121037.56+315706.0  & $0.39018$ & $0.07812^\mathrm{\star D}$ & N & N & N & $2$ & $2$ & $13.66$ \\
SDSS J121114.56+365739.5  & $0.17075$ & $0.07779^\mathrm{D}$  & N & N & N & $2$ & $2$ & $14.10$ \\
SDSS J123604.02+264135.9  & $0.20789$ & $0.06163^\mathrm{H}$ & Y & Y & N & $2$ & $1$ & $14.23$ \\
SDSS J124511.25+335610.1  & $0.71170$ & $0.00222^\star$  & N & N & N & $1$ & $1$ & $13.39$ \\
SDSS J133053.27+311930.5  & $0.24252$ & $0.03413^\mathrm{H}$  & Y & Y & N & $2$ & $2$ & $14.26$ \\
\hline
\end{tabular}
\label{tab1}
\end{threeparttable}
\end{center}
\end{table*}

\begin{table*}
\begin{center}
\begin{threeparttable}
\begin{tabular}{lcccccccc}
\hline
QSO & $z_\mathrm{em}$     & $z_\mathrm{abs}$     & Heavier Metals & Low Ions & High Ions & No. of {\HI} & No. of {\CIV} & $\log[\Sigma N(\CIV)~(\cmsq)]$ \\
(1) &     (2)      &    (3)        &     (4)      &      (5)     &      (6)      &      (7)  & (8) & (9)    \\
\hline
SDSS J134206.56+050523.8  & $0.26387$ & $0.13989^\mathrm{\star D}$  & Y & Y & N & $4$ & $2$ & $14.46$ \\
SDSS J134251.60-005345.3  & $0.32396$ & $0.07155^\mathrm{\star D}$  & Y & Y & N & $2$ & $2$ & $13.89$ \\
SDSS J135712.61+170444.1  & $0.15084$ & $0.08364^\mathrm{DF}$  & Y & N & N & $2$ & $1$ & $13.79$ \\
SDSS J135712.61+170444.1  & $0.15084$ & $0.09763^\mathrm{DF}$  & Y & Y & Y & $6^\mathrm{C}$ & $5$ & $14.06$ \\
SDSS J161916.54+334238.4  & $0.46853$ & $0.14133^\mathrm{\star DH}$  & Y & N & N & $2$ & $1$ & $13.75$ \\
VII Zw 244  & $0.13089$ & $0.00235^\mathrm{H}$ & Y & Y & N & $1$ & $1$ & $13.38$ \\
\hline
\end{tabular}
\begin{tablenotes}
\item[](1) QSO name; (2) Emission redshift is based on the intrinsic $\Lya$ emission, or taken from NASA/IPAC Extragalactic Database (NED) if $\Lya$ isn't covered by COS; (3) Absorber redshift; (4) Presence of metal species other than {\CII} and {\CIV}; (5) Detection of at least one low ion ({\CII}, {\SiII}, {\NII}); (6) Detection of at least one high ion ({\NV}, {\OVI}); (7) No. of {\HI} components; (8) No. of {\CIV} components; (9) Combined $N(\CIV)$ of components based on Voigt profile fits.
\newline $^\star$The $N(\HI)$ is well constrained for these absorbers for at least one component.~$^+$The $\Lya$ is saturated but the $N(\HI)$ is well constrained (see Sec.~\ref{nhidet}).   
\newline $^\mathrm{A}$The $b(\HI)$ and $N(\HI)$ are adopted from \citet{Sembach2004};~$^\mathrm{B}$The emission redshift is taken from \citet{Rovero2016};~$^\mathrm{C}$The {\HI} is heavily contaminated with Galactic $\CII~\lambda1334$;~$^\mathrm{D}$The systems overlap with \citet{Burchett2015};~$^\mathrm{E}$The systems overlap with \citet{Cooksey2010};~$^\mathrm{F}$The systems overlap with \citet{Danforth2016};~$^\mathrm{G}$The saturated {\CIV} component in this absorber is excluded here;~$^\mathrm{H}$The systems are around targeted galaxies.
\end{tablenotes}
\end{threeparttable}
\end{center}
\end{table*}

\begin{table}
\caption{Summary of the line measurements}
\begin{center}
\begin{threeparttable}
\begin{tabular}{lc}
\hline
Median redshift & $0.062$ \\
Maximum $\Delta Z$ for {\CIV} & $20.2$\\
Systems & $69$\\
{\CIV} components & $127^\star$\\
{\HI} components & $168$\\
Systems with heavier metals than carbon & $52~(75\%)$\\
Systems with low ions & $32~(46\%)$\\
Systems with high ions & $16~(23\%)$\\
$b(\CIV)~(\kms)$ & $4 - 44$\\
$b(\HI)~(\kms)$ & $4 - 81$\\
$\log[N(\CIV)~(\cmsq)]$ & $12.31 - 14.69$\\
$\log[N(\HI)~(\cmsq)]$ & $12.91 - 18.09$\\
\hline
\end{tabular}
\begin{tablenotes}
\item[] Note: The column density ranges mentioned here are for \textit{individual} components. The range of $b$ and $N$ listed for {\HI} encompass all systems in the sample, though reliable profile fits for {\HI} are available only for $32\%$ of the absorbers (22/69) with the $\Lya$ saturation affecting measurements in the remaining systems. $^\star$This includes the saturated {\CIV} component. 
\end{tablenotes}
\label{tab2}
\end{threeparttable}
\end{center}
\end{table}

\begin{figure}
    \begin{center}
        \includegraphics[width=220pt,height=160pt,trim={0.3cm 0cm 0.15cm 0.3cm},clip=true]{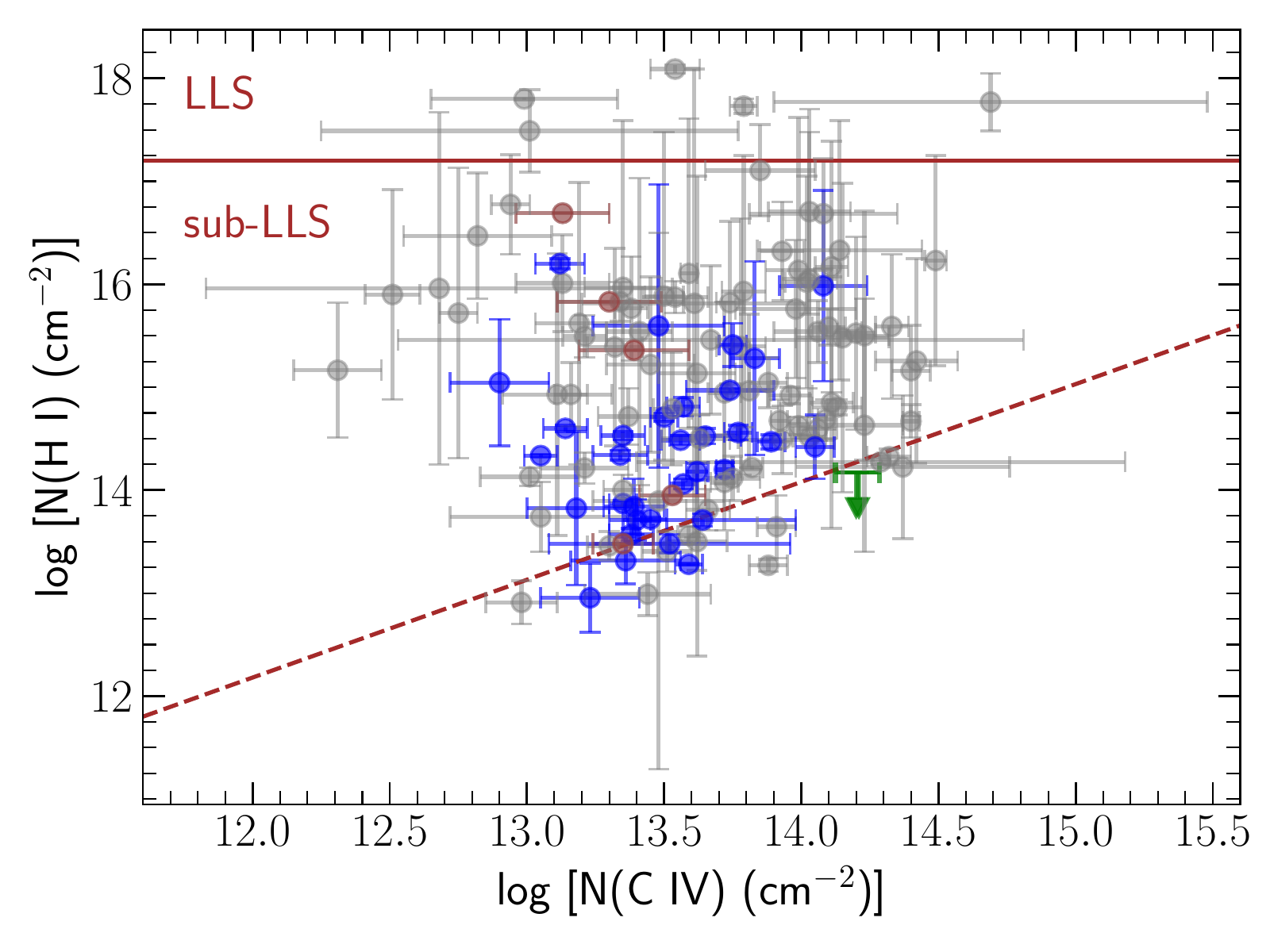}
        \includegraphics[width=220pt,height=160pt,trim={0.3cm 0.4cm 0.15cm 0.3cm},clip=true]{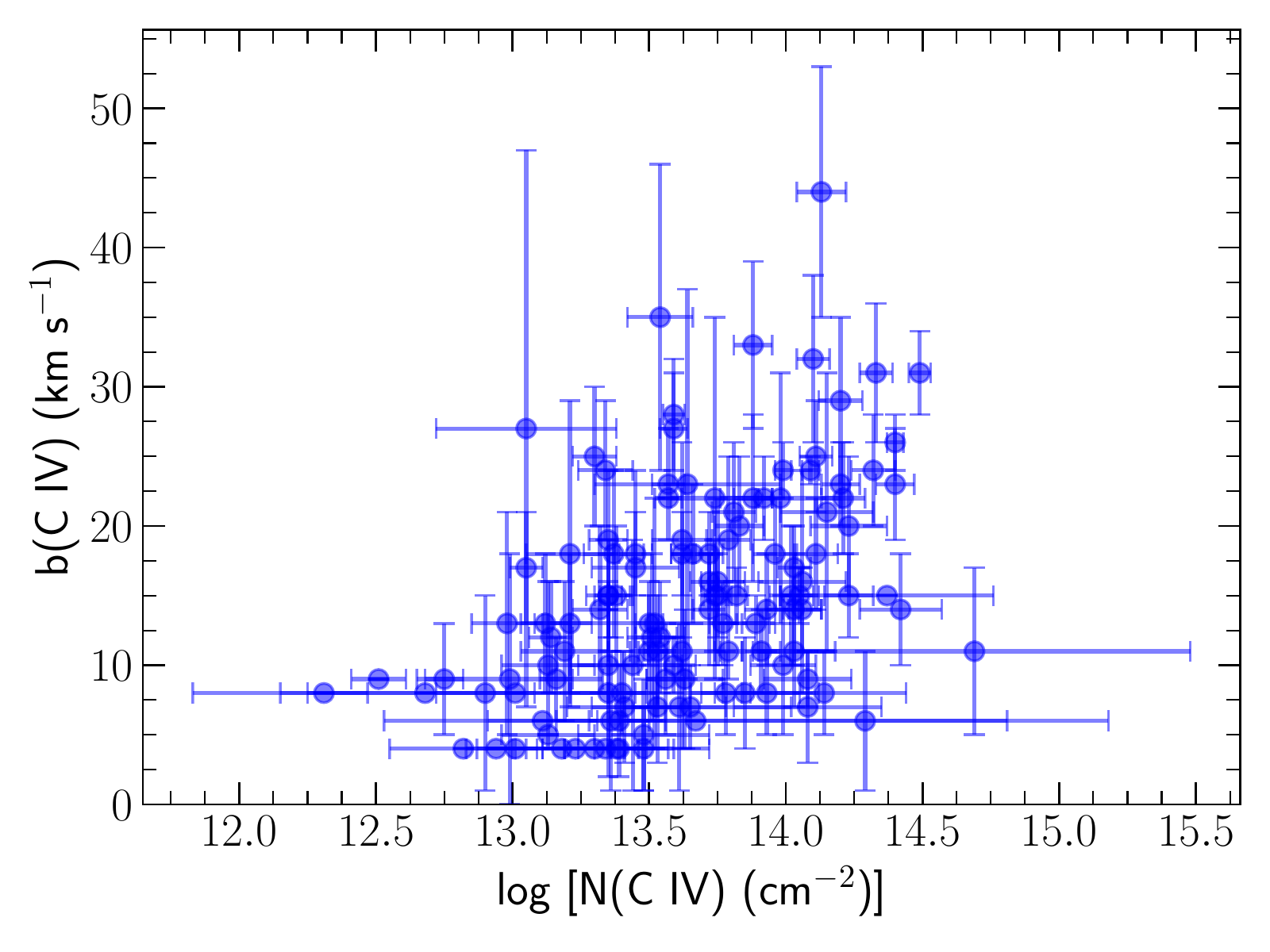}
    \caption{Top: The $N(\HI)$-$N(\CIV)$ scatter plot. The blue points correspond to the sub-sample of 22 systems with secure {\HI} measurements, while the components from other absorbers are shown in grey. The green upper limit with error-bar for $N(\CIV)$ corresponds the $z_\mathrm{abs}=0.00218$ absorber towards SBS 1108+560 that does not have a {\HI} detection. The dashed line represents $N(\HI) = N(\CIV)$, and the solid line corresponds to $N(\HI) = 10^{17.2}~\cmsq$ that distinguishes Lyman Limit Systems (LLS) from sub-LLS. The brown points correspond to free-fits of {\HI} in the $z_\mathrm{abs} = 0.09763$ system towards SDSS J135712.61+170444.1 with $\Lya$ that is heavily contaminated with Galactic {\CII}~$1334~{\AA}$. Bottom: The distribution of Doppler $b$-parameter and column density for the {\CIV} components, which show large scatter.}
    \label{3}
    \end{center}
\end{figure}

\begin{figure}
    \begin{center}
        \includegraphics[width=215pt,height=158pt,trim={0.3cm 0cm 0.15cm 0.3cm},clip=true]{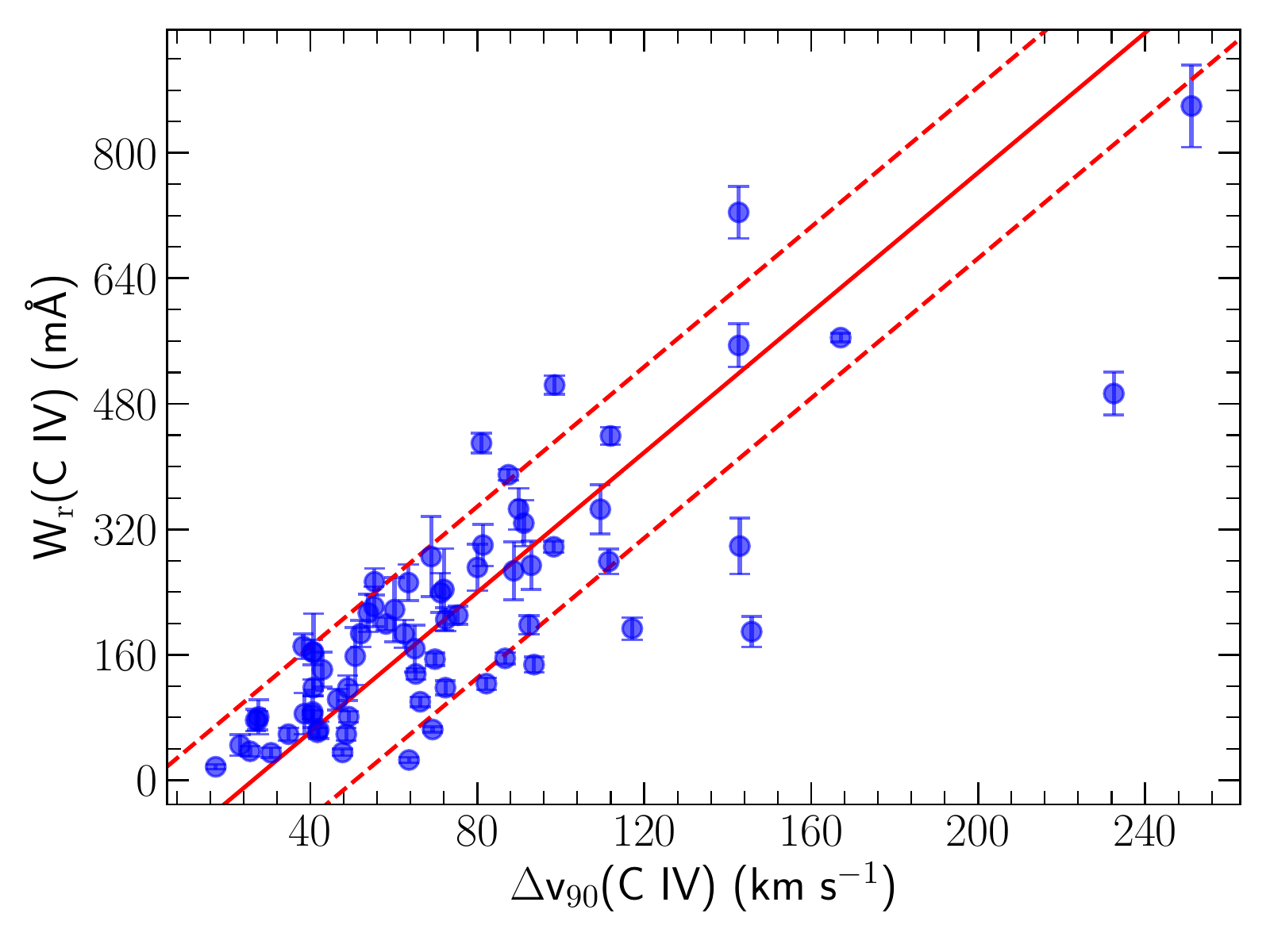}
        \includegraphics[width=215pt,height=158pt,trim={0.3cm 0cm 0.15cm 0.3cm},clip=true]{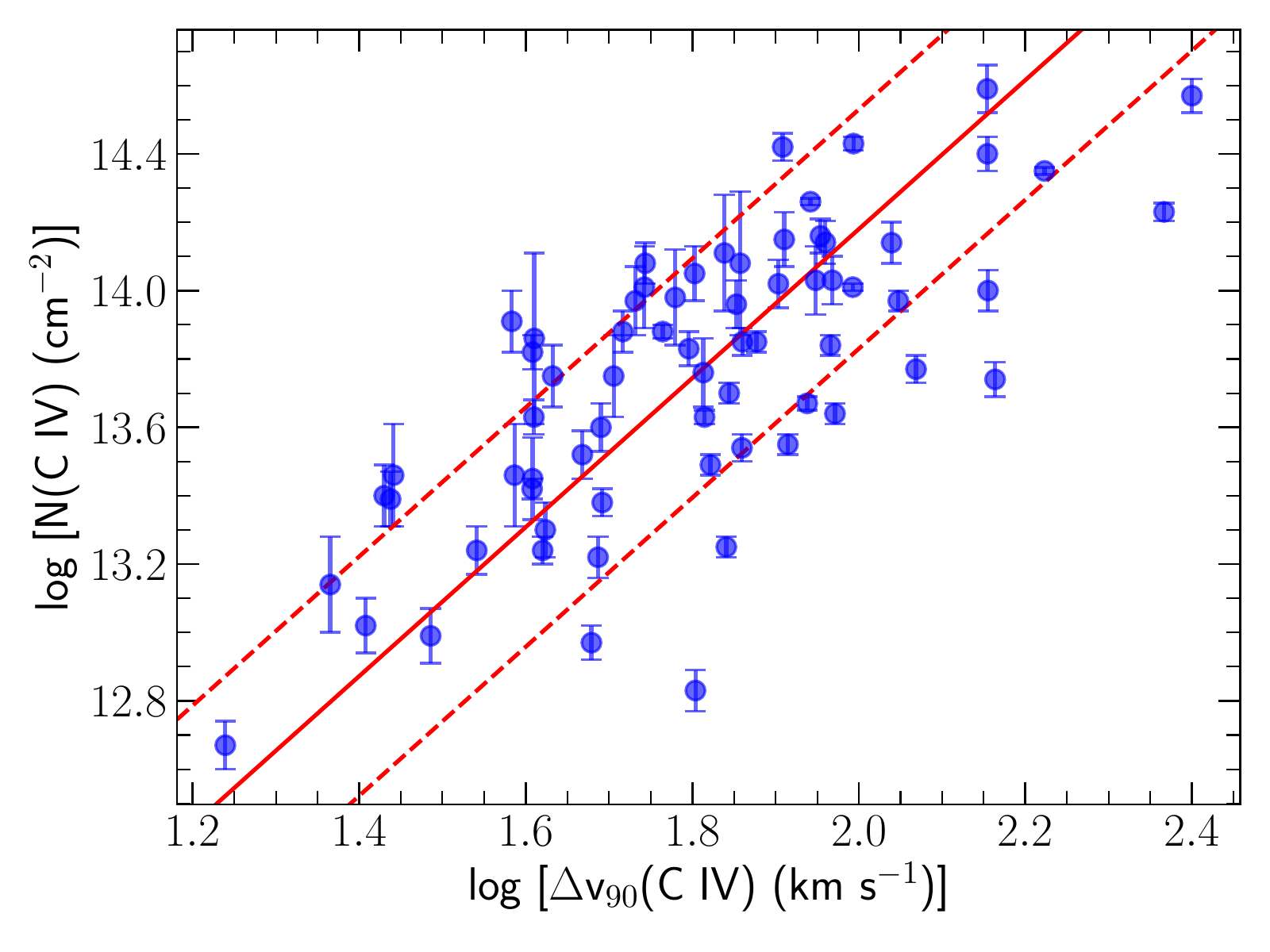}
        \includegraphics[width=215pt,height=158pt,trim={0.3cm 0cm 0.15cm 0.3cm},clip=true]{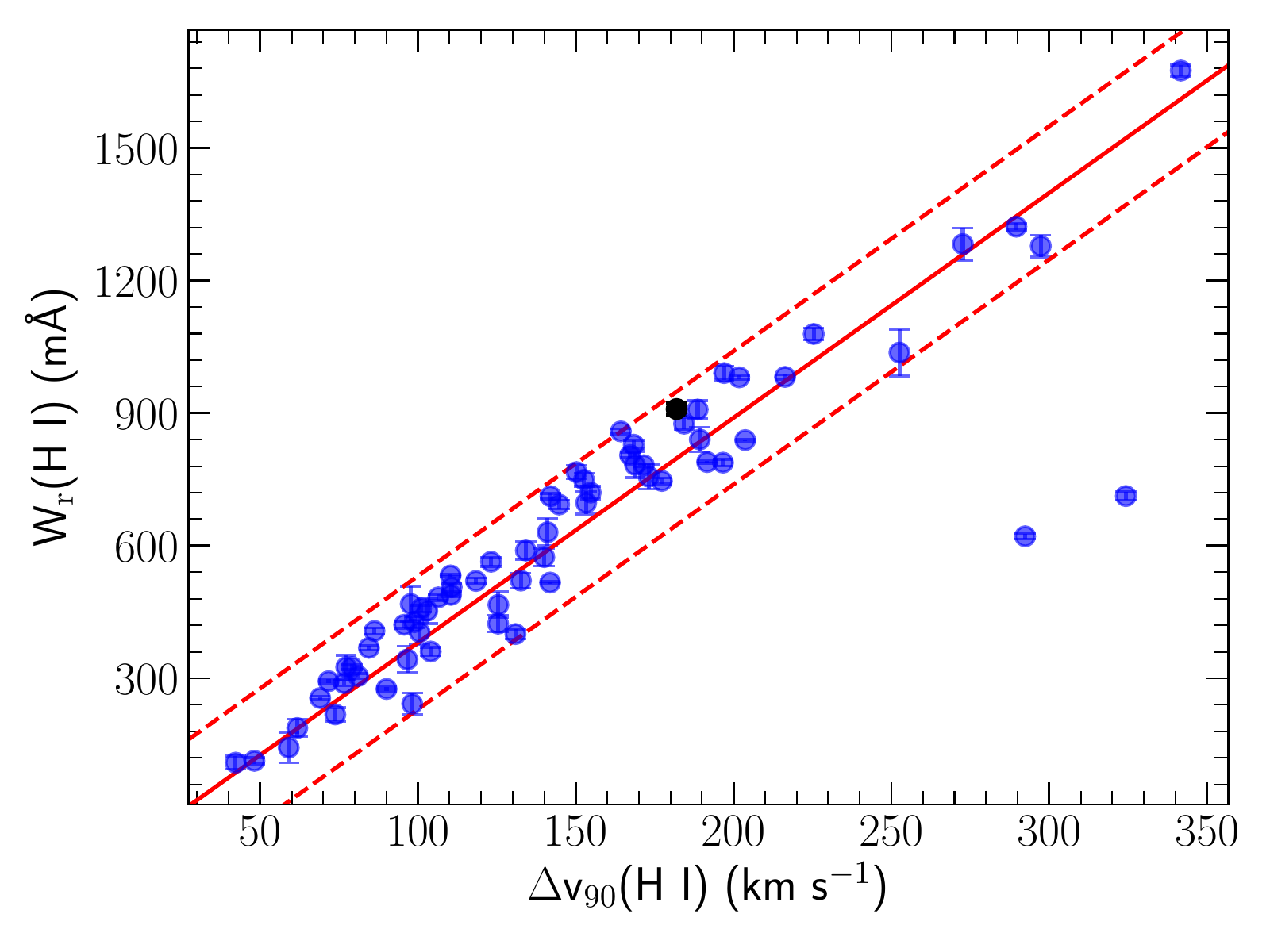}
        \includegraphics[width=215pt,height=158pt,trim={0.3cm 0.4cm 0.15cm 0.3cm},clip=true]{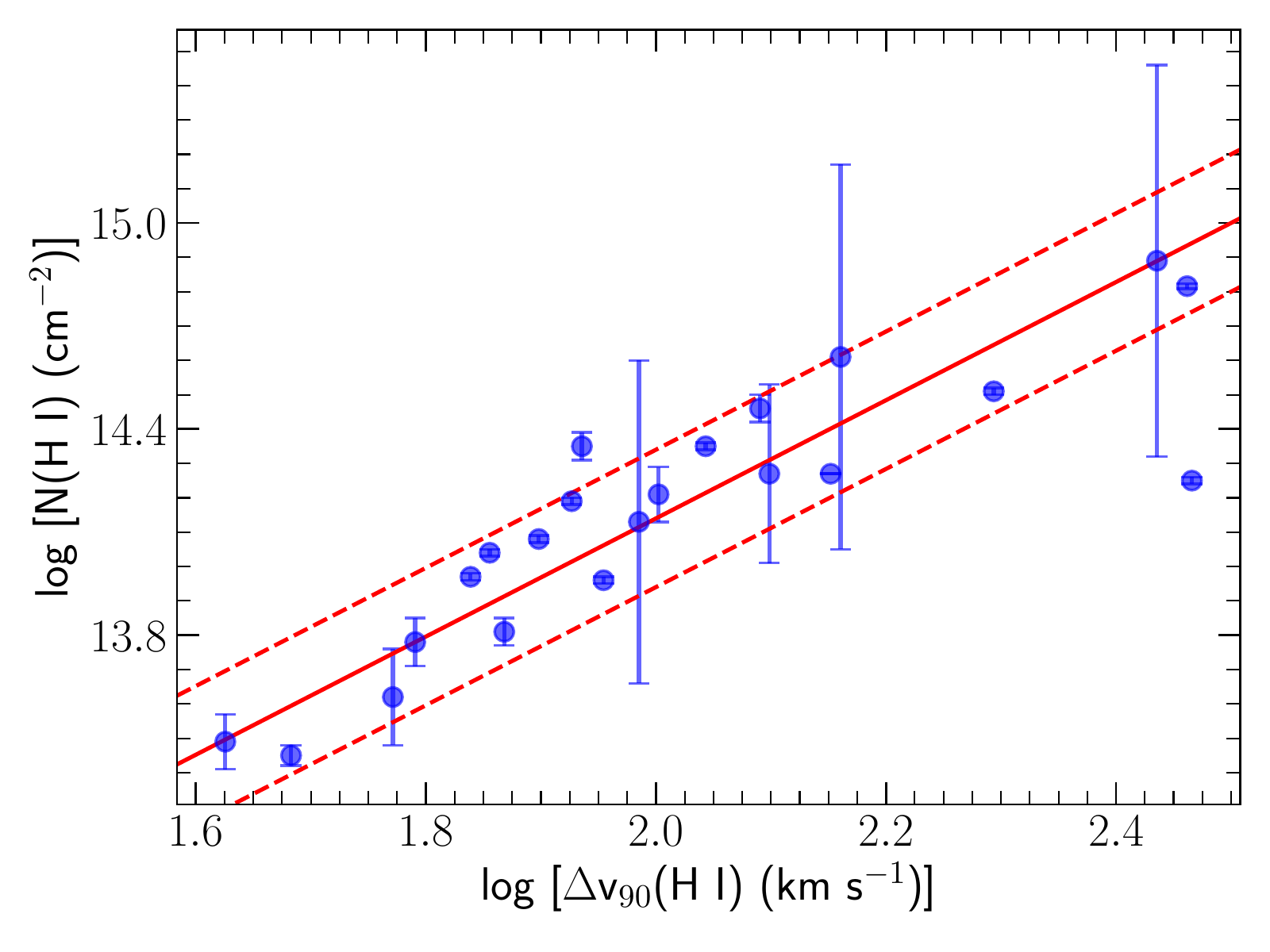}
    \caption{Correlations between line strengths of {\HI} and {\CIV} with the $\Delta v_{90}$ based on AOD integration for the entire profile. The $W_\mathrm{r}(\HI)$ of $\Lya$ in $z = 0.09763$ absorber (black circle) is heavily contaminated by Galactic {\CII}~$1334$~{\AA}. The bottom panel only shows the sample with secure {\HI}.}
    \label{4}
    \end{center}
\end{figure}

\begin{figure*}
    \begin{center}
        \includegraphics[width=240pt,height=180pt,trim={0.45cm 0cm -0.43cm 0.3cm},clip=true]{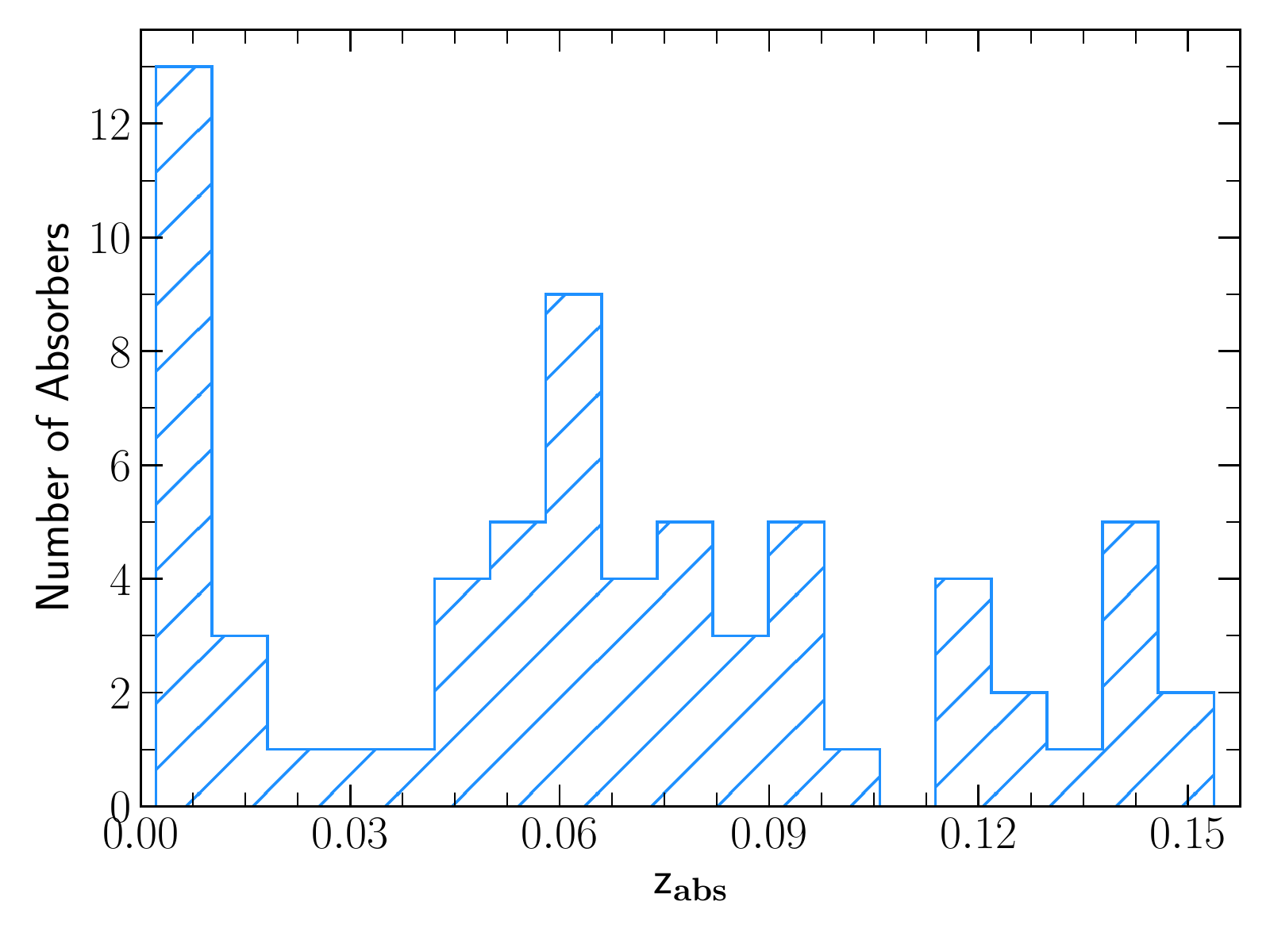}\quad
        \includegraphics[width=240pt,height=180pt,trim={0.38cm 0cm 0.3cm 0.3cm},clip=true]{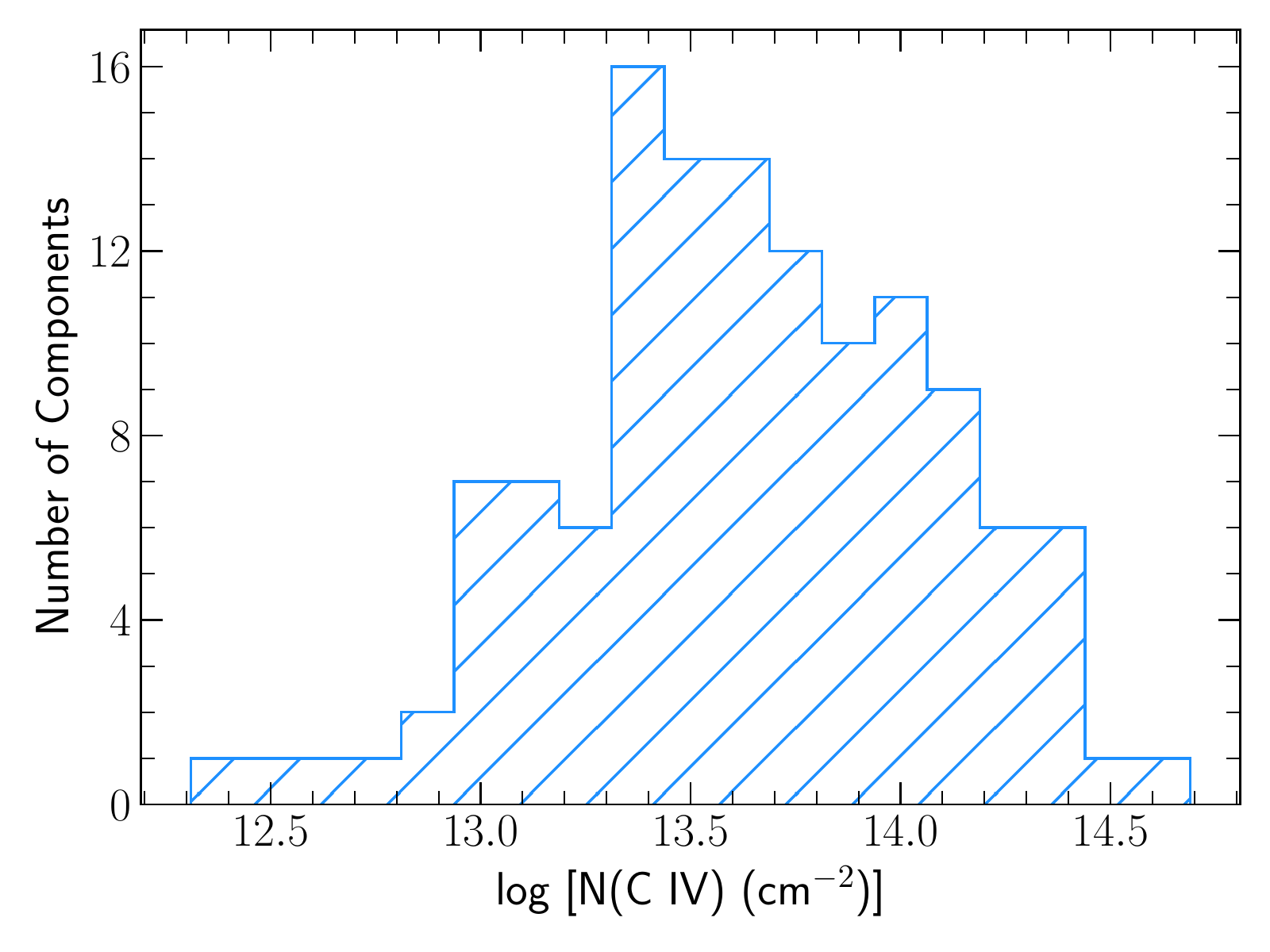} \quad 
        \includegraphics[width=240pt,height=180pt,trim={0.38cm 0.4cm 0.3cm 0.3cm},clip=true]{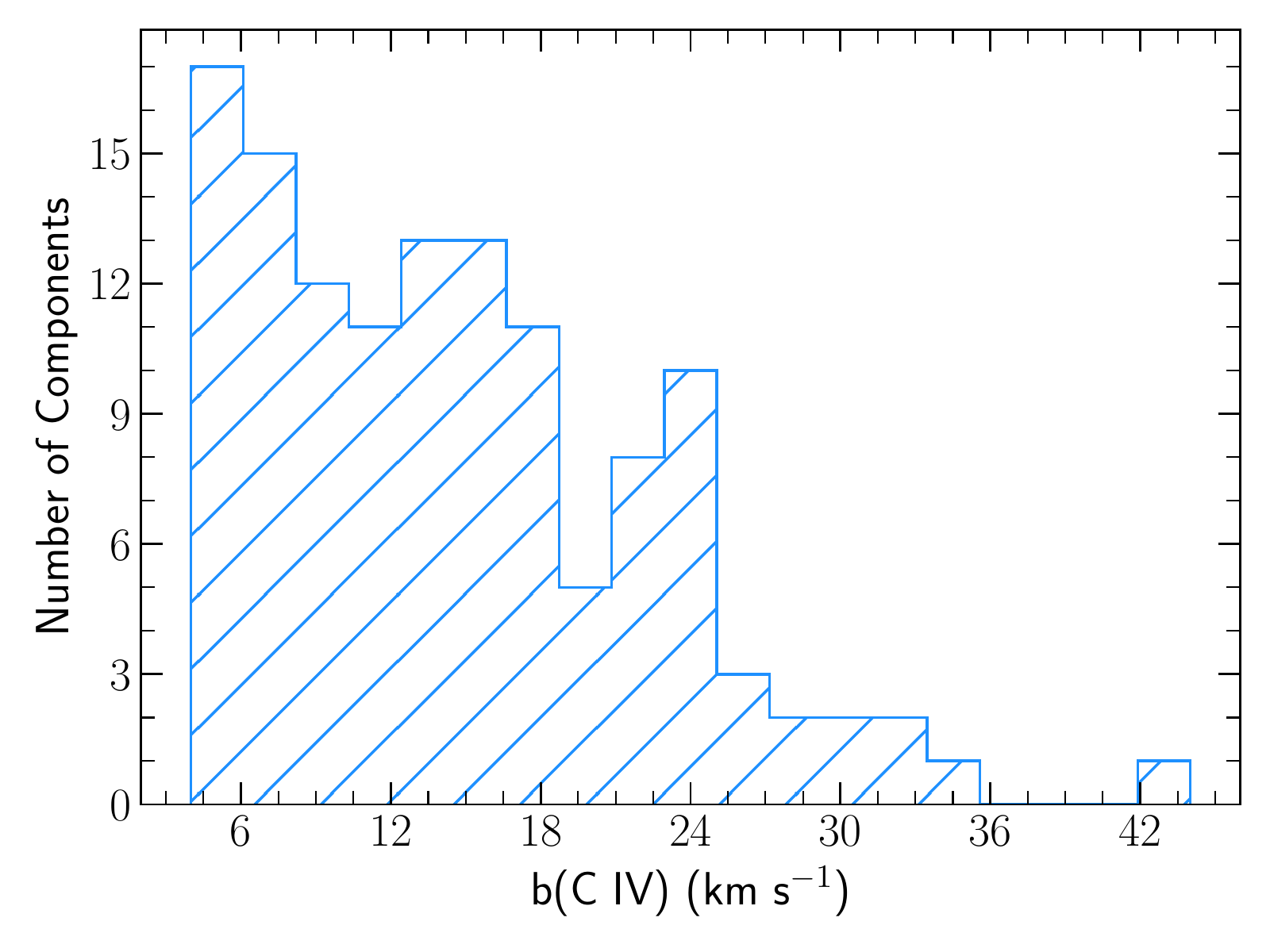} \quad
        \includegraphics[width=240pt,height=180pt,trim={0.38cm 0.4cm 0.3cm 0.3cm},clip=true]{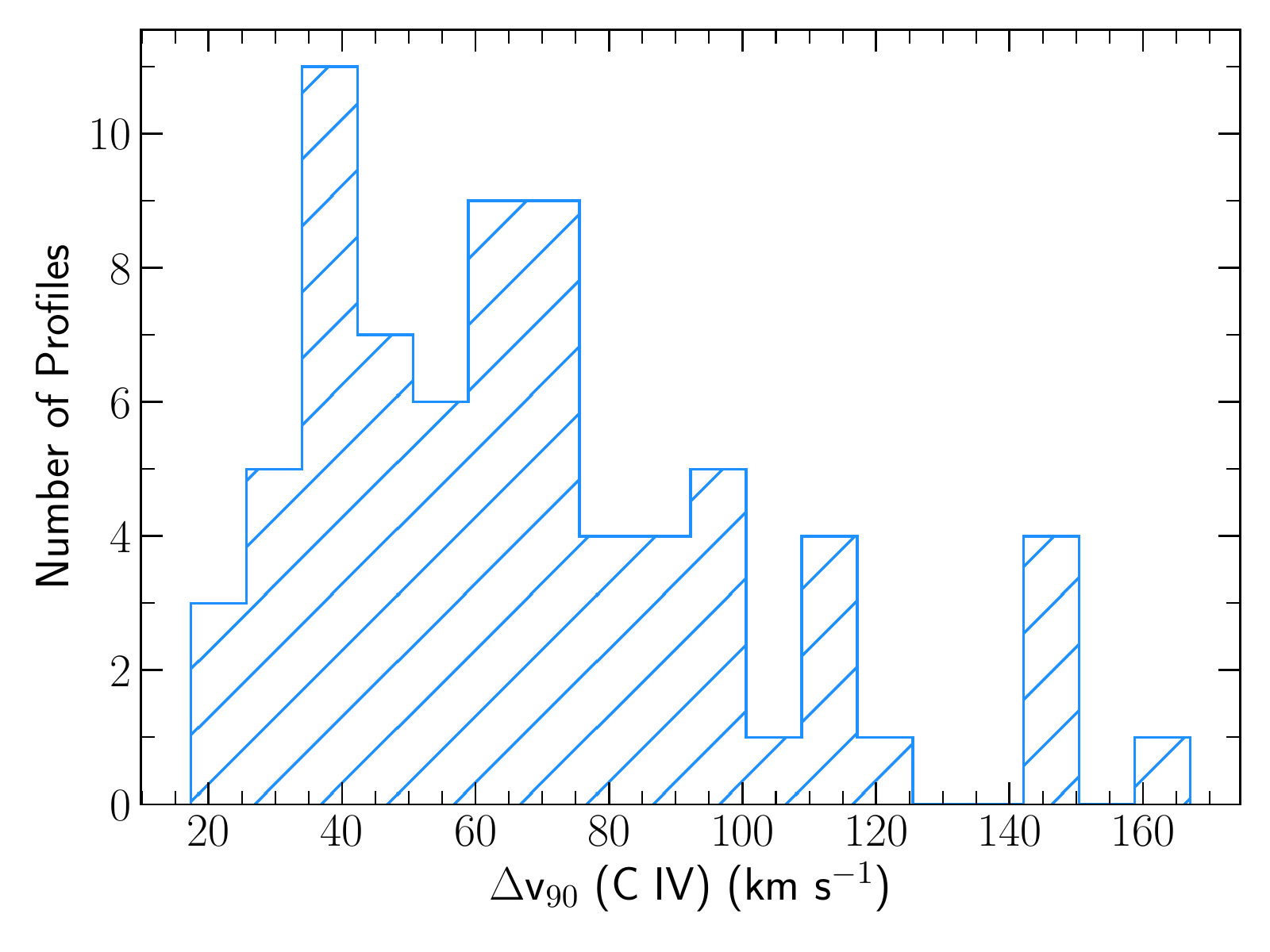}
    \caption{Top left-hand panel: The distribution of absorber redshifts. Top right-hand panel: The column density distribution of {\CIV} components. Bottom left-hand panel: The distribution of Doppler parameter for {\CIV} components. Bottom right-hand panel: The distribution of velocity width containing $90\%$ of the flux in {\CIV} absorption line profiles.}
    \label{5}
    \end{center}
\end{figure*} 

\begin{figure}
    \begin{center}
        \includegraphics[width=220pt,height=160pt,trim={0.38cm 0.1cm 0.3cm 0.3cm},clip=true]{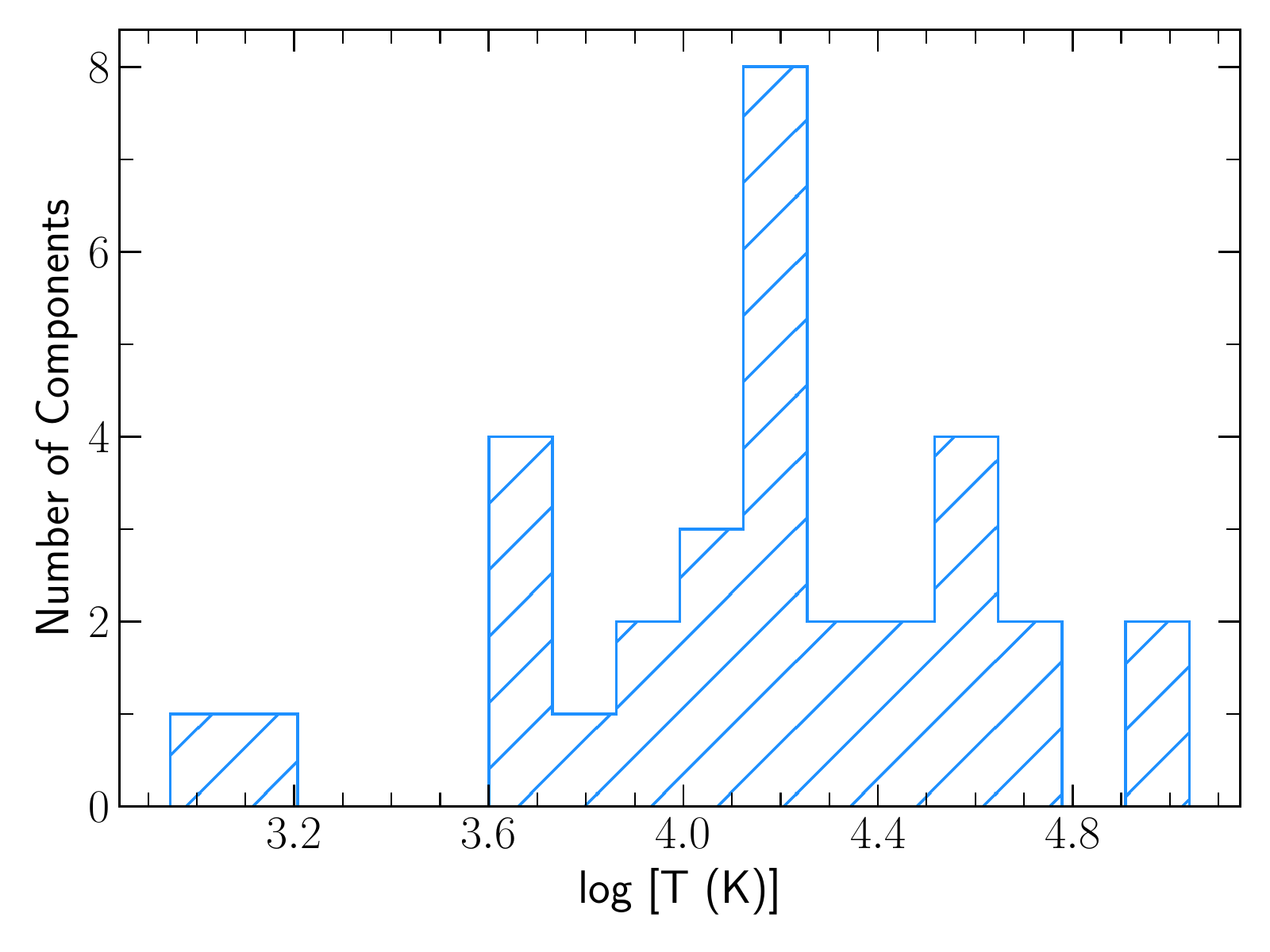}
        \includegraphics[width=220pt,height=160pt,trim={0.38cm 0.2cm 0.3cm 0.3cm},clip=true]{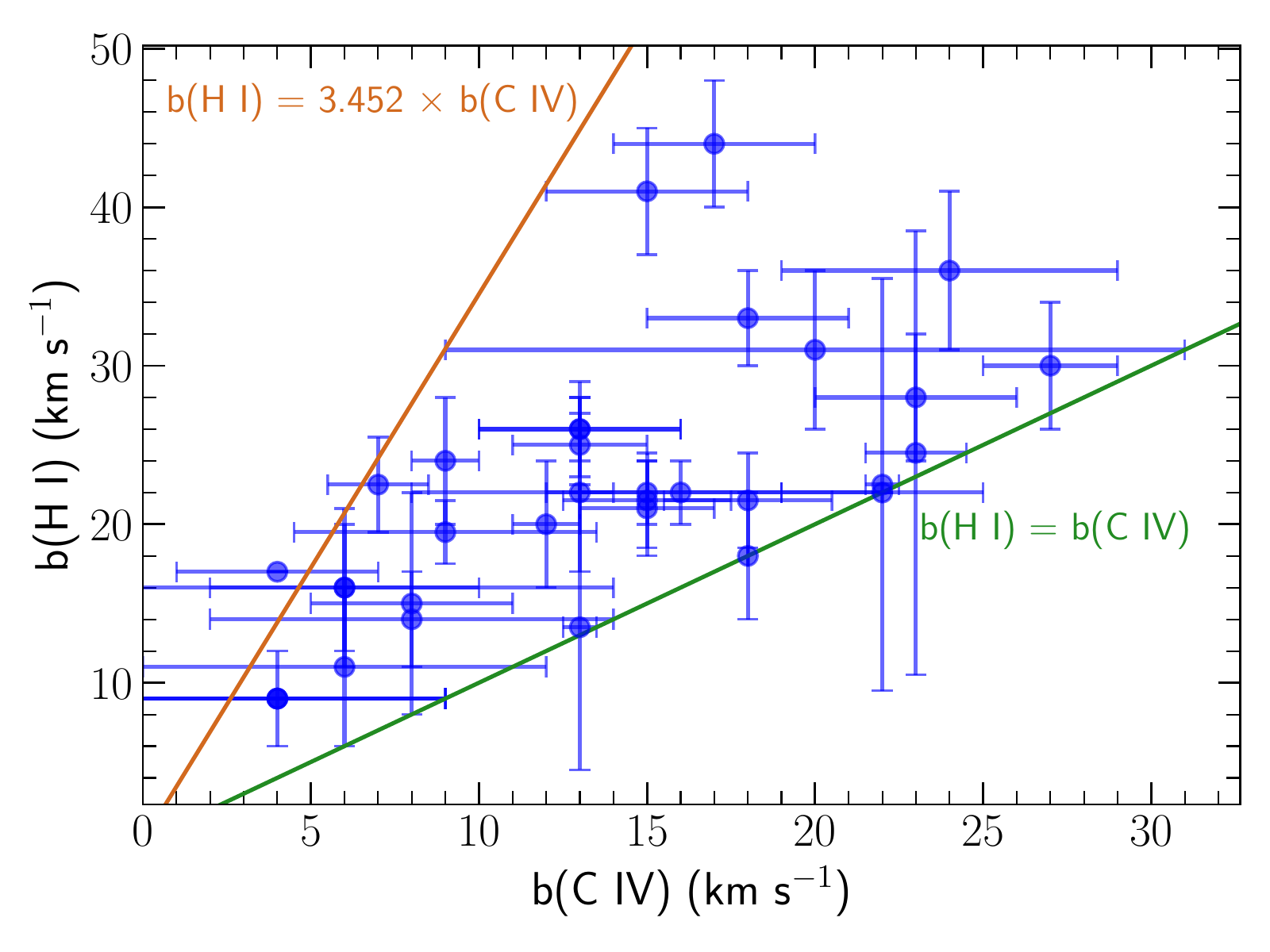}
        \includegraphics[width=220pt,height=220pt,trim={0cm 0.3cm 0cm 0.2cm},clip=true]{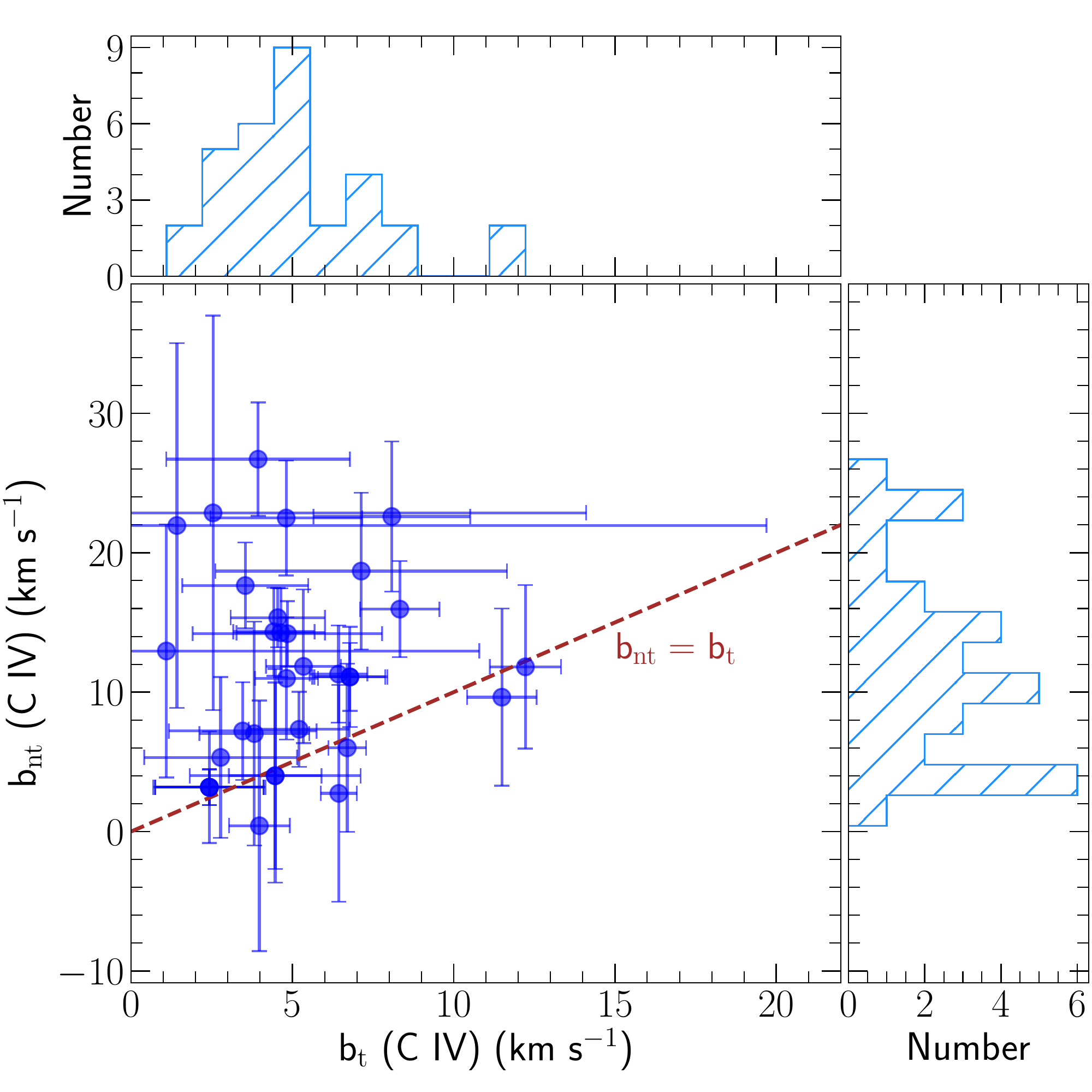}
    \caption{Top: Temperatures in the {\CIV} absorbers inferred from the $b$-values of $(\HI)$ and $(\CIV)$. Here we consider only the sub-sample with secure {\HI} measurements. Middle: The $b(\HI)$ - $b(\CIV)$ relation for the sub-sample with secure {\HI}. The steeper solid-brown line represents the relation between $b(\HI)$ and $b(\CIV)$ when the line broadening is purely thermal, and the solid-green line marks line broadening that is purely non-thermal. Bottom: The $b_\mathrm{nt}$-$b_\mathrm{t}$ scatter plot with histograms for {\CIV} in the middle panel. The dashed line represents $b_\mathrm{nt} = b_\mathrm{t}$.}
    \label{6}
    \end{center}
\end{figure}

\subsection{Absorption properties and their correlations}\label{comps}

As shown in Table~\ref{tab2}, $75\%$ of the systems have elements that are heavier than carbon, possibly indicating that {\CIV} absorbers trace highly enriched gas. A majority of the systems (57/69, $83\%$) have a kinematically simple {\CIV} profile with only one or two absorbing components. Complex absorption profiles with a large number of components in either {\HI} or {\CIV} are rare in this sample. However, there can be very narrow components that went undetected due to the limited resolution of COS. The entire sample is comprised of $127$ (including the saturated {\CIV} component) and $168$ kinematically distinct components in {\CIV} and {\HI}, respectively. \citet{Savage2014} carried out an unbiased survey of {\OVI} absorbers and showed that $90\%$ of the systems have two or fewer {\OVI} components. Thus, the most common {\CIV} and {\OVI} absorptions tracing CGM and IGM occur in kinematically simple systems at the resolution of COS. 

In Fig.~\ref{3}, we compare the {\HI} and {\CIV} column densities in each component. We find no clear correlation between the two column densities, suggesting that the population of {\CIV} absorbers are possibly tracing a range of physical conditions, if the {\CIV} and {\HI} are from the same gas phase. This can also be deduced by comparing the range of column densities for {\CIV} and {\HI} (see Table~\ref{tab2}). The {\CIV} column densities for the sample spans two orders of magnitude whereas the corresponding {\HI} values are spread over four orders of magnitude. This lack of correlation is perhaps an indication of metals being confined to patchy zones in a wide range of {\HI} environments. It is interesting to note that there are a few absorbers where the $N(\CIV) \gtrsim N(\HI)$, indicated by the points near to and below the dashed line in Fig.~\ref{3}. Such systems where the {\CIV} column density is comparable to or greater than {\HI} are likely to be extremely metal-rich gas clouds in the circumgalactic and intergalactic space similar to the population of {\CIV} absorbers discussed in \citet{Schaye2007}.

The $z_\mathrm{abs} = 0.13850$ system towards PG 1116+215 shows $\Lya$ and $\Lyb$ absorption but both are saturated. We therefore adopt the $b(\HI)$ and $N(\HI)$ from \citet{Sembach2004} who used higher order Lyman series lines from \textit{HST}/STIS to constrain the {\HI} column density for this absorber. The $z_\mathrm{abs} = 0.09763$ system towards SDSS J135712.61+170444.1 has $\Lya$ that is heavily contaminated with the Galactic $\CII~1334$ and therefore we do not use the $N(\HI)$ for these components in the ionization modelling. One out of 3 {\CIV} components in the $z_\mathrm{abs} = 0.08940$ system towards QSO B1435-0645 is saturated, and we have excluded it from analysis in this paper.

The $b(\CIV)$-$N(\CIV)$ relationship has a Spearman rank test\footnote{We are using Spearman rank test throughout the paper instead of Pearson test because it is agnostic to the order, unlike the latter which specifically tests for linear trends. A $p<0.003$ implies $>3\sigma$ significance level.} $p$-value of $\ll 0.001$ for null hypothesis (that there is no correlation), and the correlation coefficient of $r=0.47$ (\textit{bottom-panel}, Fig.~\ref{3}), but there is significant scatter. It is comparatively more difficult to detect lines with low $N$ and high $b$ at $\geq 3\sigma$ level due to noise. This is the reason for paucity of points in the upper-left region of Fig.~\ref{3}. It should be noted that the aforementioned trend can also be an artifact of this.

We however do find strong correlations ($r\gtrsim0.7$) between absorption strengths based on AOD measurements and $\Delta v_{90}$ for both {\CIV} and {\HI} (Fig.~\ref{4}). Here the values are based on entire absorption in the system. It should be noted that by {\HI} and {\CIV} we mean $\Lya$ and $\CIV~\lambda1548$, respectively. Since many of the $\HI$ absorptions are saturated, we only use the sample with secure {\HI} in the $N(\HI)-\Delta v_{90}(\HI)$ relation. The $W_\mathrm{r}-\Delta v_{90}$ correlations are strong for both {\CIV} ($r=0.79$) and {\HI} ($r=0.94$), and the $N-\Delta v_{90}$ is correlated similarly for both ions ($r=0.70$ for {\CIV} and $r=0.89$ for {\HI}) ($p\ll0.001$ for all). We use the \textsc{\large hyper-fit} package\footnote{\url{https://hyperfit.icrar.org}} \citep{Robo2015} to fit linear models to these relations by optimizing likelihood of data for a Gaussian distribution about the model, using the Nelder-Mead method \citep{Nelder1965}. It takes into account the uncertainties in the data as well. The dashed lines in Fig.~\ref{4} represent 1$\sigma$ scatter. The scaling relationships are
\begin{align}
W_\mathrm{r}(\CIV) = (4.46\pm 0.36)\Delta v_{90}(\CIV) - (117.21\pm 30.03), \\
W_\mathrm{r}(\HI) = (5.09\pm 0.30)\Delta v_{90}(\HI) - (128.51\pm 47.95),\\
\log N(\CIV) = (2.18\pm 0.24)\log \Delta v_{90}(\CIV) + (9.82\pm 0.43),\\
\log N(\HI) = (1.72\pm 0.20)\log \Delta v_{90}(\HI) + (10.70\pm 0.40).
\end{align}
If we exclude the two outliers in $W_\mathrm{r}(\HI)-\Delta v_{90}(\HI)$ scatter, the slope and intercept of the fit change to $5.20\pm 0.14$ and $-119.39\pm 21.03$, respectively. Such relationships confirm that observed $W_\mathrm{r}$ is a stronger function of absorption width than $N$. Given that there are more {\HI} components in a system compared to {\CIV}, it suggests that {\CIV} bearing {\HI} components may dominate equivalent widths compared to the ones without {\CIV}, which typically tend to have lower $N(\HI)$. Interestingly, slopes and intercepts in these relationships are similar even though they correspond to different elements. Broadly this suggests that the {\HI} and {\CIV} ions reside in gas of similar physical conditions. The absence of $N(\HI)-\Delta v_{90}(\HI)$ correlation in \citet{Lehner2018} is mainly because they were also including saturated systems with $N(\HI)>10^{15}~\cmsq$. Moreover, their sample was {\HI}-selected unlike our sample which is {\CIV}-selected. Absorbers with detectable amounts of metals could be probing a smaller range of astrophysical environments than a sample of {\HI} absorbers unbiased by the presence of metals.

Fig.~\ref{5} shows the distributions of the {\CIV} absorption properties. As seen in the \textit{top-left} panel, the system redshifts have a median at $\approx 0.062$. The $N(\CIV)$ values (\textit{top-right} panel) range from $10^{12.3} - 10^{14.7}$~$\cmsq$ and peak at $\sim 10^{13.4}$~$\cmsq$. One expects a decline in the number of detections with increasing column density since large {\CIV} column density systems are likely to be part of optically thick Lyman limit and damped Lyman-alpha absorbers, whose redshift number densities are lower compared to metal absorbers tracing more diffuse gas belonging to the CGM and IGM. The rise till $N(\CIV)\approx 10^{13.4}$~$\cmsq$ is an artifact of the overall sensitivity of the spectra. Pathlengths are expected to saturate for high column densities of $N(\CIV)\gtrsim10^{15.0}~\cmsq$. From Table~\ref{pathtab}, it seems that pathlengths are going to saturate around $\approx 20.2$ and therefore we can approximate completeness as $\approx 100\%$ at $N(\CIV)=10^{14.9}~\cmsq$. Based on this, the completeness level is $55\%$ at $N(\CIV) = 10^{13.4}$~$\cmsq$. Hence, the number of {\CIV} detections at the $\geq 3\sigma$ confidence is expected to increase till such $N(\CIV)$ and decline at higher column densities.

Fig.~\ref{5} also shows the distribution of Doppler $b$ parameters for the {\CIV} components obtained from the profile fitting. The median $b(\CIV)$ is $\sim 14~\kms$, corresponding to a temperature of $T \leq 1.4\times 10^5$~K; assuming no turbulent broadening. There are $74$ components with $b(\CIV) \geq 12~{\kms}$, corresponding to temperature upper limits of $T \gtrsim 10^5$~K. Six {\CIV} absorbers in our sample also show {\OVI}, which is a better tracer of warm collisionally ionized gas than {\CIV}. The bottom panel of Fig.~\ref{5} shows the distribution of the velocity widths of the {\CIV} profiles. The {\CIV} absorption is spread over a range of velocities from as low as $\approx 17~\kms$ up to $\approx 167$~$\kms$ with the median at $\approx 64$~$\kms$. There are no absorptions with $\Delta v_{90} > 170~\kms$ which implies that, unlike the population of {\CIV} absorbers in the high-$z$ Universe \citep{Lehner2014}, kinematically broad {\CIV} absorbers are rare in the present Universe.

For the secure sample, the temperatures derived from the different $b$-values of $\HI$ and $\CIV$ are such that almost all but one of the clouds have $T < 10^5$~K, with the median at $T\sim10^{4.2}$~K (Fig.~\ref{6}). The \textit{middle} panel of Fig.~\ref{6} shows the aligned {\HI} and {\CIV} components spread between two extreme scenarios of exclusively thermal and fully turbulent broadening. This would also mean that the {\CIV} absorbers (of the secure sample at least) are predominantly tracing relatively cooler ($T\sim10^4$~K) gas, with photoionization being the dominant ionizing mechanism. A comparison of non-thermal ($b_\mathrm{nt}$) and thermal ($b_\mathrm{t}$) broadening in the $\CIV$ reveals that turbulence has a higher contribution (bottom panel of Fig.~\ref{6}). The median $b_\mathrm{nt}$ and $b_\mathrm{t}$ are $11~\kms$ and $5~\kms$, respectively. This is lesser than the median $b_\mathrm{nt}$ for photoionized {\OVI} absorbers by $12~\kms$ \citep{Savage2014}. The relationship between $b_\mathrm{nt}$ and environment is explored in Sec.~\ref{galaxies}.

\subsection{Correlations between ions}

Ions of different elements with similar ionization potentials are uniformly influenced by the ionizing radiation and hence can be used to assess the relative chemical abundances in the absorber. For example, in the photoionization models, the observed column densities of {\CII}, {\OII}, {\MgII}, and {\SiII} are routinely used to estimate the relative abundances as these four species have comparable creation and destruction energies. Observed column densities of well aligned ions of the same element, or different elements, tracing the same phase will exhibit a certain amount of correlation in their observed column densities. By the same token, a weak correlation between column densities of different ions of the same element can be an indication of their preference for different gas phases. Also, a weak correlation between low ions or between high ions of different elements can be suggestive of differences in elemental abundances across the sample.

\begin{figure}
    \begin{center}
        \includegraphics[width=210pt,height=159pt,trim={0.3cm 0.3cm 0.3cm 0.3cm},clip=true]{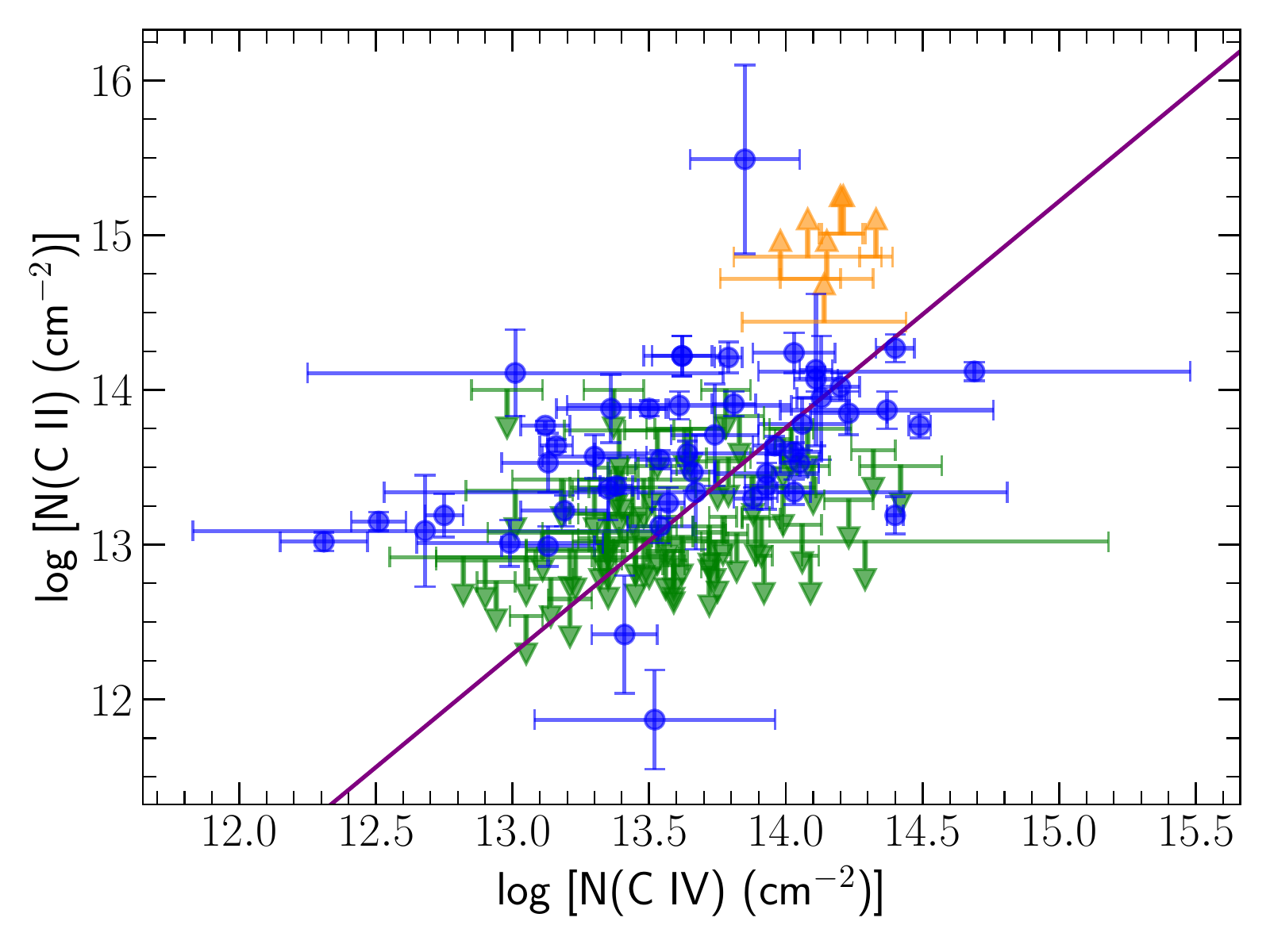}        \includegraphics[width=210pt,height=159pt,trim={0.0cm 0.3cm 0.3cm 0.3cm},clip=true]{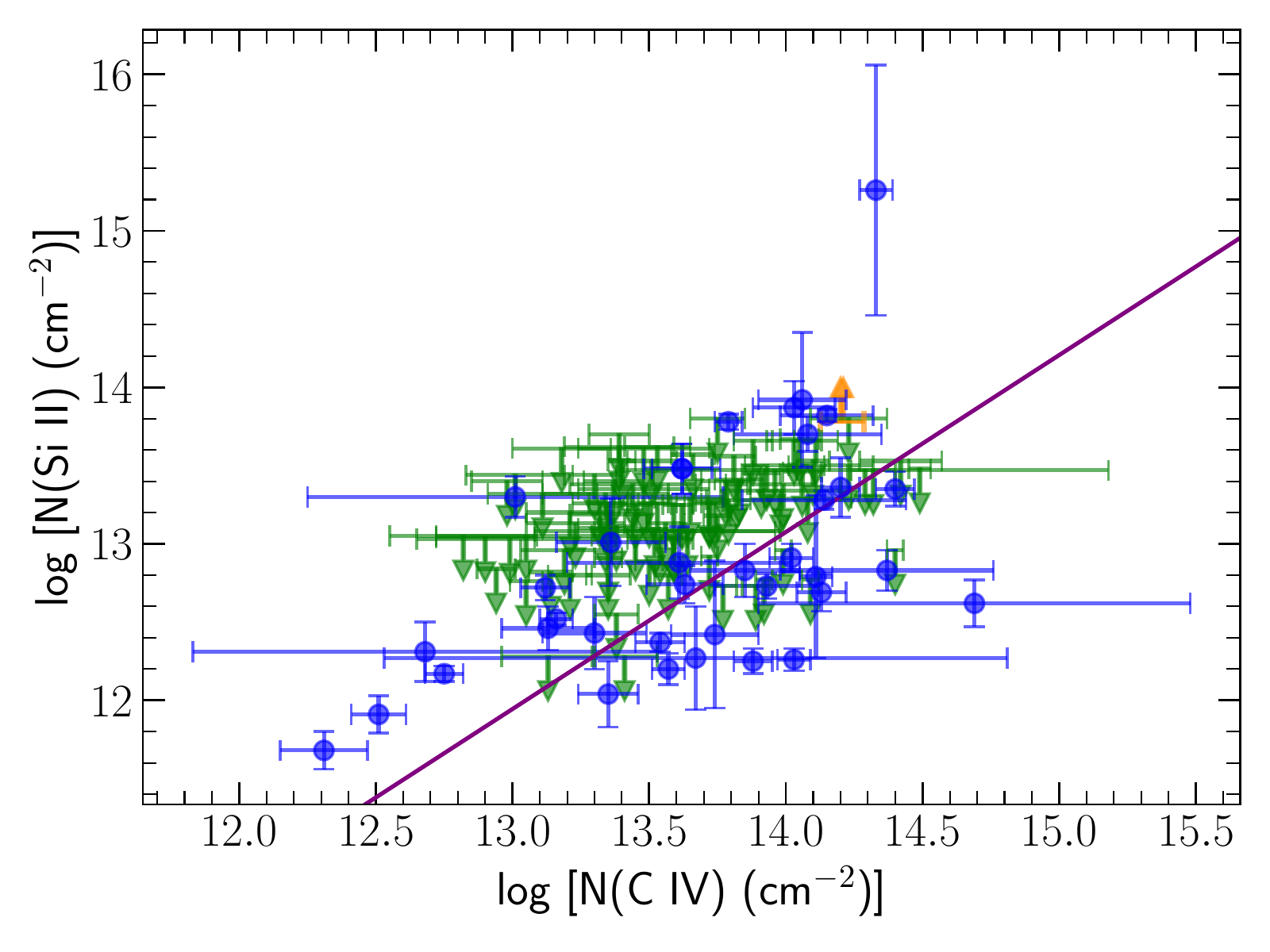}\quad
        \includegraphics[width=210pt,height=159pt,trim={0.3cm 0.3cm 0.3cm 0.3cm},clip=true]{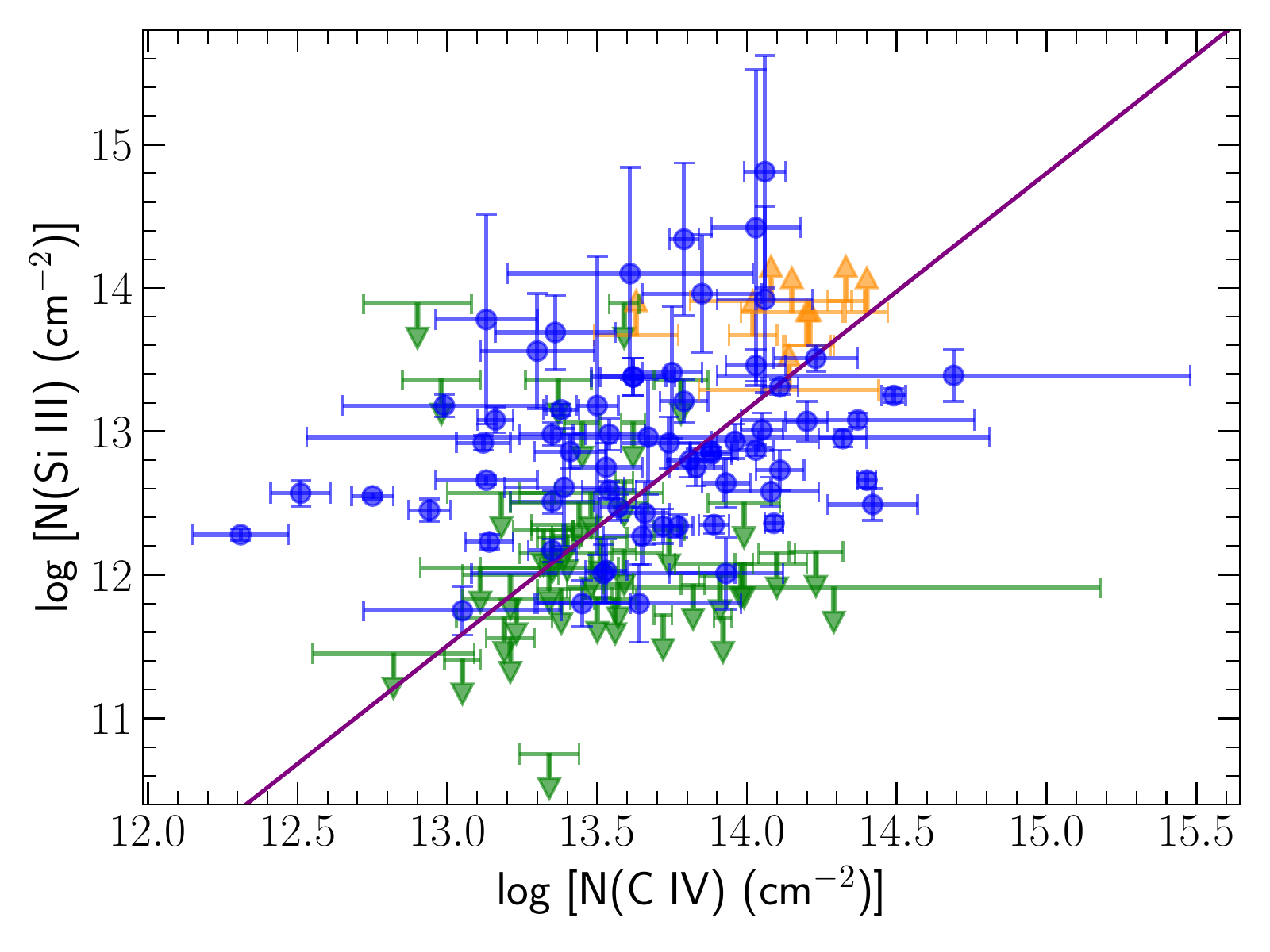}\quad
        \includegraphics[width=210pt,height=159pt,trim={0.35cm 0.3cm 0.3cm 0.3cm},clip=true]{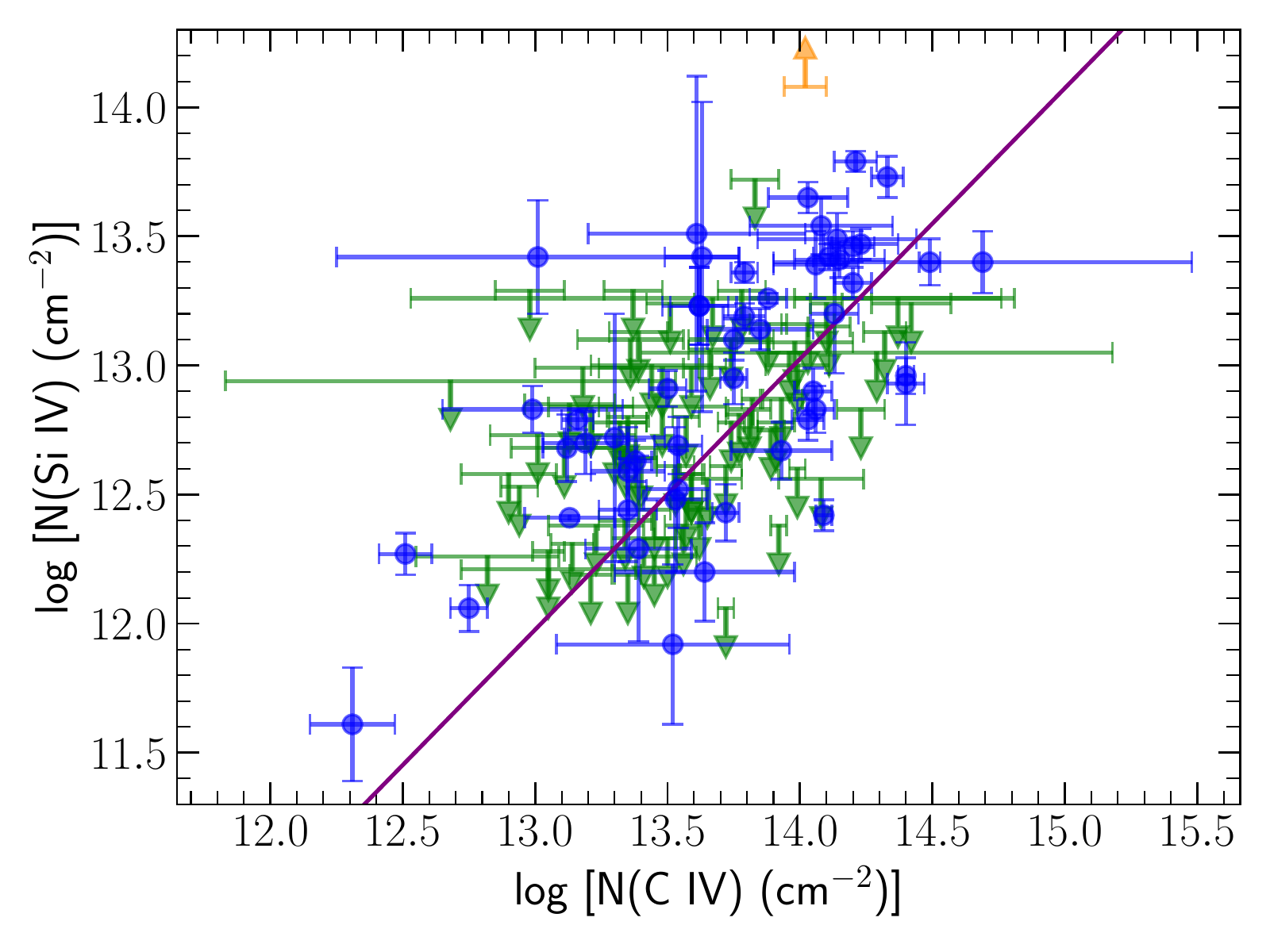}
    \caption{Column densities of different ions plotted against $\CIV$. The green arrows represent $3\sigma$ upper limits from non-detections and the  orange arrows are lower limits for saturated components. The solid line in each panel correspond to best-fit relations that are based on survival analysis, including the censored data i.e. upper and lower limits.}
    \label{7}
    \end{center}
\end{figure}

\begin{figure}
    \begin{center}
        \includegraphics[width=210pt,height=159pt,trim={0.33cm 0cm 0.3cm 0.3cm},clip=true]{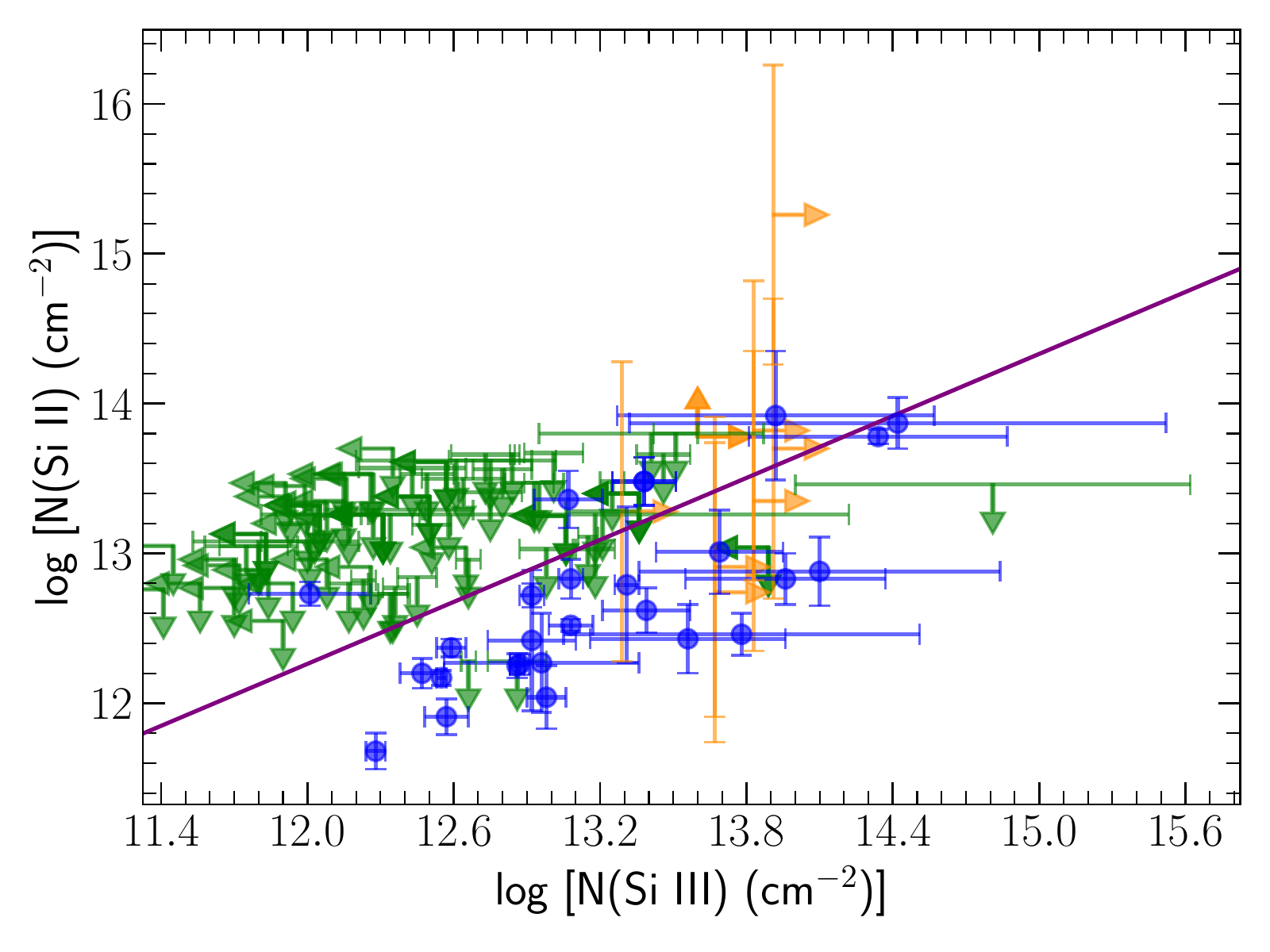}
        \includegraphics[width=210pt,height=159pt,trim={0.33cm 0cm 0.3cm 0.3cm},clip=true]{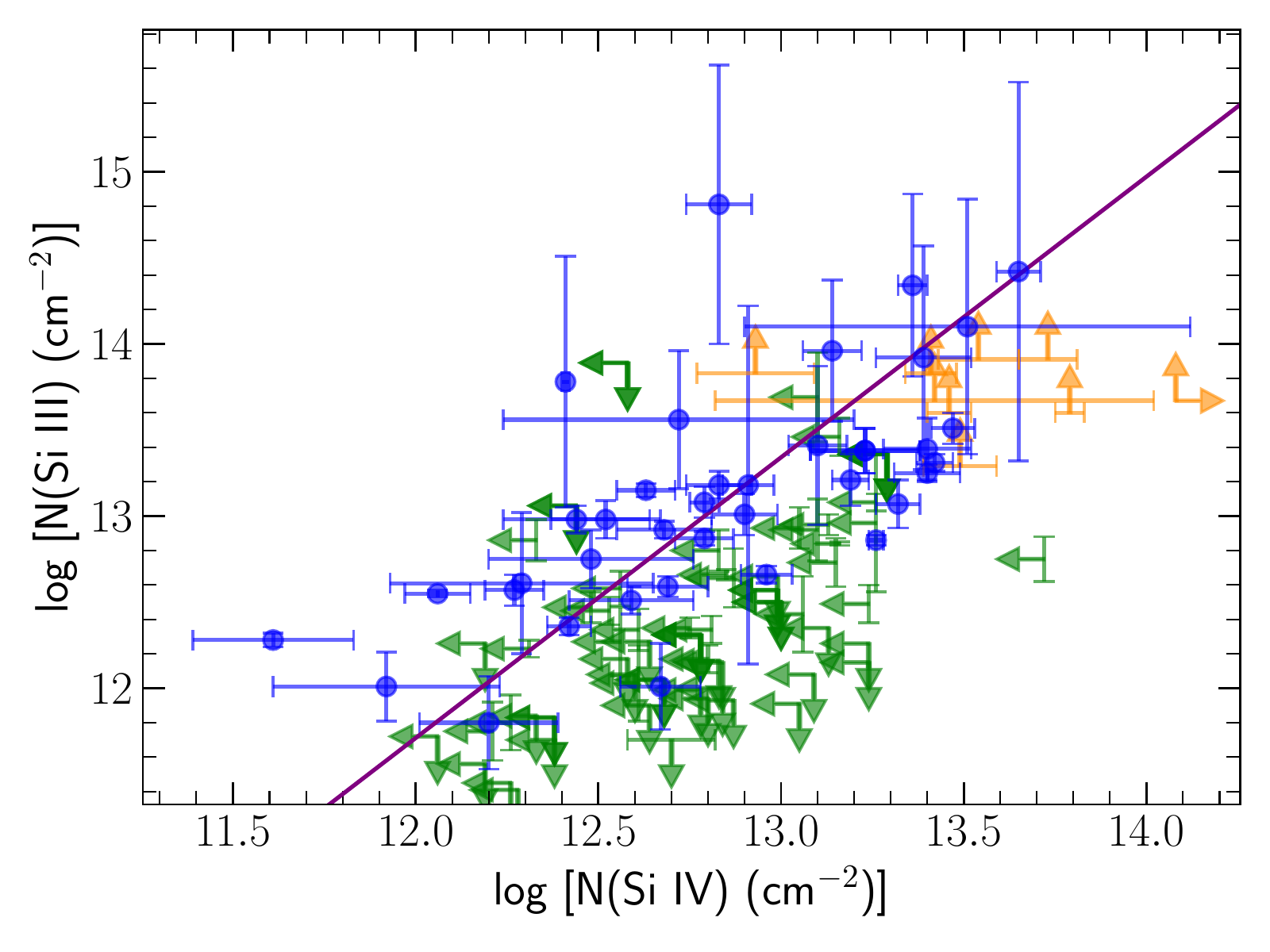}
        \includegraphics[width=210pt,height=159pt,trim={0.33cm 0.38cm 0.3cm 0.3cm},clip=true]{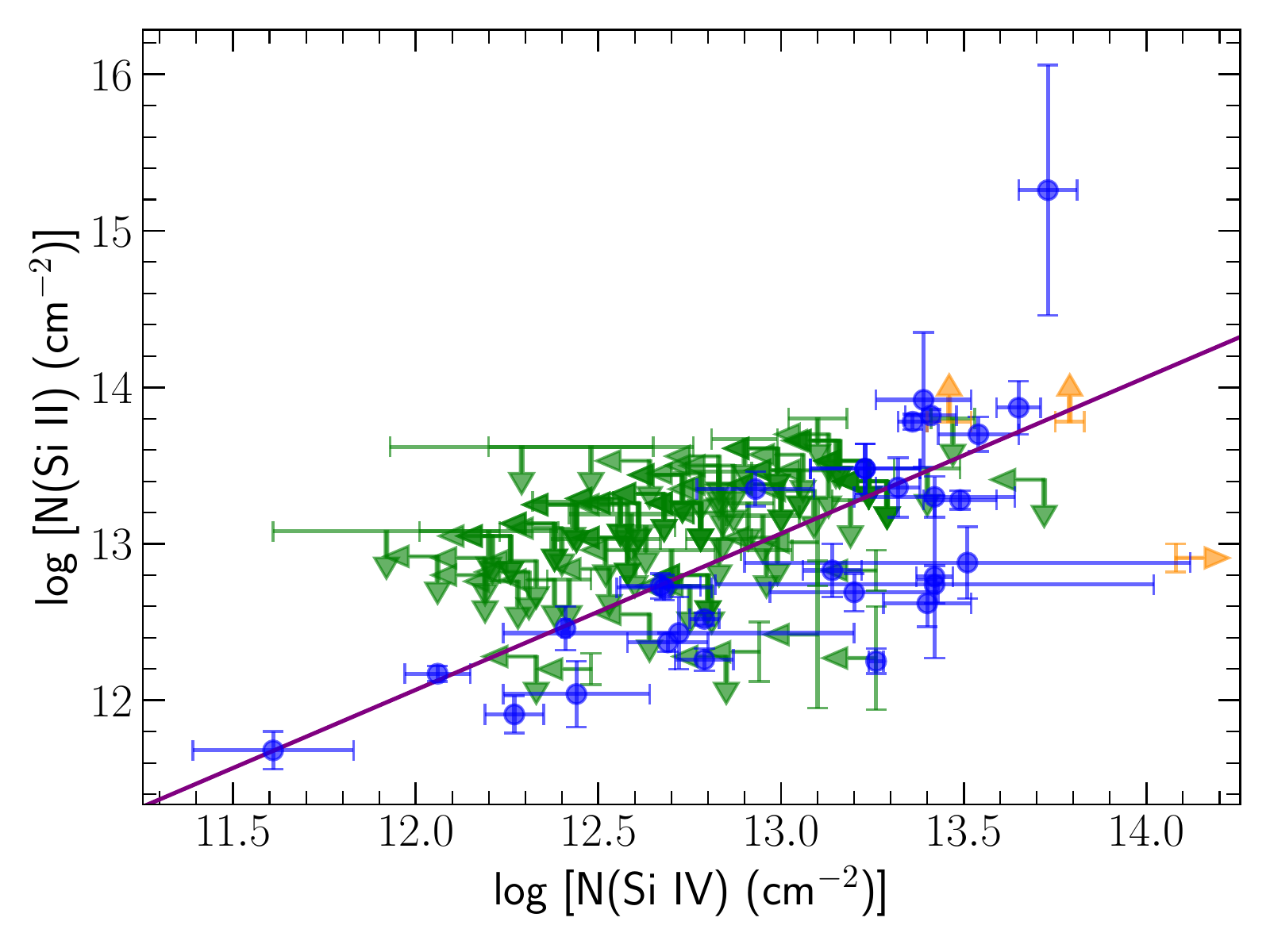}
    \caption{The column density trends of different ions plotted against $\SiIV$. The green arrows represent $3\sigma$ upper limits from non-detections and the  orange arrows are lower limits for saturated components. The solid line in each panel correspond to correlations between the column densities, including the data points which are upper and lower limits.}
    \label{8}
    \end{center}
\end{figure}

\begin{figure}
    \begin{center}
        \includegraphics[width=220pt,height=160pt,trim={0.33cm 0cm 0.3cm 0.3cm},clip=true]{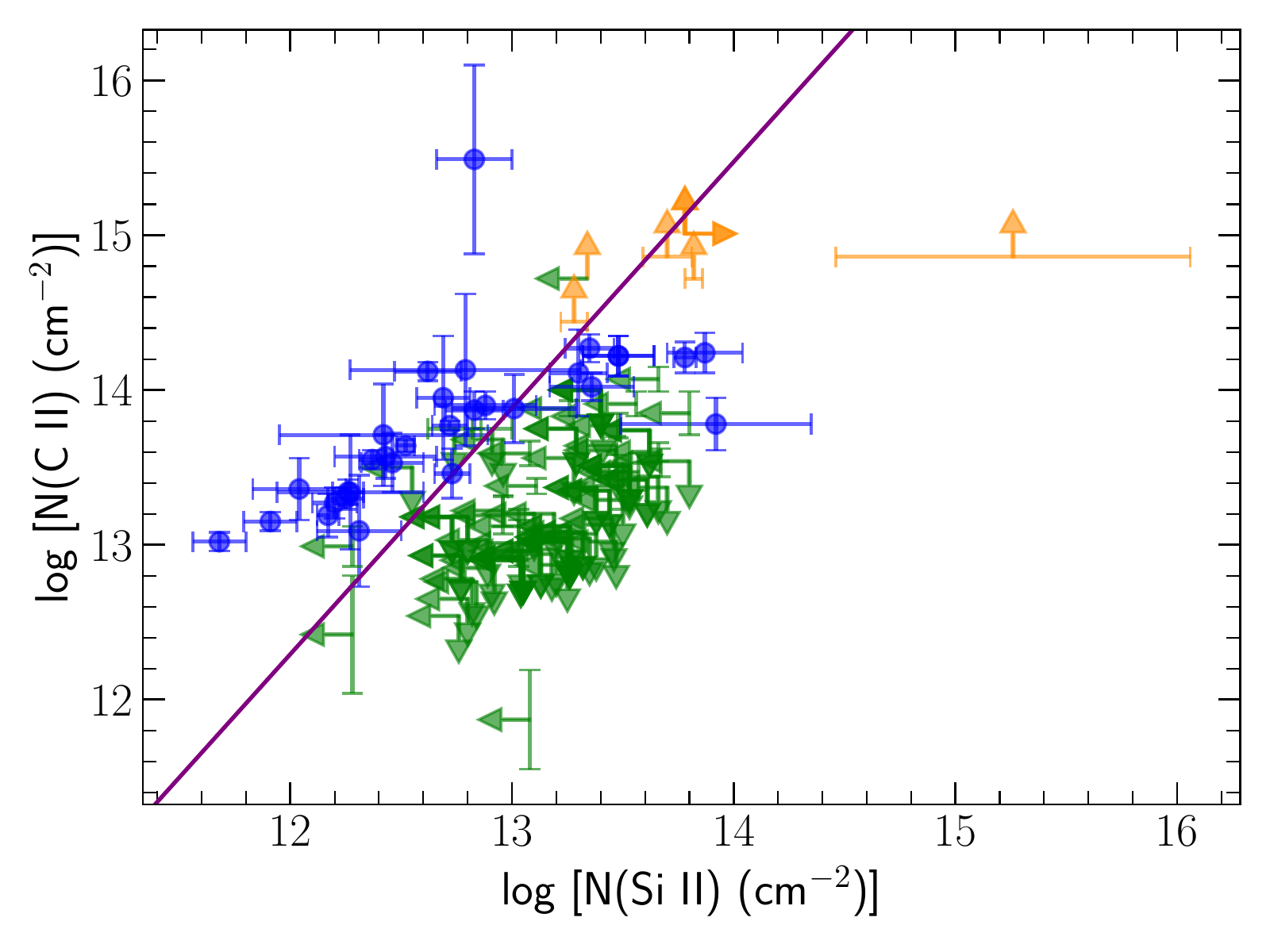}
        \includegraphics[width=220pt,height=160pt,trim={0.33cm 0.4cm 0.3cm 0.3cm},clip=true]{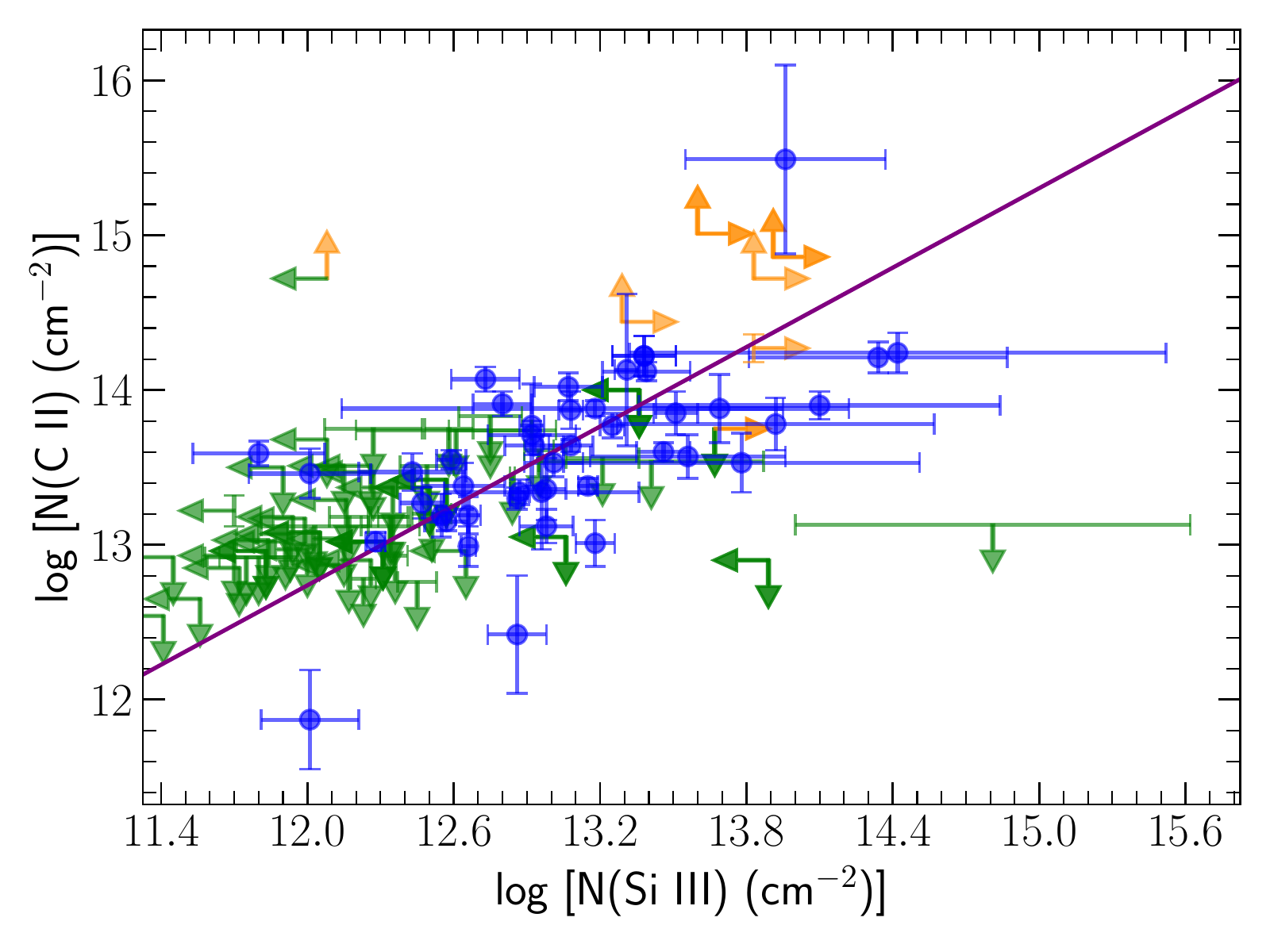}
    \caption{The column density trends of $\CII$ with $\SiII$ and $\SiIII$.The green arrows represent $3\sigma$ upper limits from non-detections and the  orange arrows are lower limits for saturated components. The solid line in each panel correspond to correlations between the column densities, including the data points which are upper and lower limits.}
    \label{9}
    \end{center}
\end{figure}

In Fig.~\ref{7}, Fig.~\ref{8} and Fig.~\ref{9} we investigate the relations amongst the C and Si ions, which have a higher detection rate in the sample compared to ions of other elements. In a number of cases, {\CII}, {\SiII} and {\SiIII} are non-detections, yielding useful upper limits on the column densities. We determine the bisector regression fits to the column density correlations by taking into account these censored values also, in addition to the detections, by using the survival analysis statistical package \textsc{\large asurv}\footnote{\url{http://python-asurv.sourceforge.net/}} \citep{Feigelson1985,Isobe1986}. The Schmitt's binning method is used \citep{Schmitt1985} since it is the only method in \textsc{\large asurv} capable of handling censoring in both the independent and dependent variables. The fits are based on convergence after 10000 iterations. Fig.~\ref{7} shows the relations between C and Si ions with {\CIV}. The linear fits imply
\begin{align}
\log N(\CII) = (1.46\pm 0.14)\log N(\CIV) - (6.72\pm 1.94),\\
\log N(\SiII) = (1.13\pm 0.13)\log N(\CIV) - (2.75\pm 1.70),\\
\log N(\SiIII) = (1.65\pm 0.18)\log N(\CIV) - (9.90\pm 2.49),\\
\log N(\SiIV) = (1.05\pm 0.08)\log N(\CIV) - (1.65\pm 1.13).
\end{align}
The least scatter is for the {\SiIV} - {\CIV} relation. This is expected since the ionization energy for {\CIV} is closest to that for {\SiIV} amongst the four ions. However, the scatter only slightly increases with increasing difference in ionization potential suggesting that the {\CIV} and lower ions may not always represent vastly different gas phases. Ionization models often show that absorbers where the low ions are a non-detection can be explained with a single {\CIV} phase. Instances where the low ions are a detection, the {\CII} (along with similar low ions) and {\CIV} can be explained by considering photoionized gas within a narrow density range \citep{Lopez1999,Manuwal2019}. In certain absorbers {\CIV} requires a separate phase, especially when the kinematic profiles of the low and high ions differ from each other \citep{Ding2003,Misawa2008}. However, unlike {\OVI} and {\NeVIII} absorbers which trace distinctly different density-temperature zones from the low ions, the predictions from ionization models for {\CII} and {\CIV} often indicate origin in a medium where ionization levels are not vastly different. A similar trend of weakening of correlation with increasing differences in ionization potential is also seen by \citet{Burchett2015}. The trends amongst the low and high ions of Si are explored in Fig.~\ref{8} and the linear fits suggest
\begin{align}
\log N(\SiII) = (0.69\pm 0.09)\log N(\SiIII) + (3.99\pm 1.10),\\
\log N(\SiIII) = (1.63\pm 0.16)\log N(\SiIV) - (7.85\pm 2.11),\\
\log N(\SiII) = (1.00\pm 0.18)\log N(\SiIV) - (0.08\pm 2.29).
\end{align}
The scatter in {\SiIII} - {\SiIV} is slightly lower than that in {\SiII} - {\SiIV} because of the closer ionization conditions between the former pair of species. The creation and destruction energies of {\CII} ($11$~eV/$24$~eV) fall in between the corresponding values for {\SiII} and {\SiIII} ($8$~eV/$16$~eV/$33$~eV). In Fig.~\ref{9} we compare the column densities of these ions with {\CII}. The survival analysis yields
\begin{align}
\log N(\CII) = (1.59\pm 0.21)\log N(\SiII) - (6.79\pm 2.62),\\
\log N(\CII) = (0.86\pm 0.11)\log N(\SiIII) + (2.47\pm 1.38).
\end{align}
The {\CII} is comparably well correlated with {\SiIII} and {\SiII}. In ionization models, all three species are often identified as originating from gas of similar density. In fact, surveys of {\MgII} absorbers at $z < 0.4$ often use {\CII} and {\SiII} as proxy doublets for {\MgII} because of their origin in a common phase \citep[e.g.][]{Narayanan2005,Muzahid2018}. 

\section{ABSORBER LINE STATISTICS}\label{stats}

Here we determine the column density distribution function and the contribution of {\CIV} absorbers to the closure density where the statistics are based on \textit{total} column densities of systems as per \textsc{\large vpfit}, as opposed to AOD based integrated column densities. We exclude the 16 systems that were found around targeted galaxies, and the $z_\mathrm{abs}=0.08940$ system towards QSO B1435-0645 that has a saturated {\CIV} component.

\subsection{Column density distribution function}
The column density distribution function $f(N(\CIV))$ is the distribution of number of absorbers $\mathcal{N}(N(\CIV))$ in each {\CIV} column density bin $\Delta N(\CIV)$ normalized by the comoving path length $\Delta X(N(\CIV))$ for that column density threshold. The normalization is to correct for the completeness at each column density. This is determined as
\begin{equation}
f(N(\CIV)) = \frac{\mathcal{N}(N(\CIV))}{\Delta N(\CIV)\Delta X(N(\CIV))}.
\end{equation}
\noindent In order to determine the comoving path lengths, we first convert each wavelength in our spectra to redshift, and then from redshift to $X$ as
\begin{equation}
X(z) = \frac{2}{3\Omega_\textrm{m}}[\sqrt{(1+z)^3\Omega_\textrm{m} + \Omega_{\Lambda}} - 1].
\label{copl}
\end{equation}
The subtraction by 1 in Eq.~\ref{copl} is to account for the fact that $X(0)=0$. The total comoving path length for each column density is determined with a procedure analogous to that for the redshift path length (Sec.~\ref{complete}). The values are provided in Table~\ref{pathtab}.

\begin{figure}
    \begin{center}
        \includegraphics[width=220pt,height=160pt,trim={0.2cm 0cm 0.35cm 0.3cm},clip=true]{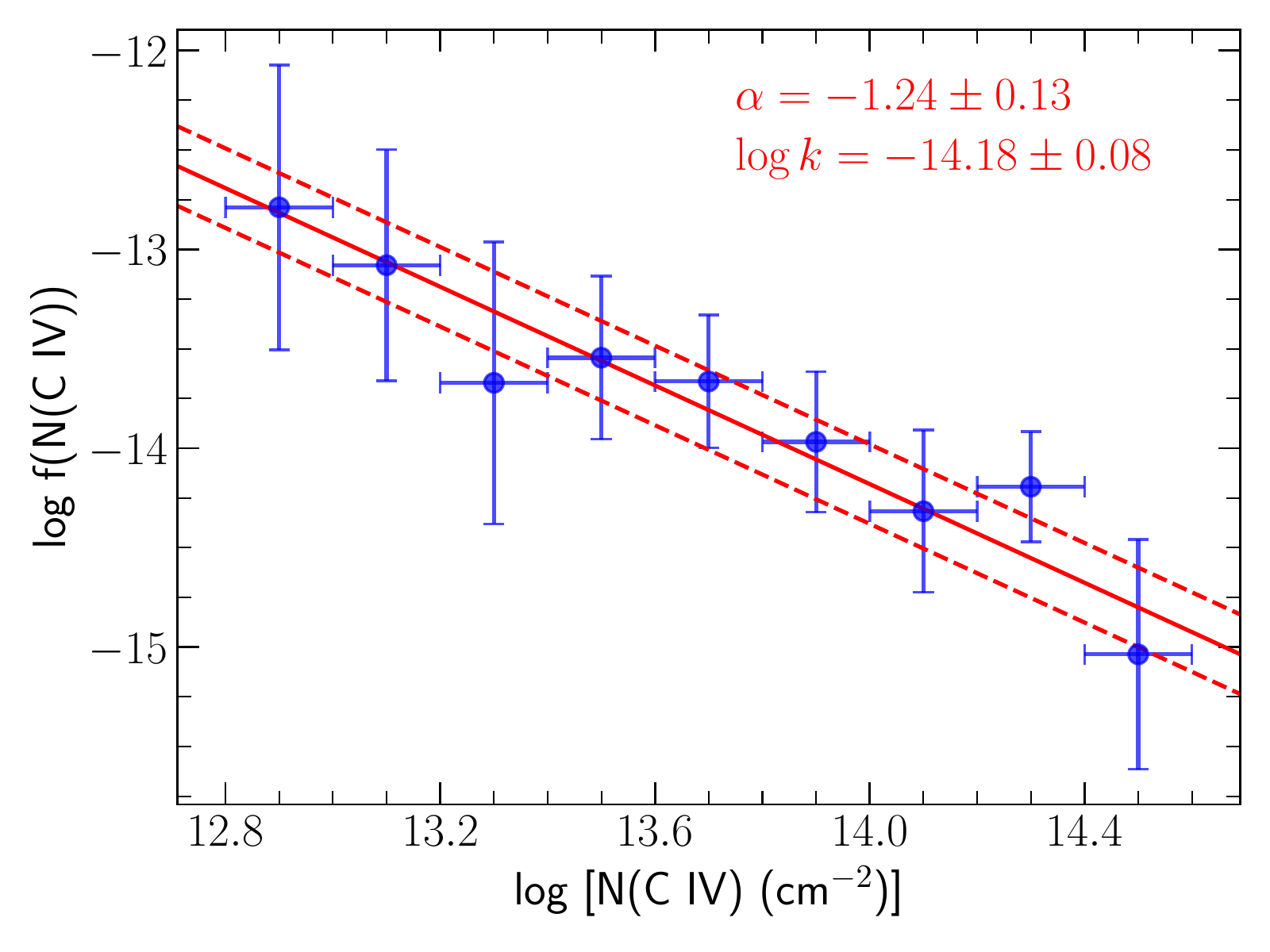}
    \caption{The column density distribution function for the {\CIV} sample. The solid-red line is the power-law fit of Eq.~\ref{cddf} and the dashed lines show its $1\sigma$ uncertainty.}
    \label{fn}
    \end{center}
\end{figure}

\begin{figure}
    \begin{center}
        \includegraphics[width=240pt,height=180pt,trim={0.2cm 0.4cm 0.2cm 0.1cm},clip=true]{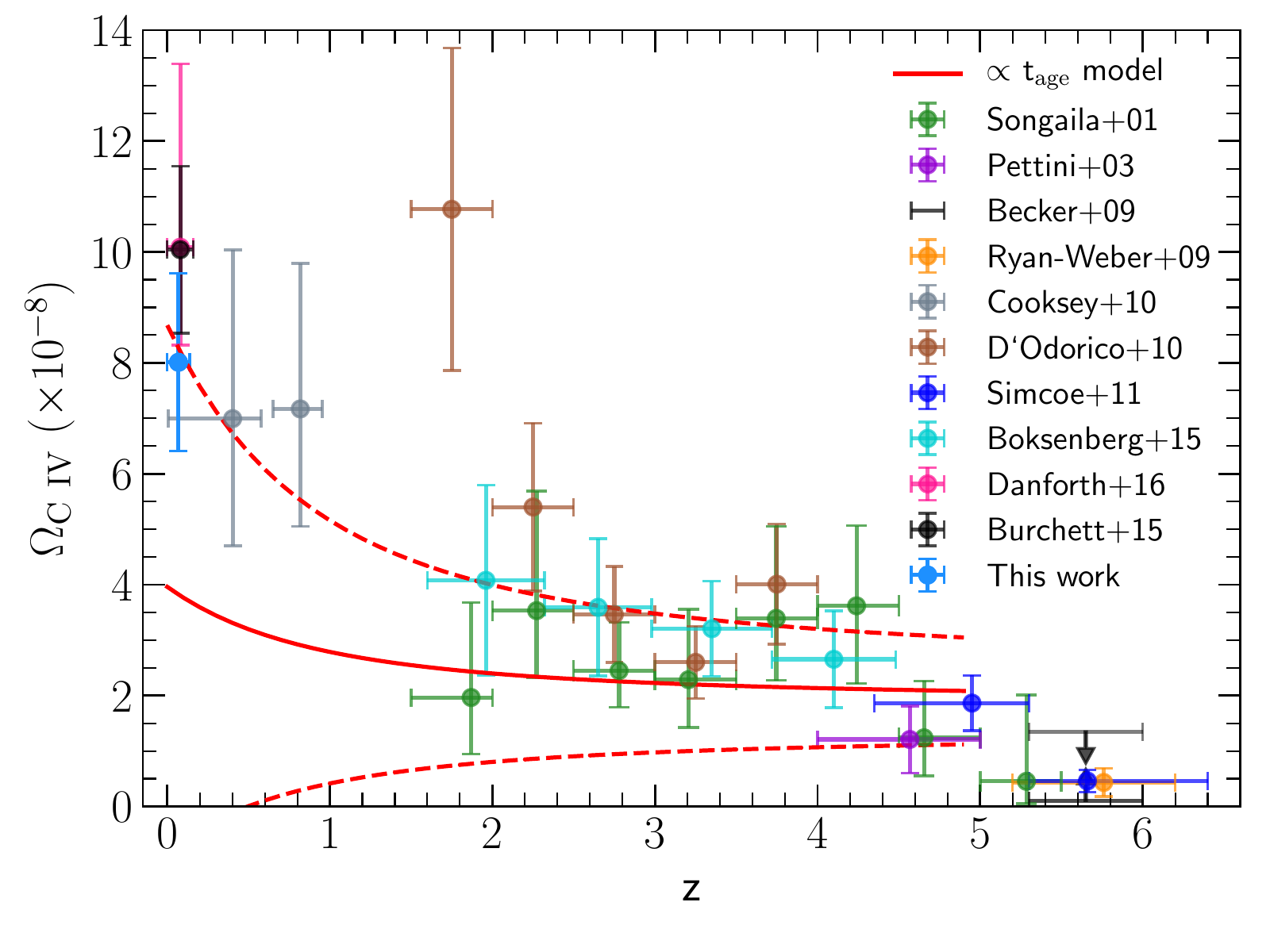}
    \caption{This shows various estimates from previous surveys along with this one (shown in light-blue) for the ratio of mass density in {\CIV} relative to the critical density for a flat universe from high-$z$ to the present. The solid-red curve is the best-fit model given by \citet{Cooksey2010} (see their Eq.~23) for linear evolution in $\Omega_{\CIVm}$ with age of the universe, based on the {\CIV} mass density values for $1 < z < 5$. The dashed curves show $1\sigma$ region for the same. The model extended to lower redshifts shows that the mass-density of {\CIV} has steadily increased from high redshifts to the present.}
    \label{omega}
    \end{center}
\end{figure}

The $f(N(\CIV))$ for our absorbers with column density bins of $\Delta \log N(\CIV) = 0.2$~dex is shown in Fig.~\ref{fn}. The error is determined by adding two sources of error in quadrature. The first is the contribution from Poisson uncertainty in $\mathcal{N}(N(\CIV))$, which is the primary source of error. Another contribution is from the uncertainty in comoving path length \citep{Danforth2008}
\begin{equation}
\delta\Delta X_\mathrm{i} = \frac{\Delta X_\mathrm{i+1} - \Delta X_\mathrm{i-1}}{8},
\label{cplunc}
\end{equation}
\noindent where $\Delta X_\mathrm{i\pm1}$ corresponds to co-moving pathlength for $\log N(\CIV) \pm 0.1$. The Spearman rank test for $f(N(\CIV))-N(\CIV)$ relation gives $r\sim-1,p\ll0.001$ and the function can be modelled accurately with a power law given by the optimum likelihood fit using \textsc{\large hyper-fit} as
\begin{equation}
f(N(\CIV))=k \left[\frac{N(\CIV)}{N_0} \right]^{\alpha},
\label{cddf}
\end{equation}
\noindent where $N_0 = 10^{14}~\cmsq$, $\log k = -14.18\pm 0.08$ and $\alpha = -1.24\pm 0.13$. This $k$ is within $1\sigma$ of $-14.06^{+0.15}_{-0.16}$ for <$z$>~$=0.40227$ from G=1+2 sample in \citet{Cooksey2010}, but not consistent with $-13.76\pm 0.07$ in \citet{Burchett2015}. The power-law index $\alpha$ is also significantly higher than $-2.07\pm 0.15$ in \citet{Burchett2015}, but consistent within $2\sigma$ with $-1.74^{+0.28}_{-0.31}$ in \citet{Cooksey2010}. However, \citet{Burchett2015} used their gamma-function fit instead of the power-law fit for analysis. The column density distribution function is a crucial parameter in the estimation of the mass density contribution from {\CIV} absorbers to the critical density. As the distribution function shows, the low column density absorbers are far more numerous compared to the high column density systems. Constraining the distribution function accurately at lower column densities hence becomes important. As shown in Fig.~\ref{fn}, the distribution function arrived at in Eq.~\ref{cddf} explains well the column density trend across three orders of magnitude.

\subsection{C~{\textsc{\large iv}} density parameter ($\Omega_{\mathbf{C}\,\textsc{\textmd iv}}$)}

The absorber line density above a certain limiting column density $N_\mathrm{lim}$ is defined as
\begin{equation}
\frac{d\mathcal{N}}{dX}(N(\CIV)\geq N_\mathrm{lim}) = \int_{N_\mathrm{lim}}^{\infty}f(N(\CIV))dN(\CIV).  
\end{equation}
We use the model for $f(N(\CIV))$ in Eq.~\ref{cddf} to determine the $d\mathcal{N}/dX$. The observed lower limit of $N(\CIV)$ in the systems is $N_\mathrm{lim} = 10^{12.9}~\cmsq$ and that gives $d\mathcal{N}/dX = 5.1\pm1.0$. The uncertainty is determined by considering the 1$\sigma$ scatter provided by \textsc{\large hyper-fit}. This value is $\sim 2.7$ times higher than the prediction for this epoch based on the linear evolution model by \citet{Hasan2020}. They mention that such discrepancies could either be due to the lack of measurements for high redshifts, or a sudden increase in {\CIV} clouds at $z\sim0.5$ over a 0.5~Gyr time period, but argue against the latter, as it is difficult to understand such a rapid evolution.

$\Omega_{\CIVm}$ is the ratio of the mass density of {\CIV} to the critical density $\rho_c$ of the Universe. We can use the column density distribution for our sample to estimate this for the redshift range of our absorbers as
\begin{equation}\label{om1}
\Omega_{\CIVm}=\frac{H_0 m_\textsc{c}}{c\rho_\mathrm{c,0}}\int_{N_\mathrm{min}}^{N_\mathrm{max}}f(N(\CIV))N(\CIV)dN(\CIV),
\end{equation}
where $H_0$ is Hubble parameter for the present Universe, $m_\textsc{c}$ is the mass of carbon atom, $\rho_\mathrm{c,0} = 3H_{0}^2/8\pi G$ is the present-day value of the critical density, $c$ is the speed of light in vacuum, and $N_\mathrm{min}$ and $N_\mathrm{max}$ are the integration limits. On taking $N_\mathrm{min},N_\mathrm{max}=10^{12.9},10^{15.0}~\cmsq$ (to be consistent with \citealt{Burchett2015} and \citealt{Danforth2016}), we arrive at $\Omega_{\CIVm}=(8.01\pm1.62) \times 10^{-8}$.

Fig.~\ref{omega} shows the value of $\Omega_{\CIVm}$ at different redshifts reported by previous surveys from $z = 6$ to the present, along with our estimate. These values are corrected for cosmology according to the scaling method described in \citet{Cooksey2010}. \citet{Becker2009} provide lower and upper limits at $z > 5.3$ instead of a measurement. Our estimate based on the fit is consistent with the value reported by \citet{Cooksey2010} within $1\sigma$. \citet{Cooksey2010} had derived a model for linear evolution of $\Omega_{\CIVm}$ with the age of the Universe based on $1<z<5$. Our estimate is well within its prediction for the low-$z$ Universe, which suggests a steady increase of mass-density across time. It is also consistent with \citet{Burchett2015} who used their column density distribution function for $z<0.16$ to estimate the density parameter, and \citet{Danforth2016} who used a differential distribution ($\partial^2{\cal N}(N)/ \partial \log N\partial z$) of their sample to arrive at the mass-density estimate.

The mass density in {\CIV} is roughly a constant from $z \sim 5 - 1$ \citep{Pettini2003,Boksenberg2015} despite the cosmic star formation rate density and the activity of bright AGNs peaking at $z \sim 2$ \citep{Boyle2000,Madau2014}. The hydrodynamic simulations of \citet{Dave2006} show that the outflow processes from star-forming galaxies which increase the metal abundances in the IGM also raise the IGM temperature pushing carbon to ionization states higher than {\CIV}. The metal enrichment and the excessive ionization due to overheating counter act each other keeping $\Omega_{\CIVm}$ nearly a constant in diffuse gas in the interval $z \sim 5 - 1$. In a follow-up analysis, \citet{Dave2009} explain the observed increasing trend from $z \sim 1$ to $z \sim 0$ as due to the enhancement of carbon abundance in the ISM coming from mass-loss in AGB stars. The mass loss also serves as fuel for star-formation in galaxies at $z < 1$ epochs. The supernova driven feedback that follows can increase $\Omega_{\CIVm}$ by as much as $70\%$ at $z \lesssim 0.5$ ($\sim 0.2 - 0.3$~dex). 

We also use the mass-density of {\CIV} relative to the closure density to determine the metallicity of diffuse gas outside of galaxies in the CGM and IGM at $z < 0.16$ from the relation given in \citet{Ryan2009},
\begin{equation}
    Z_{\mathrm{IGM}} = \frac{\Omega_{\CIVm}}{\Omega_\mathrm{b}} \frac{1}{f_{\CIVm}} \frac{1}{A_\mathrm{C}},
\end{equation}
\noindent where $f_{\CIVm}$ is the ionization fraction of carbon and $A_\mathrm{C}$ the mass fraction of metals in carbon. For convenience, we use IGM as subscript, though in reality the sample of {\CIV} absorbers are a mix of both CGM and IGM gas. Adopting values of $\Omega_\mathrm{b} = 0.0463~{\pm}~0.0024$ \citep{Hinshaw2013} for the baryonic mass density determined from the angular power spectrum of the CMB as measured by WMAP, $f_{\CIVm} = 0.3$ corresponding to the peak in the ionization fraction of {\CIV} in photoionized gas as obtained from \textsc{\large cloudy}, and $A_\mathrm{C} = 0.182$ \citep{Caffau2011}, we obtain $Z_{\mathrm{IGM}} = (3.17~{\pm}~0.66) \times 10^{-5}$ that is comparable with the value of $4.1 \times 10^{-5}$ determined by \citet{Shull2014} from measurements on {\CIII} and {\CIV} in the $z \leq 0.4$ Universe. The mass-fraction of metals in the photosphere of the Sun is $Z_{\odot} = 0.0153$ \citep{Caffau2011}, yielding $Z_{\mathrm{IGM}} = (2.07~{\pm}~0.43) \times 10^{-3}~Z_{\odot}$, an order of magnitude more than the metal abundance in the IGM at high redshifts ($z \gtrsim 5$, \citealt{Ryan2009}). The slow build-up of metals in gas constituting the CGM and IGM is associated with feedback in the form of supernova and AGN driven winds over long timescales, and also processes such as tidal interactions, ram-pressure stripping and mergers which are common in galaxy over-density environments. 

\section{IONIZATION MODELLING}\label{models}
\begin{figure}
    \begin{center}
        \includegraphics[width=250pt,height=185pt,trim={0cm 0cm 0cm 0cm},clip=true]{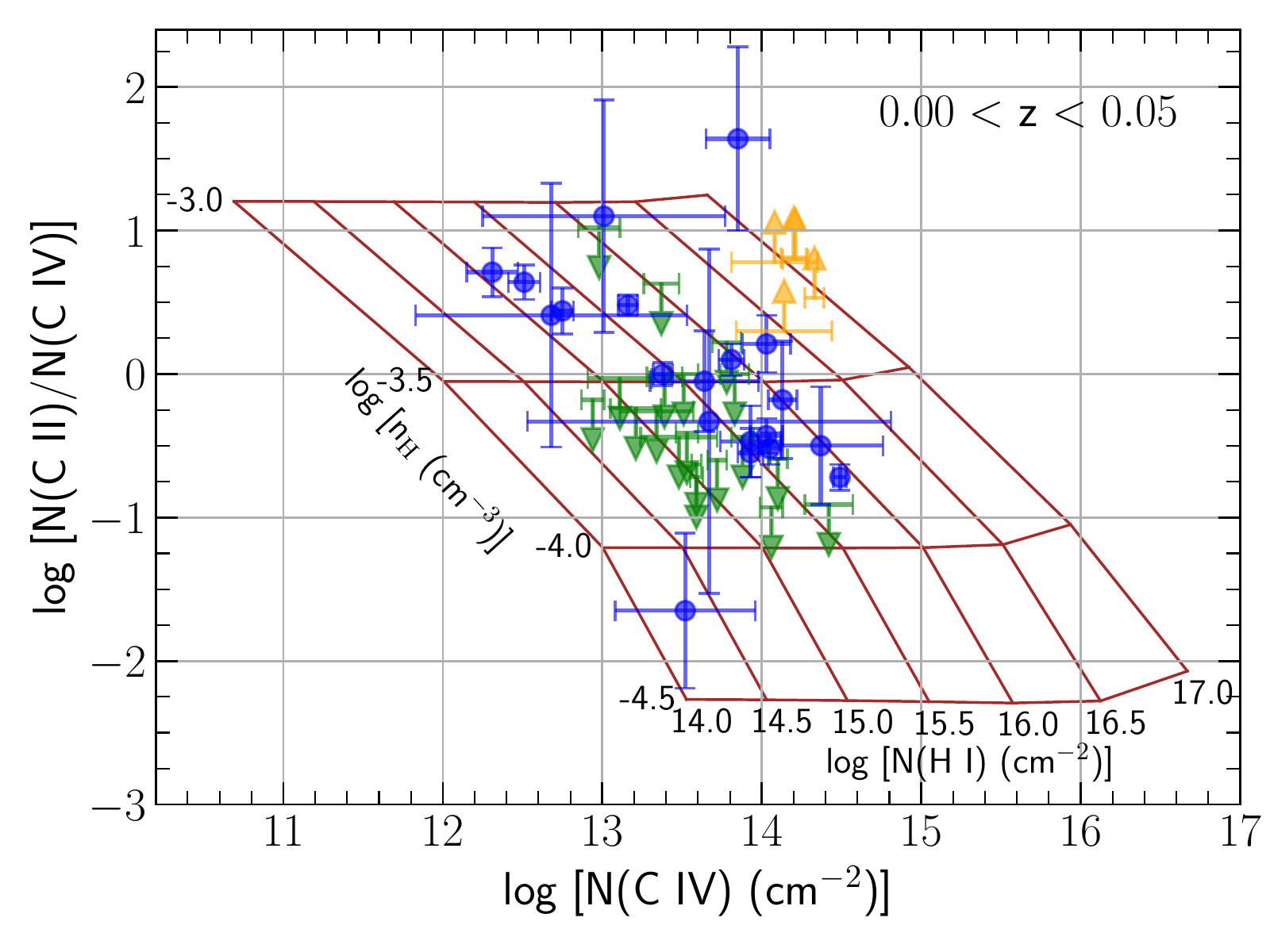}
        \includegraphics[width=250pt,height=185pt,trim={0cm 0cm 0cm 0cm},clip=true]{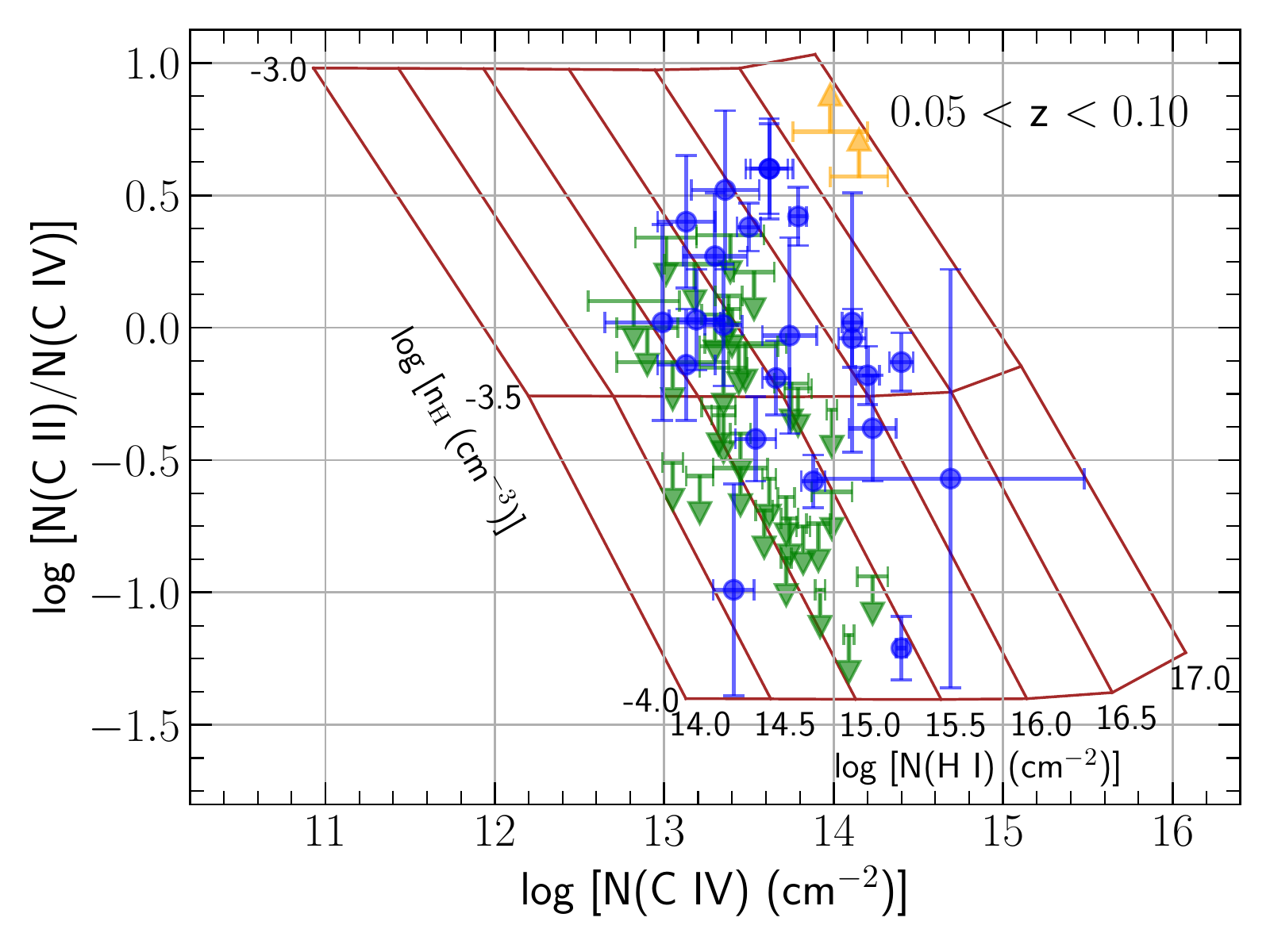}
        \includegraphics[width=250pt,height=185pt,trim={0cm 0cm 0cm 0cm},clip=true]{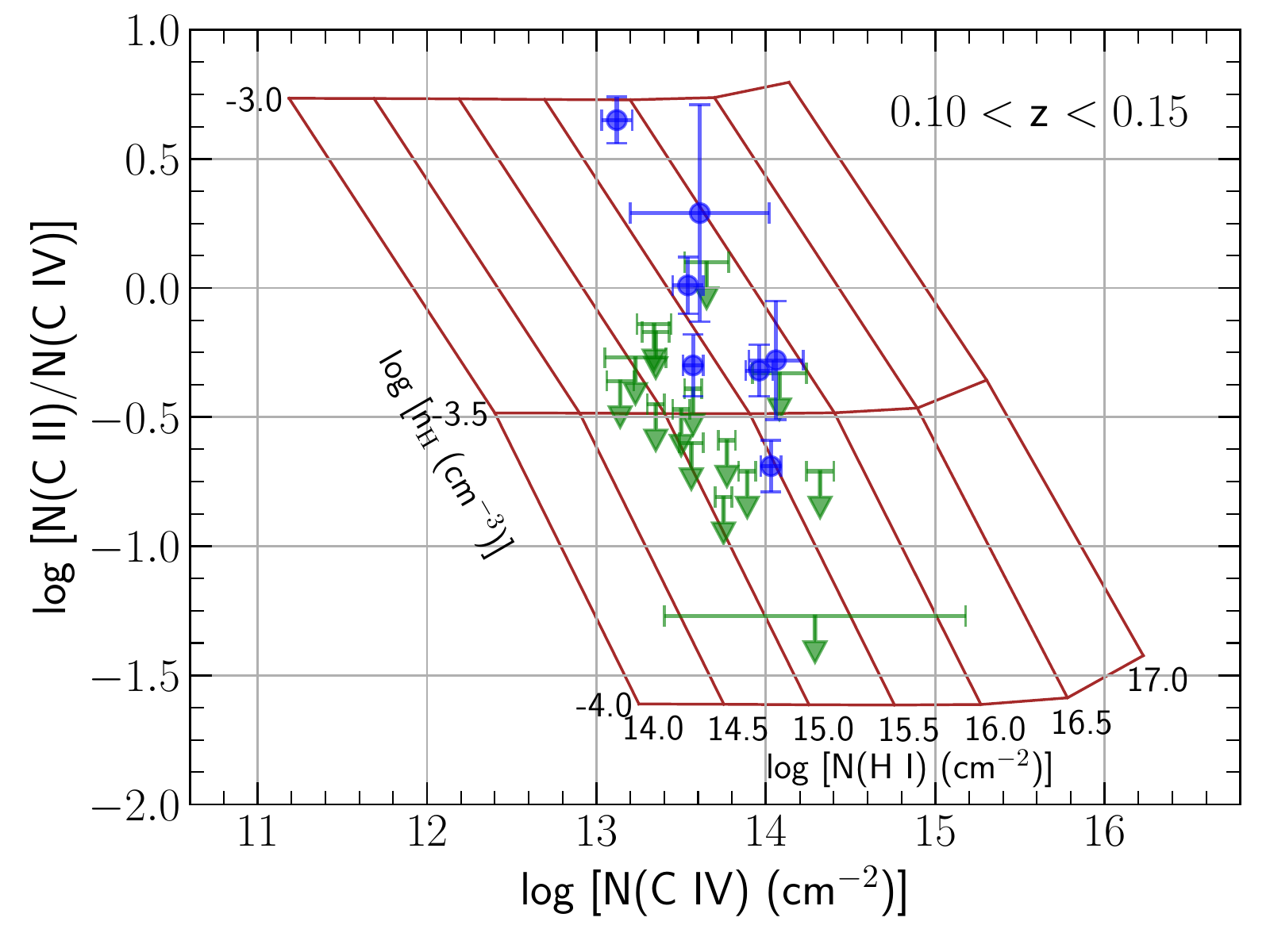}
    \caption{The observed column density ratios of $\CII$ to $\CIV$ is shown in the three panels with the full sample segregated by redshift intervals. The blue data points correspond to systems in which {\CII} is a detection, the green downward pointing arrows represent systems in which {\CII} column density is an upper limit based on $3\sigma$ non-detection, and the two orange upward pointing points in the first two panels are systems in which {\CII}~$1334$~{\AA} line is saturated. The brown grid lines represent the predicted {\CII} to {\CIV} column density ratio from photoionization equilibrium models for different densities and {\HI} column densities (in $\log$ units), the values of which are labeled next to each grid line.}
    \label{11}
    \end{center}
\end{figure}

\begin{figure}
    \begin{center}
        \includegraphics[width=250pt,height=185pt,trim={0cm 0cm 0cm 0cm},clip=true]{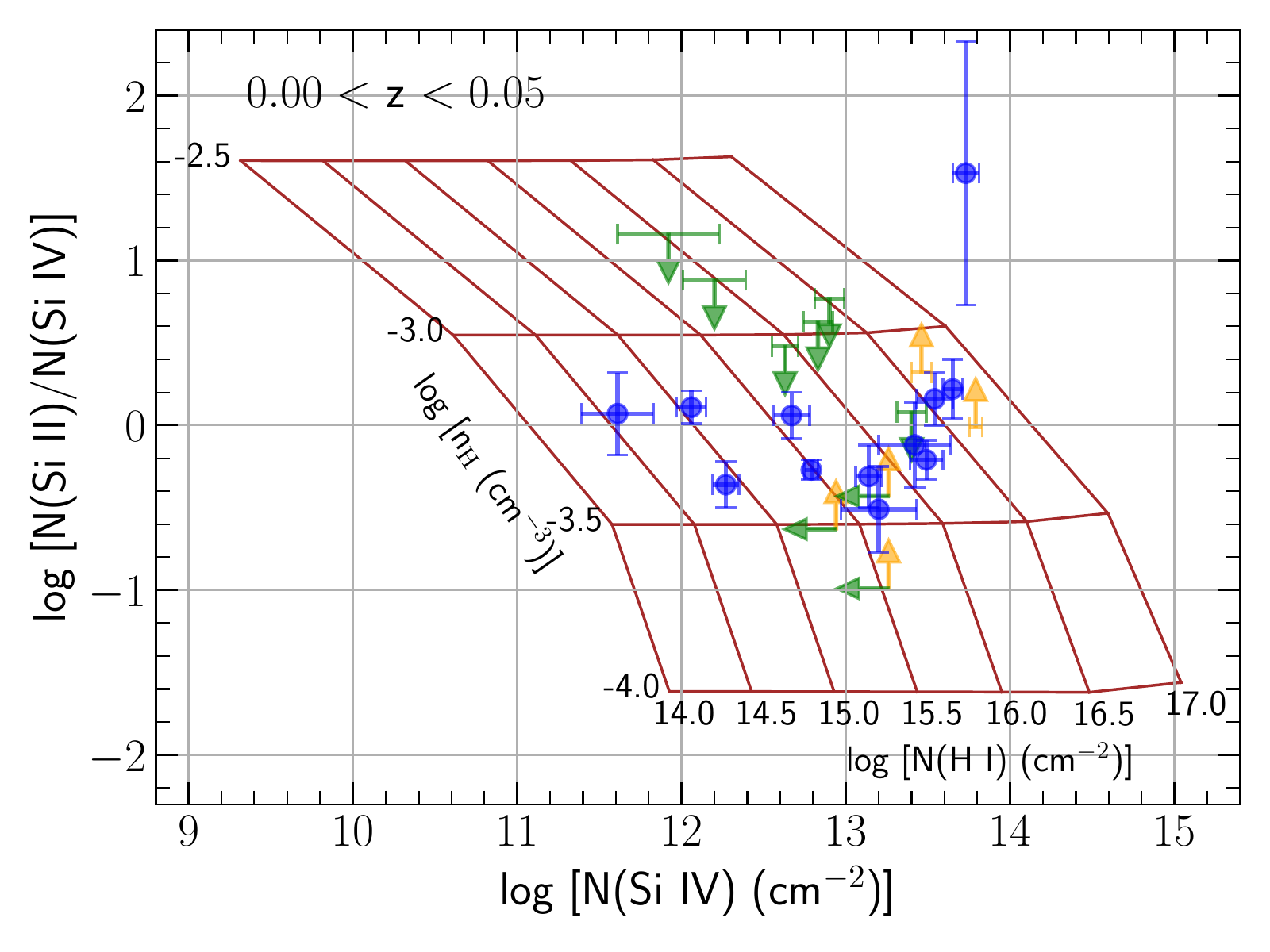}
        \includegraphics[width=250pt,height=185pt,trim={0cm 0cm 0cm 0cm},clip=true]{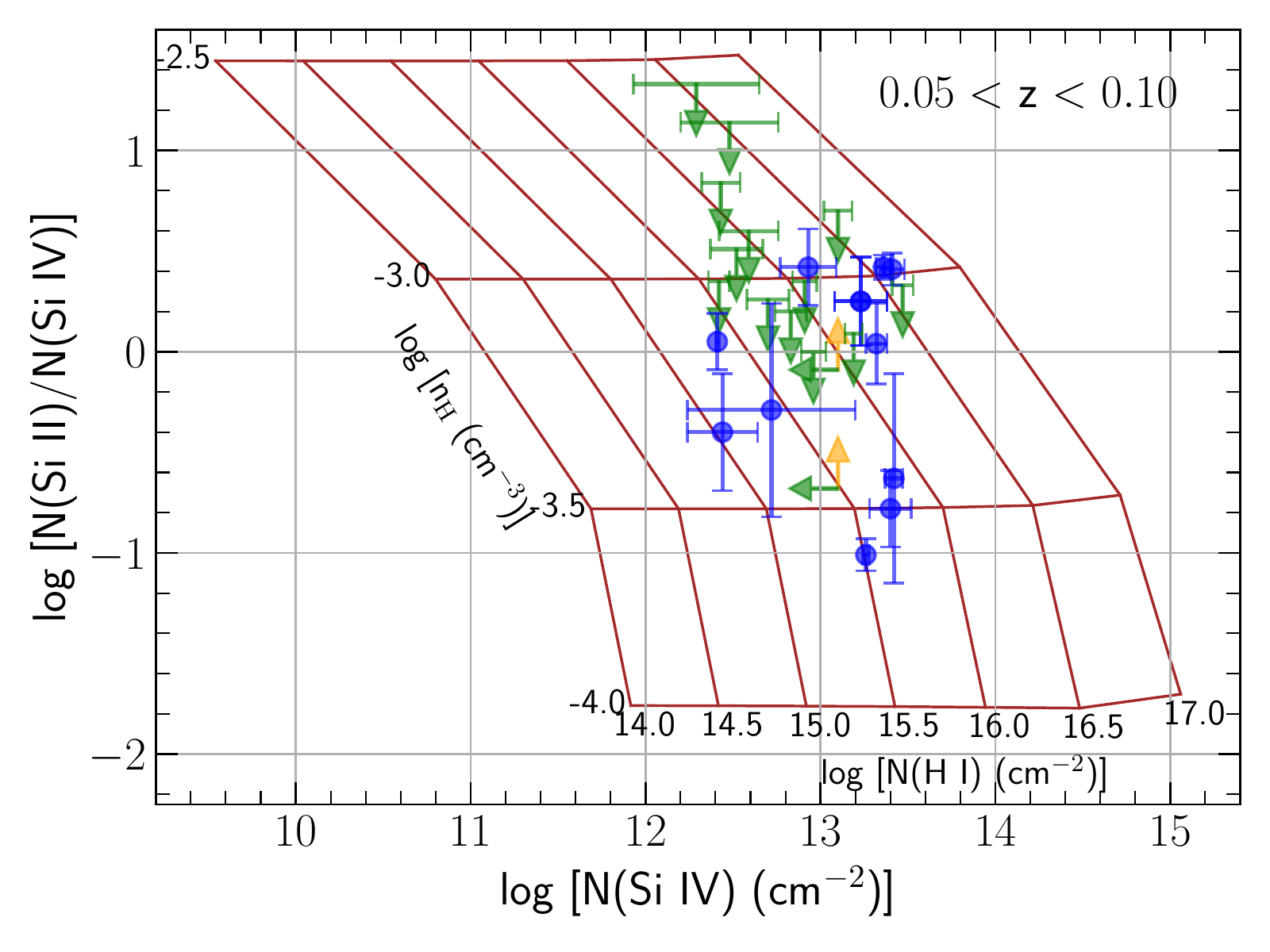}
        \includegraphics[width=250pt,height=185pt,trim={0cm 0cm 0cm 0cm},clip=true]{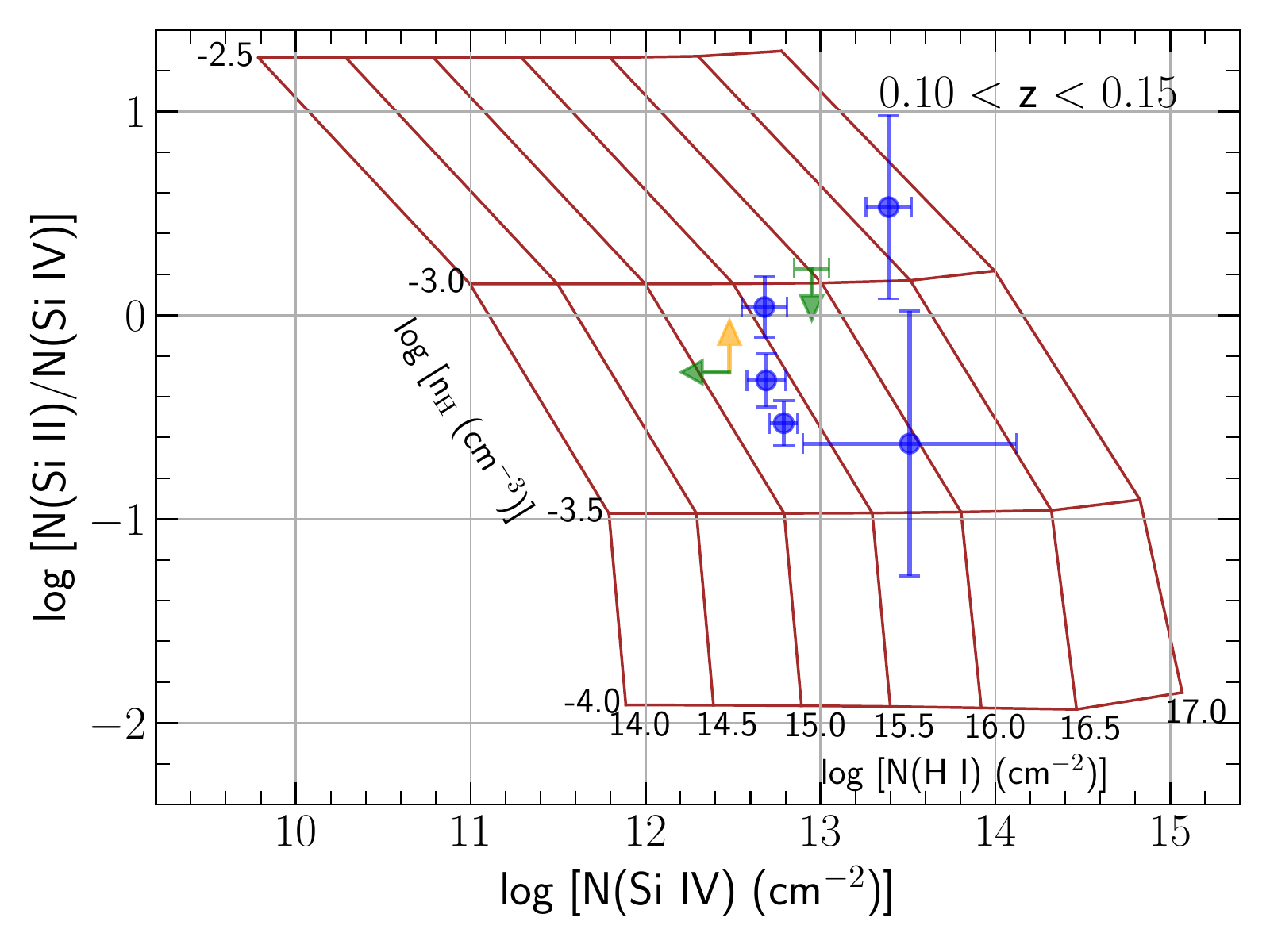}
    \caption{The observed column density ratios of $\SiII$ and $\SiIII$ with $\SiIV$ are shown in the various panels, segregated by redshift intervals. The blue data points correspond to systems in all ions are detections, the green downward or leftward pointing arrows represent systems in which one of the ions is a non-detection, and the orange upward pointing points are systems in which {\SiII} or {\SiIII} is saturated. The grid lines represent the predicted column density ratios from photoionization equilibrium models for different densities and {\HI} column densities (in $\log$ units); the values are labeled next to each grid line.}
    \label{12}
    \end{center}
\end{figure}

\begin{figure*}
    \begin{center}
        \includegraphics[width=250pt,height=185pt,trim={0cm 0cm 0cm 0cm},clip=true]{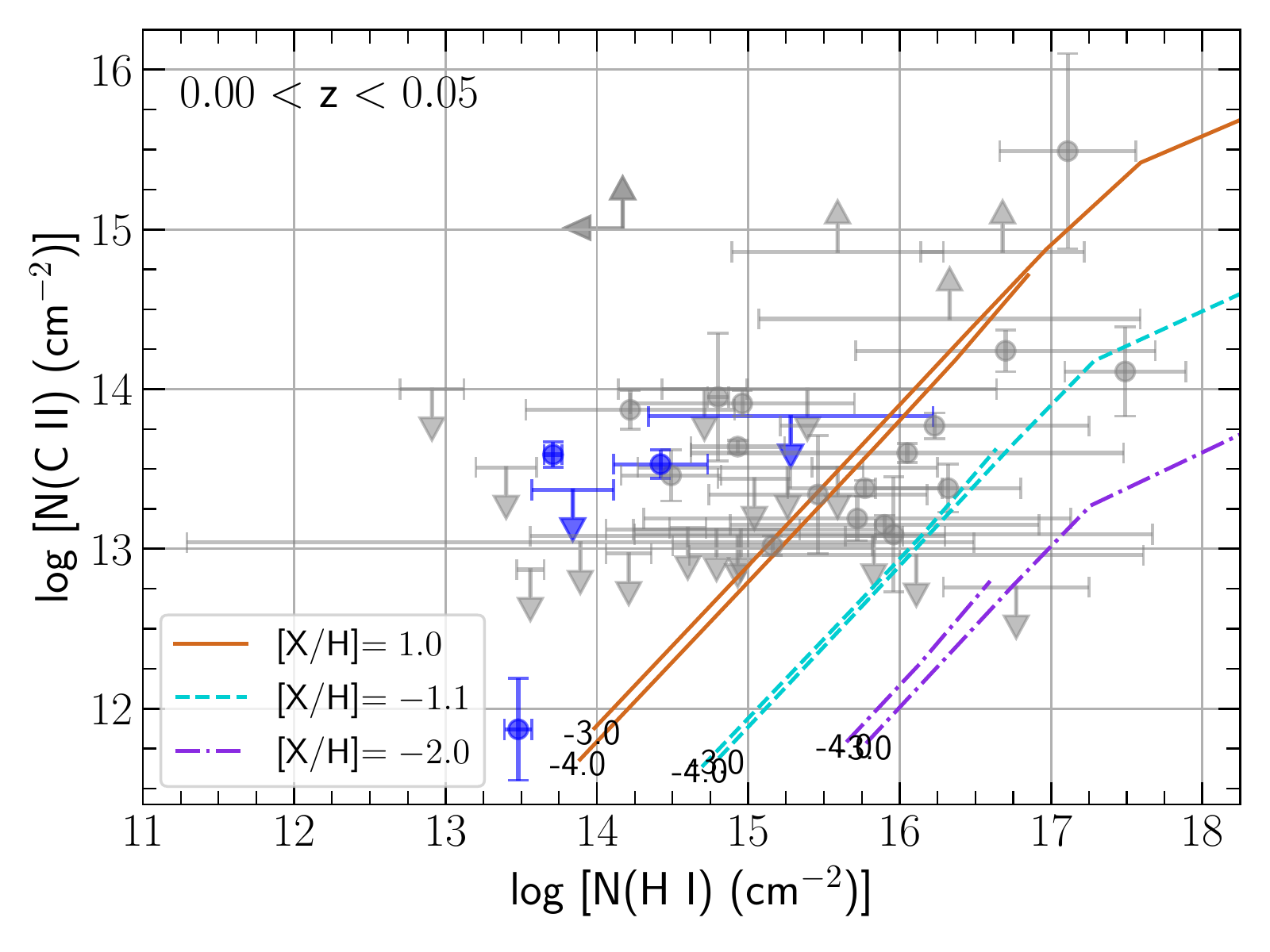}
        \includegraphics[width=250pt,height=185pt,trim={0cm 0cm 0cm 0cm},clip=true]{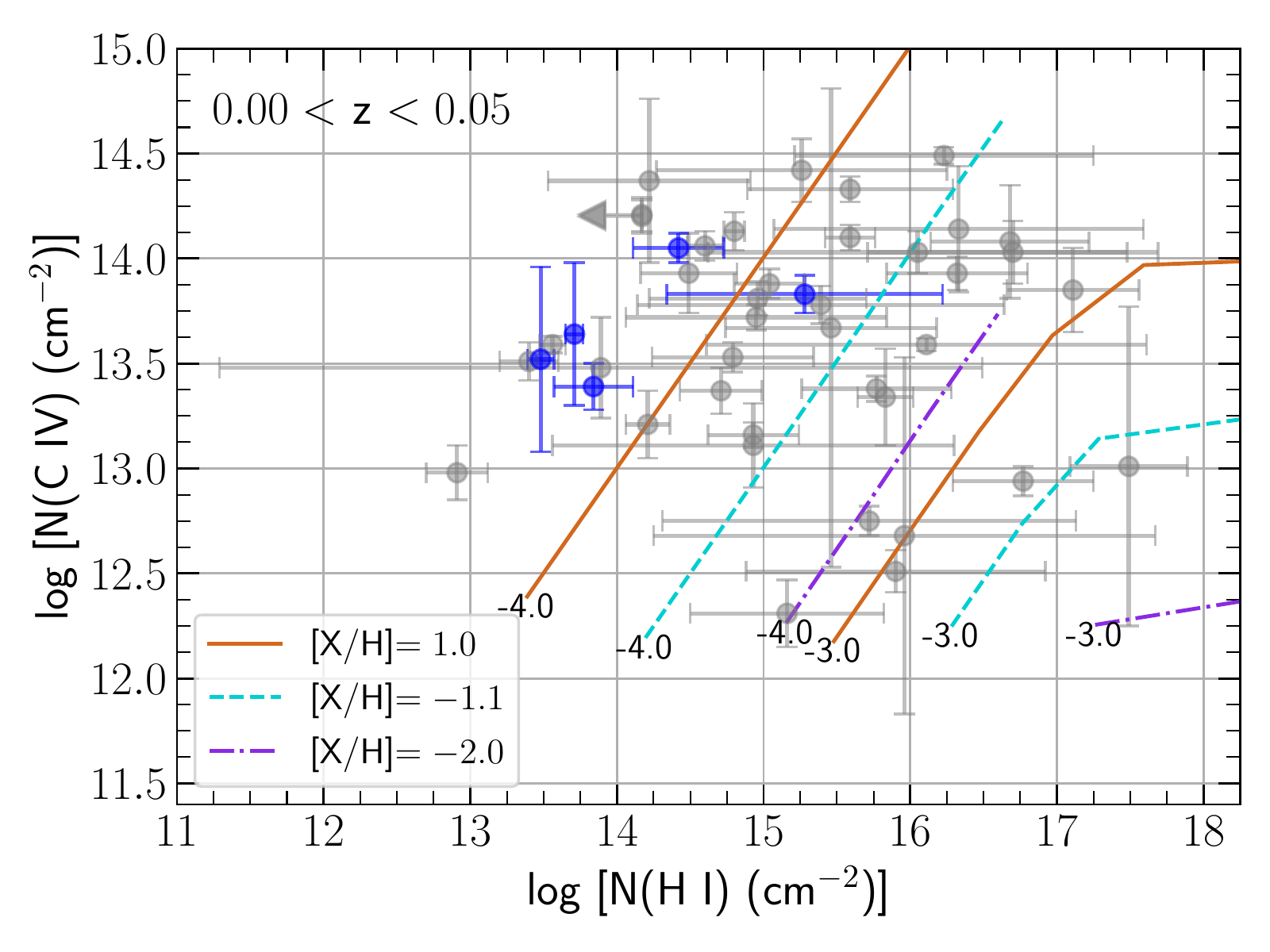}
        \includegraphics[width=250pt,height=185pt,trim={0cm 0cm 0cm 0cm},clip=true]{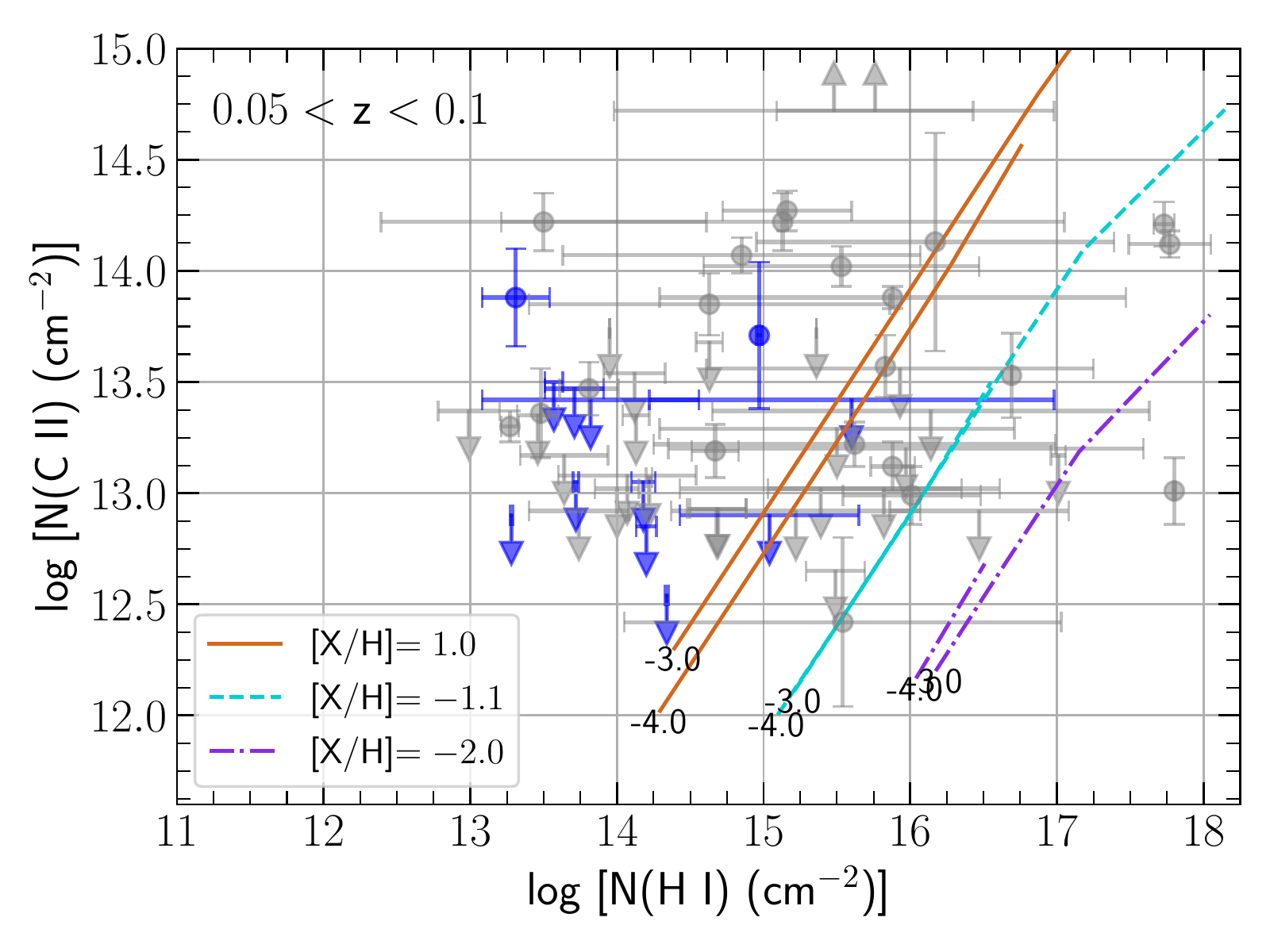}
        \includegraphics[width=250pt,height=185pt,trim={0cm 0cm 0cm 0cm},clip=true]{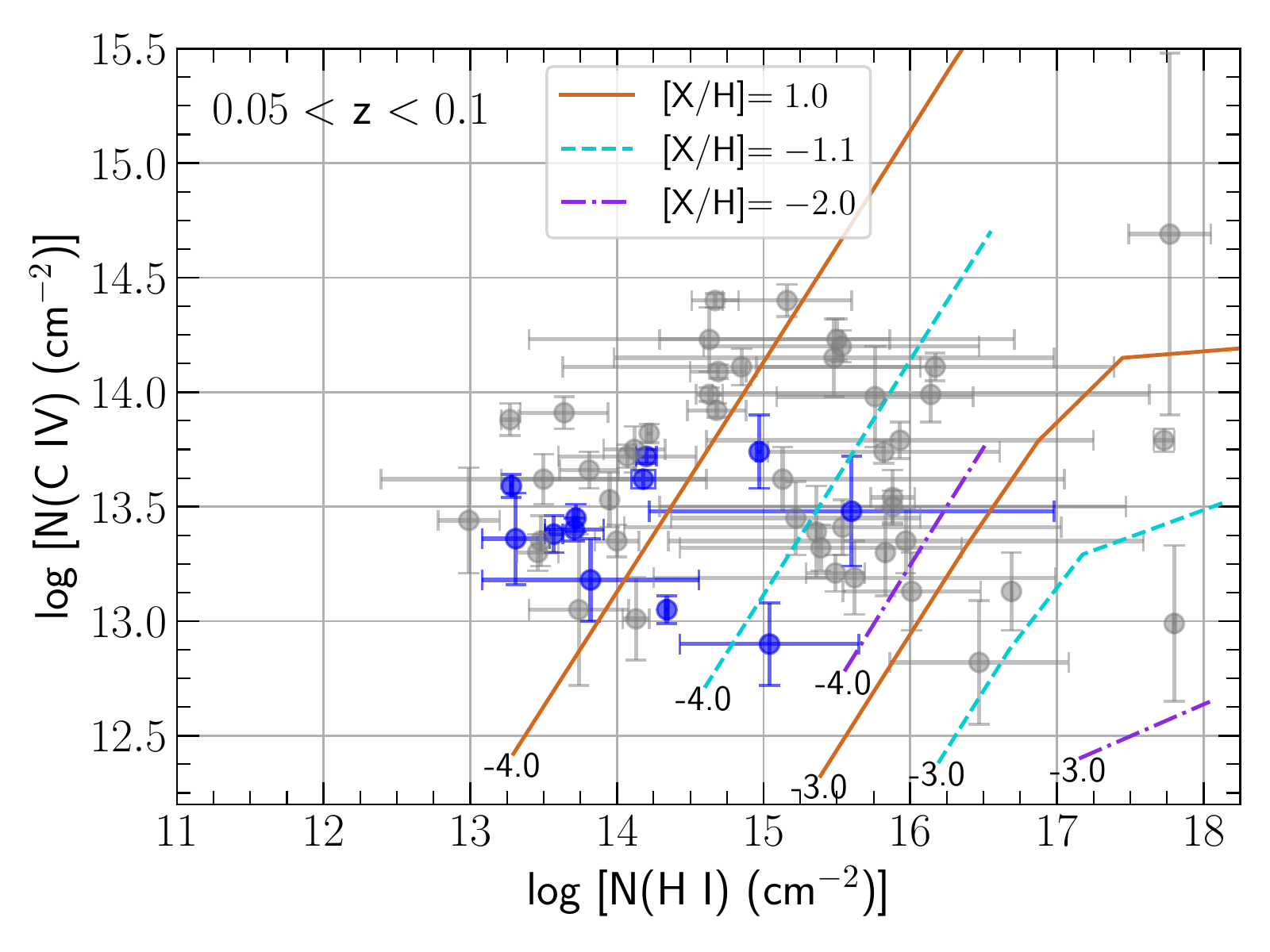}
        \includegraphics[width=250pt,height=185pt,trim={0cm 0cm 0cm 0cm},clip=true]{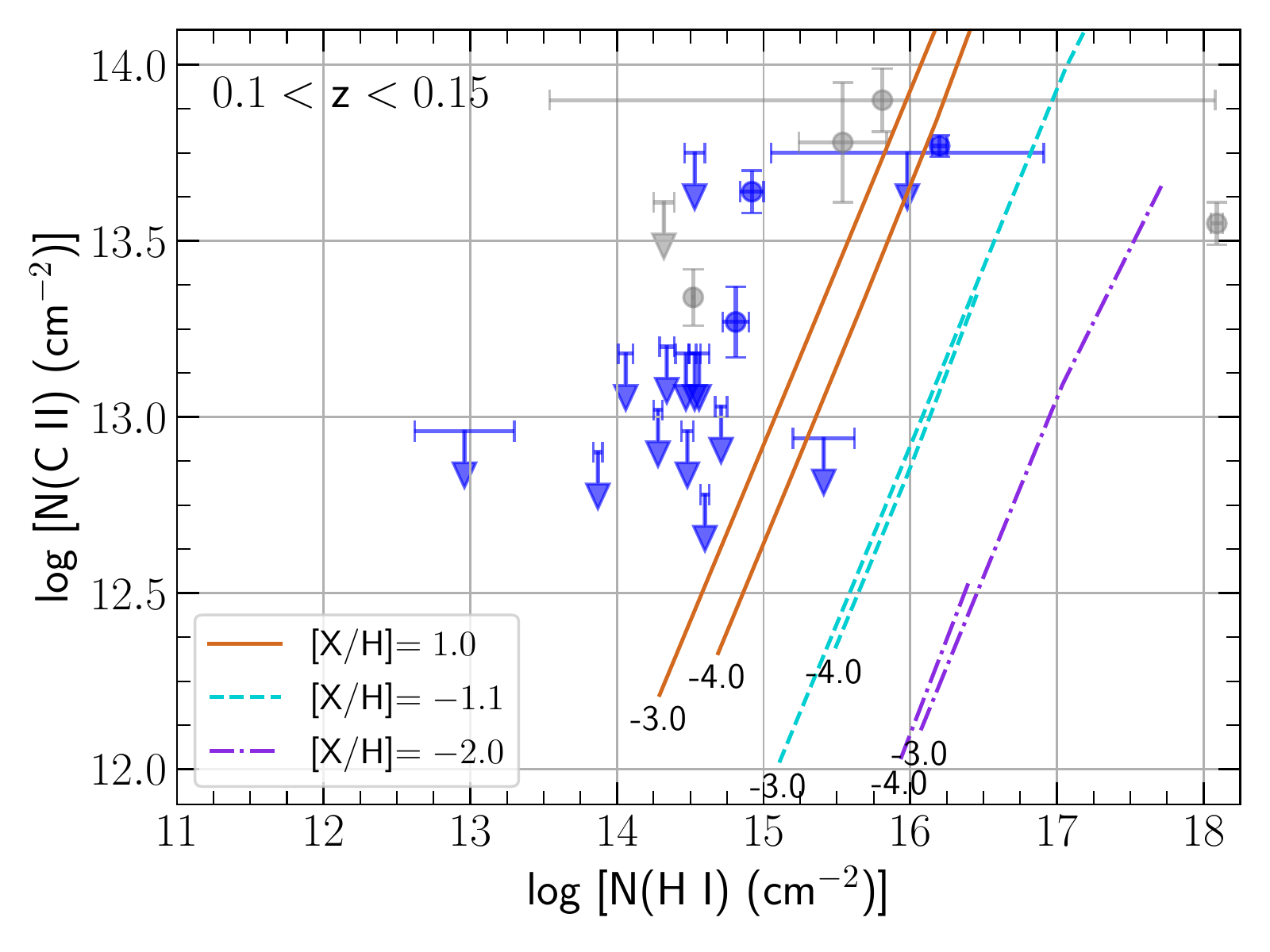}
        \includegraphics[width=250pt,height=185pt,trim={0cm 0cm 0cm 0cm},clip=true]{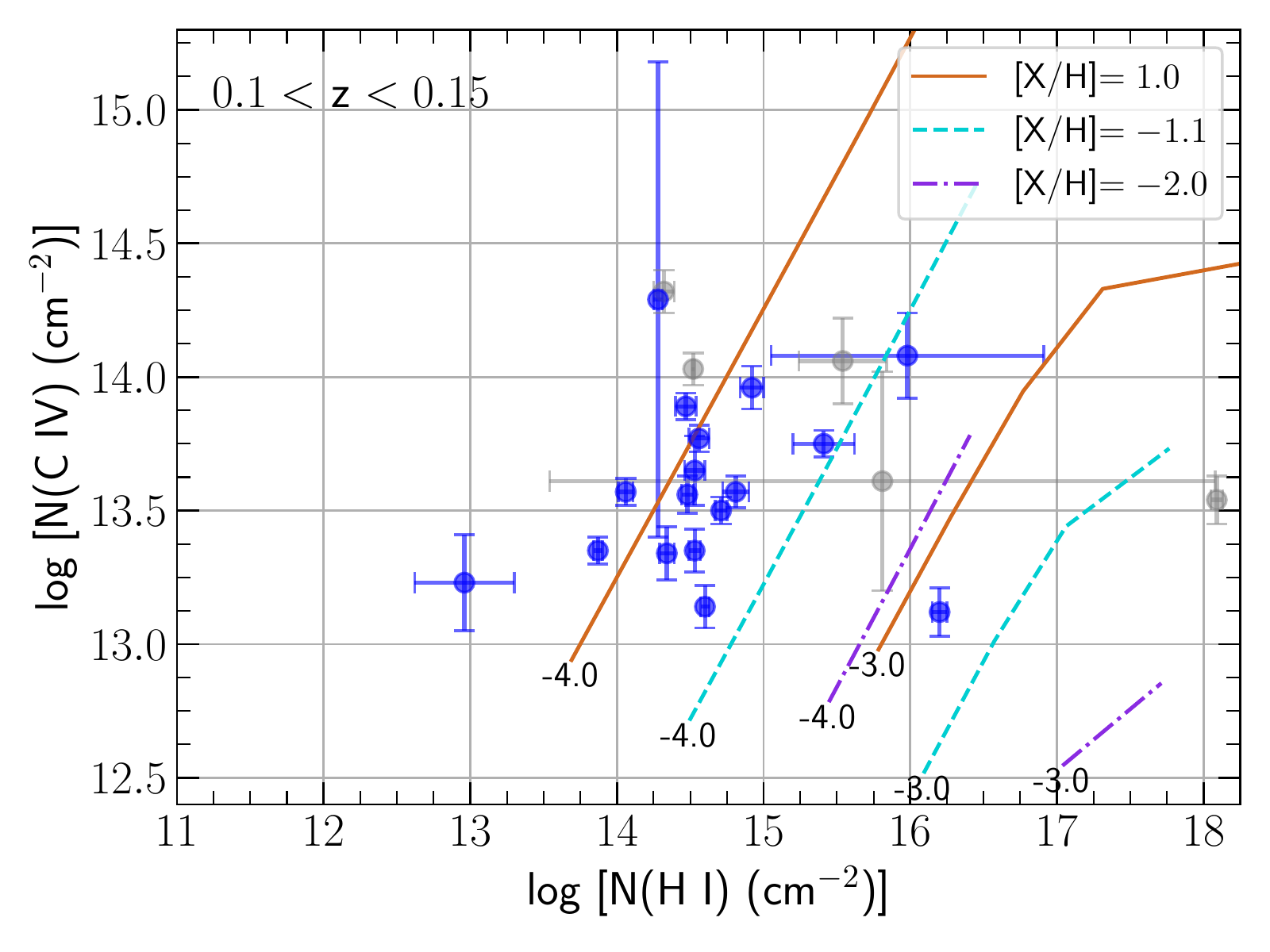}
    \caption{The panels shows the observed $\CII$ and $\CIV$ column densities against the corresponding $\HI$. The entire sample is divided into three redshift intervals. The blue data points represent the sub-sample that have the secure {\HI} measurements, and the grey points are systems in which the {\HI} measurements are based on saturated lines, with the uncertainty due to saturation taken into effect as explained in Sec.~\ref{nhidet}. The downward and upward pointing arrows are observationally determined upper and lower limits on the metal ion column densities based on non-detection and line saturation, respectively. The dashed-dot-purple, dashed-turquoise and solid-brown curves are photoionization equilibrium model predictions for metallicities of [X/H]~$=-2.0,-1.1,1.0$, respectively, and two different gas densities for each ($\log n_\mathrm{H}$ is mentioned at the end of each curve).}
    \label{13}
    \end{center}
\end{figure*}
The absorption systems are subjected to photoionization modelling to determine chemical and ionization properties of the clouds. We use \textsc{\large cloudy} v13.03 \citep{Ferland2013}, an openly accessible spectral synthesis code\footnote{\url{https://www.nublado.org}}. The models are based on the following assumptions for the clouds: (1) static (i.e. no expansion), (2) spatially invariant temperature for a given total hydrogen number density, (3) plane-parallel geometry, (4) covering factor of unity, (5) constant hydrogen density throughout the cloud, and (6) no dust content. The clouds are simulated to be photoionized by the extragalactic UV background radiation (EBR) at their corresponding redshifts. For this study, we adopted the EBR model as provided by \citet{Khaire2019} (hereafter KS19) instead of the \citet{Haardt2012} model used in previous studies of absorption systems. The KS19 ionizing background model is based on revised estimates of {\HI} distribution in the IGM \citep{Inoue2014}, cosmic star formation rate density, FUV extinction from dust \citep{Khaire2015a} as well as recent estimates of QSO emissivities \citep{Khaire2015b}. Unlike the \citet{Haardt2012} background, this model is also consistent with the recent measurements of photoionization rates at low redshifts ($z < 0.5$) \citep{Shull2015,Gaikwad2016}. The relative abundances of heavy elements are taken as solar from \citet{Asplund2009}. 

The {\HI} column density is reliably constrained only in $32\%$ of the absorbers (22/69) with line saturation hampering measurements in the remaining systems. In these well measured cases, metallicity estimates can be arrived at through ionization models. Several of the {\CIV} absorbers also have {\CII} detected. The observed {\CII} to {\CIV} ratio can be used to constrain density, assuming a single phase. In place of analyzing each absorber separately, we resort to determining the approximate range of density and metallicity in these absorbers. The sub-sample used for determining the density ranges constitutes the systems with information on {\CII} (measurement based on a detection or upper limit based on a non-detection). In addition, we also consider {\SiII} to {\SiIV}, whose information is available in a fair number of systems. For constraining [C/H] we include all systems. The results from both of these are discussed below.   

To constrain the density range of our {\CIV} sample, we generate a grid of ionization models for [X/H]~$=0$, $-5.5~\leq \log [n_{\H}~(\cc)]~\leq -1.0$, and $14.0~\leq \log [N(\textrm{\HI})~(\cmsq)]~\leq 17.0$; in steps of $0.5$~dex in density and total hydrogen column density. The observed column densities of ions of similar elements are compared with the grid of model predictions. Three different sets of model grids are computed for EBR corresponding to redshift intervals of $\Delta z = 0.05$ between $0 \leq z \leq 0.15$. The purpose of this exercise is to identify the range of densities traced by the low ions ({\CII}, {\SiII}) and {\CIV}. The advantage of this approach is that the $N(\HI)$, which is affected by line saturation in many instances, is not a constraining parameter, as we explain below. 

The model predictions are shown in Fig.~\ref{11} and Fig.~\ref{12}. It can be seen that in absorbers where {\CII} and/or {\SiII} is detected, models with $-4.0 \lesssim \log[n_{\H}~(\cc)] \lesssim -3.0$, encompass the observed {\CII} to {\CIV} and {\SiII} to {\SiIV} ratios. When {\CII} (or {\SiII}) is a non-detection, the density can be significantly lower. The narrow range of density indicates that the low ions and {\CIV} in these absorbers may not be tracing vastly different phases. This does not imply that the absorbing medium will have uniform ionization throughout. Instead, the absorbers could be probing an unresolved multiphase structure of kinematically overlapping zones contributing to both low and high ion absorptions. Previous studies of 
{\CII} and {\SiII} absorbers with associated {\CIV} suggest a photoionized two-phase structure with the low-ions coming from a dense and compact region ($n_{\H} \gtrsim 10^{-3}~\cc$, $\sim$ parsec), and the high ions traced by {\CIV} coming from a more diffuse and extended region ($n_{\H} < 10^{-3}~\cc$, $\sim$~few hundred pc to kpc \citep[e.g.][]{Rigby2002,Ding2003,Pradeep2020}. Such a scenario could be true for some of the absorbers in our sample as well. However, unlike {\OVI} which often brings out the presence of warm collisionally ionized gas, the {\CIV} and low ions still constitute photoionized plasma of some narrow spread of densities.

We note that six absorbers in our sample have {\OVI} detected\footnote{$z_\mathrm{abs} = 0.11391$ towards 3C 263, $z_\mathrm{abs} = 0.13842$ towards LBQS 1435-0134, $z_\mathrm{abs} = 0.13850$ towards PG 1116+215, $z_\mathrm{abs} = 0.12360$ towards PG 1216+069, $z_\mathrm{abs} = 0.12119$ towards PG 1424+240, and $z_\mathrm{abs} = 0.14702$ towards PG 1424+240}, three having corresponding {\CII}. In $z_\mathrm{abs} = 0.13850$ towards PG 1116+215, the {\CII} and {\CIV} have comparable and narrow $b$-values of $\sim 11~{\kms}$ whereas the broader $b \sim 35~{\kms}$ of {\OVI} clearly suggests that it originates from a much hotter gas phase. The aligned {\CII}, {\CIV} and {\OVI} in the $z_\mathrm{abs} = 0.12119$ towards PG 1424+240 have $b$-values of $\sim 15~\kms$, $\sim 23~\kms$ and $\sim 21~\kms$, respectively, which suggests that {\CIV} and {\OVI} might be tracing a similar phase which is hotter than what is traced by {\CII}. The $z_\mathrm{abs} = 0.14702$ towards PG 1424+240 has {\OVI} components that are aligned with corresponding {\CII} and {\CIV}. For the lower velocity component, $b(\CII)\sim16~\kms$, $b(\CIV)\sim12~\kms$ and $b(\OVI)\sim51~\kms$, which clearly shows {\OVI} being in a much hotter phase than the other two species. For the other component though, the three ions seem to trace similar phases based on $b(\CII)\sim16~\kms$, $b(\CIV)\sim17~\kms$ and $b(\OVI)\sim18~\kms$. We would like to remind here again that absorptions narrower than the resolution, associated with cooler gas, may have been missed. Modelling of individual absorbers in greater detail is needed to unveil the presence of such multiple gas phases.

The photoionization grid plots in Fig.~\ref{11} and Fig.~\ref{12} were generated for a fiducial metallicity of solar. The $N(\HI)$ given by the models for a certain {\CII} to {\CIV} ratio, and {\SiII} to {\SiIV} ratio corresponds to this assumed solar metallicity. The true metallicity being different does not impact the density solution obtained from the grid plots. A metallicity that is lower or higher than solar would only cause the grid to shift to the left or right, respectively, parallel to the $N(\HI)$ axis, leading to no significant change in the density solution. 

To estimate the range of [C/H] for the {\CIV} sample, a similar suite of ionization models were generated for metallicities ranging from $-2.0\le $~[X/H]~$\le1.0$ in steps of 0.3~dex, each for total hydrogen column densities in the range $15.0\le\log[ N(\H)~(\cmsq)]\le20.0$ in steps of 0.5~dex, and densities of $\log[n_{\H}~(\cc)] = -5.0, -4.0,-3.0$. These suite of models were separately computed for EBR corresponding to the three redshift bins. The panels in Fig.~\ref{13} show the models that are relevant for the data. It can be seen that in the sub-sample with secure {\HI} measurement (blue filled circle points), metallicities have to be predominantly [C/H]~$\gtrsim 0$ and in many cases significantly super-solar ([C/H]~$\gtrsim  +1.0$) to explain the observed {\CIV} and {\CII}. A similar trend is also seen for the Si ions (figure not included). This is expected as the \textit{secure} sample constitutes systems that suffer little saturation in $\Lya$ and/or $\Lyb$ and therefore possess the smallest {\HI} column densities amongst the full sample. These highly metal enriched systems are interesting as they could be direct tracers of supernova or AGN outflow material. Absorbers with higher {\HI} column densities ($N(\HI) \gtrsim 10^{15}~{\cmsq}$) tend to have lower metallicities in the range of $-2.0 < $~[C/H]~$< 0$. The lower limit on the metallicity is not well determined due to the uncertainty in the {\HI} in the highly saturated systems. Nonetheless, it is evident that the metal ion column densities for the overall sample span a range of $\sim 2$~dex, whereas the {\HI} has a spread of $\sim 5$~dex or more. Thus, for a given metal ion column density, depending on the amount of {\HI} along the sightline, the metallicity estimate can come out as high or low. The metallicity differences could be a result of incomplete mixing of metals with the ambient CGM or IGM environment. \citet{Schaye2007} infer from their study of {\CIV} absorbers that metals displaced from galaxies are predominantly confined to patchy zones. Simulations have also suggested artificial overlap of lines from {\CIV} clouds and $\Lya$ forest due to high velocity widths of the former \citep{Cen2011}. Lines of sight studies using binary and lensed quasar systems have found transverse sizes as low as $300$~pc for {\CIV} systems \citep{Rauch2001}. If this is the case, then absorption line studies may yield metallicities that are anywhere in a broad range depending on the filling factor of metals in the overall diffuse medium. 

\section{GALAXY ENVIRONMENTS}\label{galaxies}

In this section we describe the distribution and properties of galaxies in the extended environments surrounding the absorbers. The {\CIV} absorption strength for CGM clouds around $z\le 0.1$ galaxies is found to decline with increasing impact parameter \citep{Bordoloi2014} with the covering fraction being nearly $100\%$ within $\rho/R_\mathrm{vir} = 0.2$ for strong {\CIV} absorbers. \citet{Burchett2016} find more {\CIV} absorption from clouds within virial radii of galaxies with $M_\mathrm{r}<-19$, with a higher incidence around galaxies that occupy low density environments. 

Using archival data from Data Release 16 of the Sloan Digital Sky Surveys (SDSS) \citep{Ahumada2019}, we search for galaxies within projected separation $\rho < 1.5$~Mpc and line-of-sight velocity separation $|\Delta v| < 1000$~$\kms$ with respect to the systems, values that are characteristic of the radius and the radial velocity dispersion of galaxies in rich clusters, but larger than what is typical of groups \citep{Bahcall1999}. Nevertheless, clusters and groups of galaxies are understood to be natural extensions of even larger scale structures and therefore a search window that is not truncated at the cluster scale will be useful to explore the large scale environment around these absorbers. We use Ned Wright's Cosmology Calculator\footnote{\url{http://www.astro.ucla.edu/~wright/CosmoCalc.html/}} \citep{Wright2006} for converting between $\rho$ and angular separation assuming a $\Lambda$CDM universe. The virial radius of a galaxy ($R_\mathrm{vir}$) is arrived at using the scaling relationship $R_\mathrm{vir} = 250(L/L^\star)^{0.2}$~kpc given by \citet{Prochaska2011}. For this, the $L/L^\star$ is calculated using $M_\mathrm{r}$ of the galaxy and the Schecter absolute magnitude $M^\star_\mathrm{r}=-20.44+5\log h$ \citep{Blanton2003lf}, where $h=0.696$ for our adopted cosmology. We take note of the fact that, compared to the more detailed ``halo abundance matching'' method of estimating $R_\mathrm{vir}$ by combining theoretical galaxy halo mass functions with observed galaxy luminosity functions, the Prochaska et al. scaling relationship yields larger virial radii for galaxies with sub-$L^\star$ luminosity (see Fig. 1 of \citealt{Stocke2013} for a comparison). While determining the absolute magnitudes of the galaxies, appropriate K-corrections are applied using the analytical expression given by \citet{Chilingarian2010}\footnote{\url{http://kcor.sai.msu.ru/}}.

\begin{figure}
   \begin{center}
        \includegraphics[width=214pt,height=156pt,trim={0.3cm 0cm 0.15cm 0.3cm},clip=true]{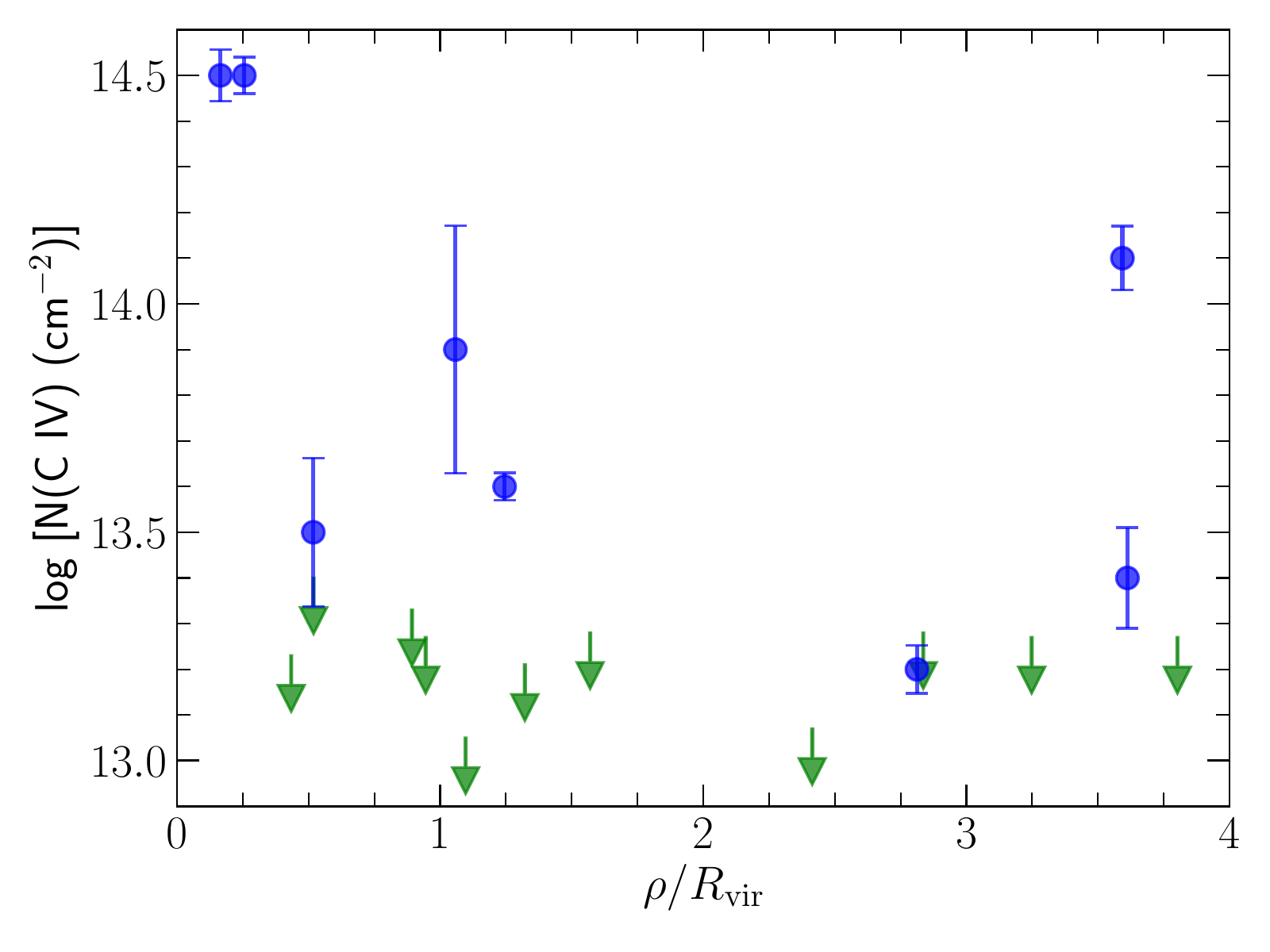}\quad
        \includegraphics[width=214pt,height=156pt,trim={0.3cm 0cm 0.15cm 0.3cm},clip=true]{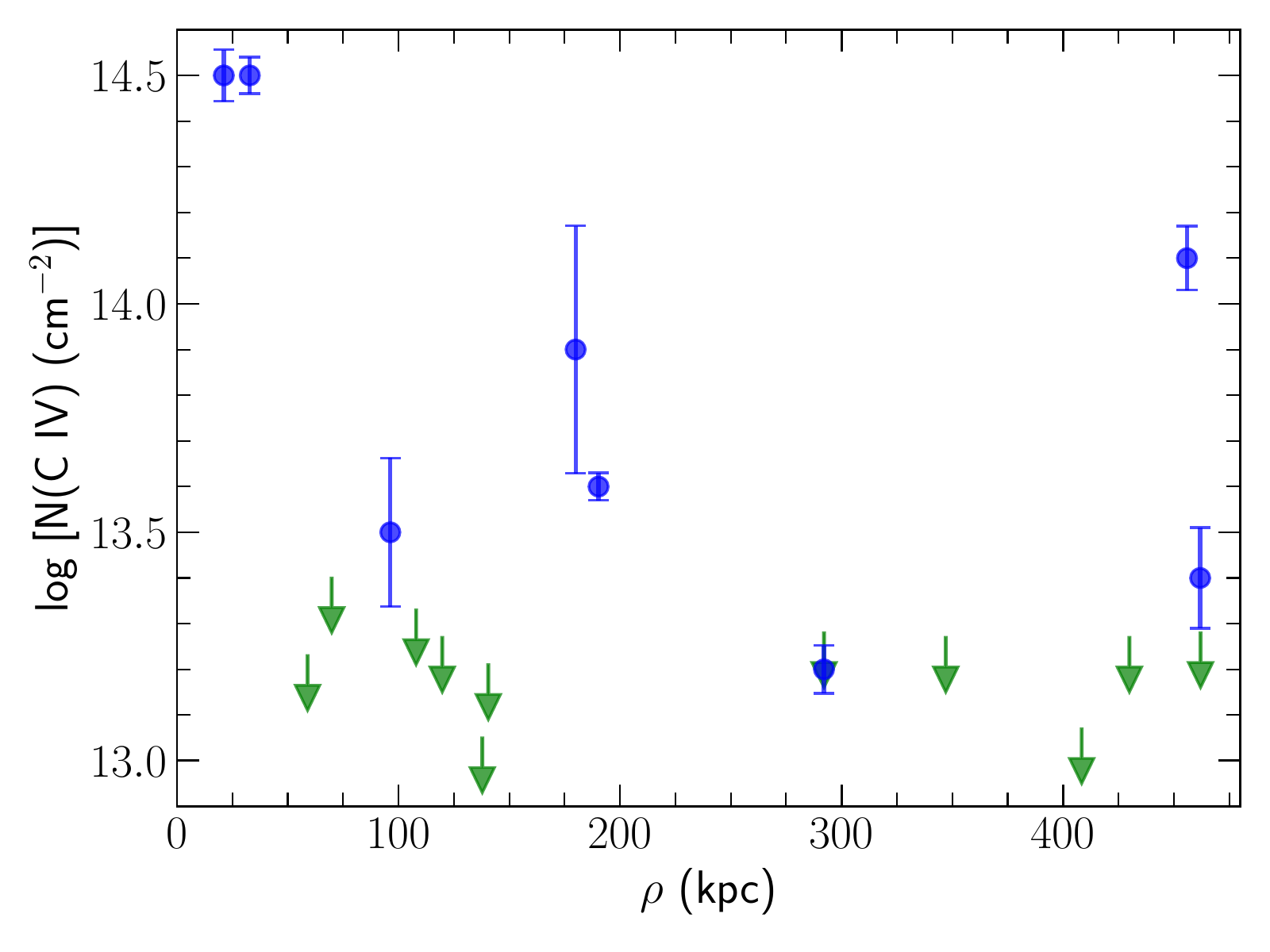}\quad
    \caption{Total {\CIV} column density in the absorbers at $z_\mathrm{abs}<0.015$ against the normalized impact parameter (top panel) and impact parameter (bottom panel). The x-values in the top panel correspond to galaxies with $L>0.01L^\star$ that are closest in $\rho/R_\mathrm{vir}$. Those in the bottom panel correspond to galaxies closest in $\rho$. The upper limits are the integrated column densities for {\CIV}~$\lambda1548$ determined over $102~\kms$ centered at redshifts of the galaxies with $L>0.01L^\star$ that lie within $\rho<500$~kpc of targeted sightlines.}
    \label{16}
    \end{center}
\end{figure}

\begin{table*}
\caption{Covering fractions of the {\CIV} absorbers for $\log~[N(\textrm{\CIV})~(\cmsq)]\geq13.5$.}
\begin{center}
\begin{threeparttable}
\begin{tabular}{lccccc}
\hline
Selection Bin & Luminosity Threshold & Redshift Range & Number of Detections & Total Number & Covering Fraction \\
\hline
$\rho/R_\mathrm{vir}<1$ & $L>0.01L^\star$ & $z<0.015$ & $3$ & $7$ & $43\%$ \\
$1<\rho/R_\mathrm{vir}<2$ & $L>0.01L^\star$ & $z<0.015$ & $2$ & $5$ & $40\%$ \\
$2<\rho/R_\mathrm{vir}<3$ & $L>0.01L^\star$ & $z<0.015$ &$0$ & $2$ & $0\%$ \\
$3<\rho/R_\mathrm{vir}<4$ & $L>0.01L^\star$ & $z<0.015$ &$1$ & $3$ & $33\%$ \\
$\rho/\mathrm{kpc}<160$ & $L>0.01L^\star$  & $z<0.015$ &$3$ & $9$ & $33\%$ \\
$\rho/\mathrm{kpc}<100$ & $L>0.01L^\star$ & $z<0.015$ &$3$ & $5$ & $60\%$  \\
$100<\rho/\mathrm{kpc}<200$ & $L>0.01L^\star$ & $z<0.015$ &$2$ & $6$ & $33\%$ \\
$200<\rho/\mathrm{kpc}<300$ & $L>0.01L^\star$ & $z<0.015$ &$0$ & $1$ & $0\%$ \\
$300<\rho/\mathrm{kpc}<400$ & $L>0.01L^\star$ & $z<0.015$ &$0$ & $1$ & $0\%$ \\
$400<\rho/\mathrm{kpc}<500$ & $L>0.01L^\star$ & $z<0.015$ &$1$ & $4$ & $25\%$ \\
N$_\mathrm{gal}\leq7$ & $L>0.13L^\star$ & $z<0.051$ & $7$ &$9$ & $78\%$\\
N$_\mathrm{gal}>7$ & $L>0.13L^\star$ & $z<0.051$ & $0$ & $5$ & $0\%$\\
\hline
\label{covfrac}
\end{tabular}
\end{threeparttable}
\end{center}
\end{table*}

We begin by exploring the absorber-galaxy connection. Although SDSS footprint covers regions around 55/69 absorbers, these systems span a wide redshift range and it becomes harder to detect fainter galaxies at higher distances. The galaxy spectroscopic data is $99\%$ complete down to the $r$-band apparent magnitude $m_\mathrm{r} \leq 17.77$ \citep{Strauss2002}. Since it is important to include faint dwarf galaxies in the analysis, we impose a galaxy luminosity cut of $L>0.01L^\star$ (consistent with \citealt{Burchett2016}). This also implies a redshift cut of $z_\mathrm{abs}<0.015$ for a uniform completeness of $99\%$ above our adopted luminosity threshold for the whole redshift range. Only 8 absorbers in our sample satisfy this criteria. For each of these absorbers, we find the associated galaxy using two approaches where we either select by $\rho/R_\mathrm{vir}$ or $\rho$. We consider all galaxies with $L>0.01L^\star$ that are separated at $|\Delta v|<600~\kms$, and then select the one closest in $\rho/R_\mathrm{vir}$ for the former, and $\rho$ for the latter. In cases where we find two such galaxies with same $\rho/R_\mathrm{vir}$ or $\rho$, we select the one with smaller $|\Delta v|$. We also use the known targeted sightlines to determine upper-limits for non-detections. For this, we identify targets with $z<0.015$, search for $L>0.01L^\star$ galaxies within our adopted window around the redshifts, find the galaxies amongst these that are separated at $\rho<500$~kpc with respect to the sightline, and determine integrated {\CIV}~$\lambda1548$ column density over $\pm51~\kms$ (as $102~\kms$ is the mean $\CIV$ absorption width in our sample) about the galaxies' spectroscopic redshift. 

Fig.~\ref{16} shows no significant trend between the {\CIV} column density and $\rho$ or $\rho/R_\mathrm{vir}$ of the nearest galaxy, similar to the findings of \citet{Chen2001} and \citet{Borthakur2013}. Although, this also bears the caveat of low number statistics. On considering a detection threshold of $\log[N(\textrm{\CIV})~(\cmsq)]\geq13.5$, we obtain covering fractions for different radial bins of $\rho/R_\mathrm{vir}$ and $\rho$. The values are summarized in Table~\ref{covfrac}. While the covering fraction for $\rho/R_\mathrm{vir}<1$ ($43\%$) is within the uncertainty of the value reported by \citet{Burchett2016}, we find a significantly higher value for $1<\rho/R_\mathrm{vir}<2$ at $40\%$. For the galaxies selected by $\rho$, we find that our covering fractions are consistent with \citet{Burchett2016} for all bins except $\rho<100$~kpc, where we find a value that is higher than what is allowed by the uncertainty. There is a hint of decline in covering fraction with impact parameter and $\rho/R_\mathrm{vir}$, which is in line with previous metal absorber-galaxy studies \citep[e.g.][]{Chen2001,Borthakur2013,Bordoloi2014,Liang2014}. However, it is seen to increase for $3<\rho/R_\mathrm{vir}<4$ and $300<\rho/\mathrm{kpc}<400$. This suggests that strong {\CIV} absorbers are not necessarily coincident with galaxies above our adopted luminosity threshold, but we cannot rule out the presence of undetected faint dwarf galaxies.

\begin{figure}
    \begin{center}
        \includegraphics[width=220pt,height=160pt,trim={0.3cm 0cm 0.15cm 0.3cm},clip=true]{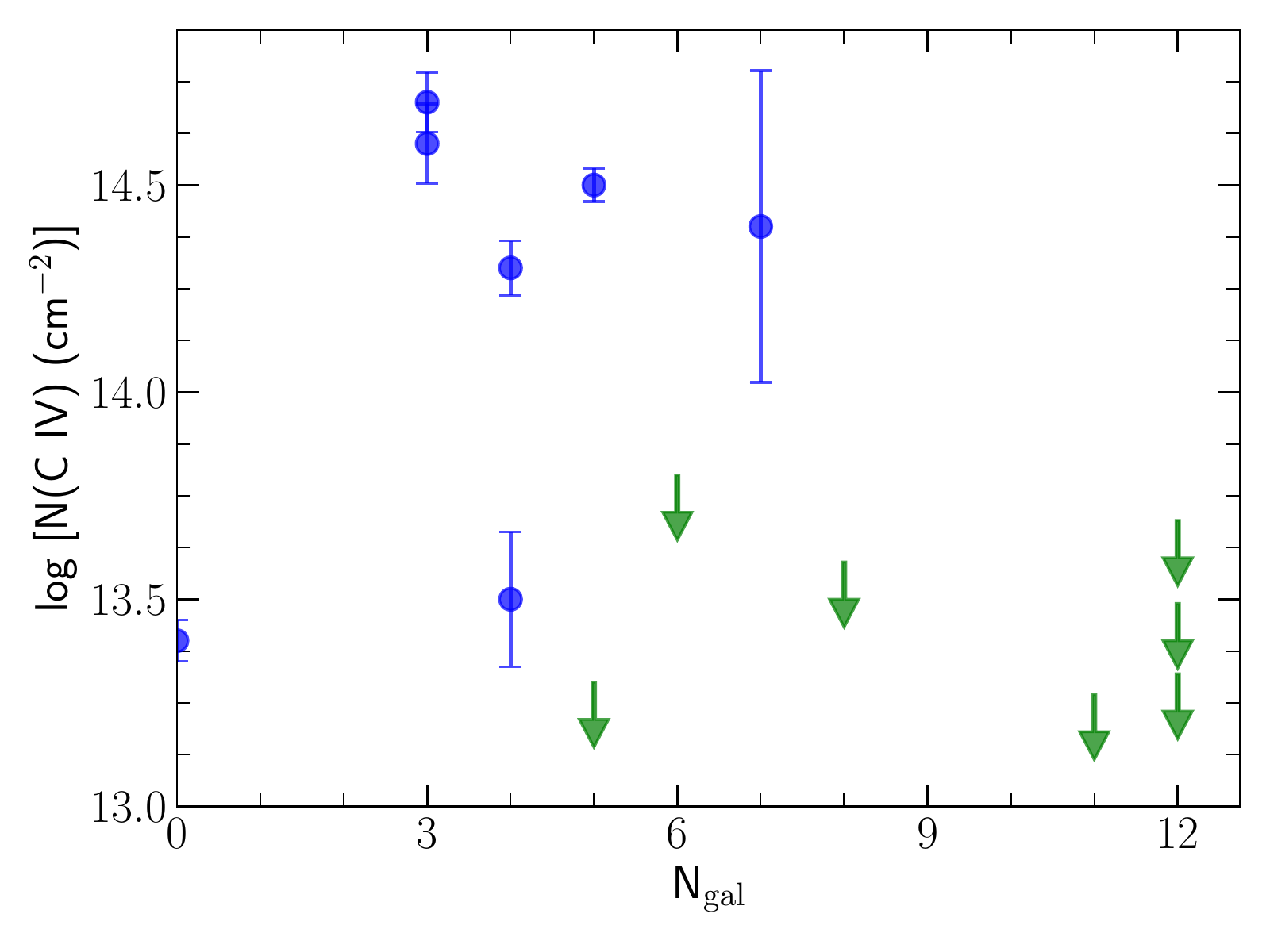}
    \caption{Total {\CIV} column density in the absorbers at $z_\mathrm{abs}<0.051$ against the number of $L>0.13L^\star$ galaxies within $\rho<1.5$~Mpc and $|\Delta v|<1000~\kms$. The detections here are systems where there is at least one $L>0.13L^\star$ galaxy within $\rho<160$~kpc and $\Delta v<600~\kms$. The upper limits are the integrated column densities for {\CIV}~$\lambda1548$ determined over $102~\kms$ centered at redshifts of the galaxies with $L>0.13L^\star$ that lie within $\rho<160$~kpc of targeted sightlines.}
    \label{15}
    \end{center}
\end{figure}

We examine the dependence of absorber strength on its extended environment by looking at relationships against number of galaxies within the search window adopted by us i.e. $\rho < 1.5$~Mpc and $|\Delta v| < 1000~\kms$. We impose a galaxy luminosity threshold of $L>0.13L^\star$, corresponding to the $M_\mathrm{r}<-19$ cut adopted by \citet{Burchett2016}. We refer to number of such galaxies within the search window as $N_\mathrm{gal}$. The luminosity cut translates to a redshift cut of $z_\mathrm{abs}<0.051$ for $99\%$ completeness above $L>0.13L^\star$ over this redshift range. Amongst the systems that satisfy this criteria, we only select those that have at least one $L>0.13L^\star$ galaxy within $|\Delta v|<600~\kms$ and $\rho<160$~kpc (typical $R_\mathrm{vir}$ of a $0.13L^\star$ galaxy). The upper limits are determined by identifying targets at $z<0.051$, searching for $L>0.13L^\star$ galaxies around the redshift, and determining the integrated column density over $102~\kms$ region in the sightline around the redshift of any galaxy that has $\rho<160$~kpc. Fig.~\ref{15} shows that no absorber is found with $N_\mathrm{gal}>7$ and {\CIV} seems to favour lower density environments, which is qualitatively consistent with \citet{Burchett2016}. In fact, the covering fraction for $\log~[N(\textrm{\CIV})~(\cmsq)]\geq13.5$ shows a sharp decline from $78\%$ for $N_\mathrm{gal}\leq7$ to $0\%$ for $N_\mathrm{gal}>7$ (see Table~\ref{covfrac}). This seems to suggest that presence of {\CIV} in circumgalactic space somehow depends on the large scale environment.

\section{SUMMARY}\label{summary}

We have presented results from a survey of 69 intervening {\CIV} absorbers at $z < 0.16$ identified in the far-UV spectra of 223 quasars from the Hubble Spectroscopic Legacy Archive. The main results are as follows

\begin{enumerate}
    
    \item About $83\%$ (57/69) of the absorption systems have a kinematically
    simple {\CIV} profile with only one or two absorbing components. Complex absorption profiles with more than two components in {\CIV} or in {\HI} are rare at \textit{HST}/COS resolution. 
    
    \item The $\Delta v_{90}$ velocity spreads of {\CIV} and {\HI} profiles have similar correlations and scaling relations with their respective rest-frame equivalent widths and column densities indicating that the {\HI} and {\CIV} ions trace gas of similar kinematic conditions.  
    
    \item For a sub-sample of 22 absorbers with well constrained {\HI} parameters, the ionization model independent temperatures derived using $b(\HI)$ and $b(\CIV)$ suggest $T < 10^5$~K, with the peak of the distribution at  $T \sim 10^{4.2}$~K, suggesting that {\CIV} absorbers might be predominantly tracing photoionized plasma with nearly equal thermal and non-thermal broadening. However, in the full sample, there are 74 {\CIV} components with $b \geq 12~{\kms}$ corresponding to temperature upper limits of $T \geq 10^5$~K, where ionizations via collisions become efficient. 
    
    \item The contribution of {\CIV} absorbers to the closure density is estimated as $\Omega_{\CIVm} = (8.01\pm1.62) \times 10^{-8}$ for $z < 0.16$; consistent with previous studies. The $\Omega_{\CIVm}$ is also inline with the steady, but slow increase with decreasing redshift seen for C IV absorbers from $z=5$ to the present. The {\CIV} absorbers constitute only a meager fraction of the baryons traced by narrow photoionized {\HI} in the low-$z$ Universe ($\Omega_{\HIm} = 0.013 {\pm} 0.005$,~ \citealt{Shull2012}). 
    
    \item From the mass-density of {\CIV}, we estimate a metallicity of $(2.07\pm~0.43)\times~10^{-3}$~Z$_\odot$ for the diffuse gas in the $z < 0.16$ Universe, which is an order of magnitude more than the metal abundance in the IGM at high redshifts ($z \gtrsim 5$). 
    
    \item {\CIV} absorbers with detections of low ions ({\CII}, {\SiII}) are consistent with photoionized gas with density constrained to a narrow range of $-4.0 \lesssim \log[n_{\H} ({\cc})] \lesssim -3.0$. In absorbers where the low ions are non-detections, the densities can be lower. In those former cases, instead of the {\CII} and {\CIV} coming from a single-phase photoionized plasma, it is more likely that the absorber is tracing an unresolved multiphase structure contributing both low and high ions.
    
    \item The column densities of {\CIV} (and lower ions) components span a range of $\sim 2$~dex, whereas those of the \textit{associated} {\HI} components have a spread of $\sim 5$~dex or more, resulting in metallicities that span a wide range from $-2.5 \lesssim$~[C/H]~$ \lesssim +1.5$, with the lower {\HI} column density absorbers having super-solar values. The range \textit{could} be an indication that we are detecting metal enriched clouds released into the CGM and IGM through galactic outflows before they get homogeneously mixed with the surrounding environment with diverse $\HI$.
    
    \item For $z<0.015$, we find a tentative evidence for decline of covering fraction for strong {\CIV} absorption ($N(\CIV)>10^{13.5}~\cmsq$) with both the impact parameter and the normalized impact parameter. Although, a rise in the covering fraction at high separations suggests that strong {\CIV} absorption is not necessarily coincident with $L>0.01L^\star$ galaxies.
    
    \item For $z<0.055$, we do not find any absorber in regions having $>7$ galaxies with $L>0.13L^\star$, even though there is at least one galaxy within $160$~kpc of the sightline. The covering fraction for $N(\CIV)>10^{13.5}~\cmsq$ systems falls from $78\%$ to $0\%$ for $N_\mathrm{gal}\leq7$ to more denser regions. The presence of metal-rich clouds in the circumgalactic medium of galaxies seems to be dependent on the large scale environment.

\end{enumerate}

\section*{ACKNOWLEDGEMENTS}
We acknowledge the work of people involved in the design, construction and deployment of the COS onboard \textit{HST}. We wish to extend our thanks to all those who had carried out data acquisition through FUV observations towards the sightlines used in this survey. This research has made use of the HSLA database, developed and maintained at STScI, Baltimore, USA \citep{Peeples2017}. The spectra analyzed in this paper are based on observations obtained with the NASA/ESA \textit{HST}, which is operated by the Association of Universities for Research in Astronomy, Inc. under NASA contract NAS5-2655. We are grateful to Gary Ferland and collaborators for developing the \textsc{\large cloudy} photoionization code. AM would like to thank Vikram Khaire for helping with generation of the KS19 EBR models that are used in this study. AM would also like to appreciate the efforts of \citet{Hunter2007} in development of the \textsc{\large matplotlib} package for \textsc{\large python} that was used to produce plots in this paper. AN acknowledges support for this work provided by SERB through grant number EMR/2017/002531 from the Department of Science \& Technology, Government of India.

\section*{Data Availability}
The spectral data used in this paper can be accessed through the Hubble Spectroscopic Legacy Archive\footnote{\url{https://archive.stsci.edu/missions-and-data/hst-spectroscopic-legacy-archive-hsla}}. The galaxy data are available at website for Data Release 16 of the Sloan Digital Sky Survey\footnote{\url{http://skyserver.sdss.org/dr16/en/tools/toolshome.aspx}}. The system plots and measurement tables are provided as supplementary material online at MNRAS.
\newline\newline Please note: Oxford University Press is not responsible for the content or functionality of any supporting materials supplied by the authors. Any queries (other than missing material) should be directed to the corresponding author for the article.




\bibliographystyle{mnras}
\bibliography{references} 

\section*{Supporting Information}
Supplementary data are available at \href{https://academic.oup.com/mnras/article/505/3/3635/6288433#supplementary-data}{\textit{MNRAS}} online.
\newline\newline Please note: Oxford University Press is not responsible for the content or functionality of any supporting materials supplied by
the authors. Any queries (other than missing material) should be directed to the corresponding author for the article.
\bsp	
\label{lastpage}
\end{document}